\documentclass[final,5p,times,twocolumn,authoryear]{elsarticle}

\usepackage{amssymb}

\usepackage{dutchcal}
\newtheorem{definition}{Definition}%
\newtheorem{theorem}{Theorem}
\newtheorem{statement}[theorem]{Statement}%
\usepackage{booktabs}
\usepackage{multirow}
\usepackage{amssymb} 
\usepackage{graphicx}
\usepackage{xcolor}
\usepackage{tikz}
\usetikzlibrary{shapes.geometric}
\usetikzlibrary{arrows,shapes,chains}
\usepackage{makecell}
\usepackage{soul} 
\usepackage{color, xcolor} 
\usepackage{ragged2e}

\usepackage{threeparttable} 
\usepackage{caption}
\usepackage{subcaption}
\usepackage{longtable}

\usepackage{hyperref}
\hypersetup{
        hidelinks,
	colorlinks=true,
	allcolors=black,
	pdfstartview=Fit,
	breaklinks=true
}
\usepackage[noabbrev,nameinlink]{cleveref}

\journal{Expert Systems with Applications}

\begin{document}

\begin{frontmatter}



\title{A Theoretical Review on Solving Algebra Problems}




\author[]{Xinguo Yu\corref{cor1}%
} 
\ead{xgyu@ccnu.edu.cn}

\author[]{Weina Cheng
}
\ead{wncheng@mails.ccnu.edu.cn}

\author[]{Chuanzhi Yang
}
\ead{canola@mails.ccnu.edu.cn}

\author[]{Ting Zhang
}
\ead{ting.zhang@ccnu.edu.cn}

\cortext[cor1]{Corresponding author at:  National Engineering Research Center for E-learning, Central China Normal University, Wuhan, 430079, China.}


\affiliation{organization={Central China Normal University},
            addressline={152 Luoyu Road,Hongshan District}, 
            city={Wuhan},
            postcode={430079}, 
            state={Hubei},
            country={China}}

\begin{abstract}
Solving algebra problems (APs) continues to attract significant research interest as evidenced by the large number of algorithms and theories proposed over the past decade. Despite these important research contributions, however, the body of work remains incomplete in terms of theoretical justification and scope. The current contribution intends to fill the gap by developing a review framework that aims to lay a theoretical base, create an evaluation scheme, and extend the scope of the investigation. This paper first develops the State Transform Theory (STT), which emphasizes that the problem-solving algorithms are structured according to states and transforms unlike the understanding that underlies traditional surveys which merely emphasize the progress of transforms. The STT, thus, lays the theoretical basis for a new framework for reviewing algorithms. This new construct accommodates the relation-centric algorithms for solving both word and diagrammatic algebra problems. The latter not only highlights the necessity of introducing new states but also allows revelation of contributions of individual algorithms – obscured in prior reviews without this approach. 

A review of AP solving algorithms (2014  to date) is subsequently enhanced by applying the STT  specifically designed to analyze individual and collective algorithms for states and transforms. Furthermore, the State Transform Analysis (STA) is a core function in the identification of progress in terms of states and transforms. Thirdly, the Perspective Confusion Comparison (PCC) is developed to extend the application of STA to add capabilities for systematic and individual evaluation of transforms, algorithms, and approaches. This is a new evaluation method of being different from other methods that can do more in-depth evaluation for decomposed algorithms for solving problems with multiple types of inputs.  Finally, this work identifies several research directions by extracting benefit from theoretical reviews. This work significantly contributes to the advancement of AP-solving by providing the mechanism for identifying and understanding the key contributors to building high-performance problem-solving algorithms at the three levels of transform, algorithm, and approach.

\end{abstract}


\begin{figure*}
     \centering
    \thispagestyle{empty}
    \tikzstyle{state} = [text centered, font=\fontsize{9}{10}\selectfont\bfseries]
    \tikzstyle{state2} = [rectangle, thick, rounded corners=0.5mm,minimum width = 6cm, minimum height=0.4cm,text centered, draw = black,font=\fontsize{9}{10}\selectfont\bfseries, fill=blue!10]
    \tikzstyle{state3} = [rectangle, thick,rounded corners, minimum width = 1cm, minimum height=0.8cm,text centered, draw = black,font=\fontsize{7}{7}\selectfont]
    \tikzstyle{state4} = [rectangle, thick,rounded corners, minimum width = 1cm, minimum height=0.5cm,text centered, draw = black,font=\fontsize{7}{7}\selectfont]
    \tikzstyle{state5} = [circle, thick,minimum height=0.6cm,text centered, draw = black,font=\fontsize{11}{12}\selectfont\bfseries, fill=blue!10]
    \tikzstyle{state6} = [rectangle, thick,rounded corners, minimum width = 1cm, minimum height=0.5cm,text centered, draw = black,font=\fontsize{7}{7}\selectfont]
    \tikzstyle{state7} = [rectangle, thick, minimum width = 18cm, minimum height=1cm,text centered, draw = black,font=\fontsize{9}{10}\selectfont\bfseries]
    \tikzstyle{state8} = [ circle, thick,minimum height=0.6cm,text centered, draw = black,font=\fontsize{11}{12}\selectfont\bfseries, fill=blue!10]
    \tikzstyle{state9} = [rectangle, thick, minimum width = 3cm, minimum height=1cm,text centered, draw = black,font=\fontsize{9}{10}\selectfont\bfseries]
    \tikzstyle{state10} = [rectangle, thick, minimum width = 5cm, minimum height=1cm,text centered, draw = black,font=\fontsize{9}{10}\selectfont\bfseries]
    \tikzstyle{arrow} = [->,>=stealth]
    \begin{tikzpicture}[node distance=1cm]
        \selectcolormodel{gray}
       \node[state ](A){  Graphical Abstract of "A Theoretical Review on Solving Algebra Problems"}; 
       \node[state2,right of = A,yshift = -0.8cm,xshift=-5.5cm](B){Aggregated Graph};
       \node[state5,right of = A,yshift = -0.8cm,xshift=-9cm](B){3};

       \draw [arrow,rounded corners, line width=8pt] (-1,-0.8) --  (0.8,-0.8);
       
       \node[state5,right of = A,yshift = -0.8cm,xshift=0.4cm](B){4};
       
       \node[state2,right of = A,yshift = -0.8cm,xshift=3.9cm](B){State Graph};
       \node[state3,right of = A,yshift = -7.5cm,xshift=-9.2cm](s0){$\mathcal{S}_{00}$};
        \node[state3,right of = s0,xshift=5.0cm](s2){$\mathcal{S}_{02}$};
        \node[state3,right of = s2,xshift=0.5cm](see){$\mathcal{S}_{os}$};
        \node[state3,right of = s0,xshift=3.5cm,yshift=0.8cm](s8){$\mathcal{S}_{07}$};
        \node[state3,right of = s0,xshift=0.5cm,yshift=2.6cm](s7){$\mathcal{S}_{06}$};
        \node[state3,right of = s7,xshift=0.5cm,yshift=1cm](s12){$\mathcal{S}_{13}$};
        \node[state3,right of = s12,xshift=0.5cm,yshift=0.8cm](s3){$\mathcal{S}_{03}$};
        \node[state3,right of = s3,xshift=-2.5cm,yshift=0.8cm](s04){$\mathcal{S}_{04}$};
        \node[state3,right of = s7,xshift=0.5cm,yshift=-1cm](s9){$\mathcal{S}_{08}$};
        \node[state3,right of = s04,xshift=2cm,yshift=0.6cm](s1){$\mathcal{S}_{01}$};
        \node[state3,right of = s0,xshift=3.5cm,yshift=-0.8cm](s4){$\mathcal{S}_{05}$};
        \node[state3,right of = s0,xshift=0.5cm,yshift=-2cm](s07){$\mathcal{S}_{06}$};
        \node[state3,right of = s0,xshift=2cm,yshift=-2.8cm](s5){$\mathcal{S}_{09}$};
        \node[state3,right of = s5,xshift=2cm](s6){$\mathcal{S}_{10}$};
        \node[state3,right of = s0,xshift=0.5cm,yshift=-3.8cm](s13){$\mathcal{S}_{14}$};
        \node[state3,right of = s13,xshift=3.5cm](s14){$\mathcal{S}_{15}$};
        \node[state3,right of = s07,xshift=0.5cm,yshift=0.6cm](s11){$\mathcal{S}_{12}$};
        \node[state3,right of = s7,xshift=0.5cm](s012){$\mathcal{S}_{11}$};
        
        \draw [arrow,rounded corners,  thick] (s0) |-  (s12);
        \draw [arrow,rounded corners,  thick] (s0) |-  (s04);
        \draw [arrow,rounded corners,  thick] (s0) |-  (s1);
        \draw [arrow,rounded corners,  thick] (s7) |-  (s12);
        \draw [arrow,rounded corners,  thick] (s7) |-  (s9);
        \draw [arrow,rounded corners,  thick] (s7) |-  (s8);
        \draw [arrow,rounded corners,  thick] (s0) |-  (s4);
        \draw [arrow,rounded corners,  thick] (s0) |-  (s07);
        \draw [arrow,rounded corners,  thick] (s07) |-  (s11);
        \draw [arrow,rounded corners,  thick] (s04) -|  (s3);
        \draw [arrow,rounded corners,  thick] (s3) -|  (s1);
        \draw [arrow,rounded corners,  thick] (s12) -|  (s3);
        \draw [arrow,rounded corners,  thick] (s9) -|  (s3);
        \draw [arrow,rounded corners,  thick] (s9) -|  (s1);
        \draw [arrow,rounded corners,  thick] (s8) -|  (s1);
        \draw [arrow,rounded corners,  thick] (s4) -|  (s2);
        \draw [arrow,rounded corners,  thick] (s11) -|  (s2);
        \draw [arrow,rounded corners,  thick] (s07) -|  (s2);
        \draw [arrow,rounded corners,  thick] (s1) -|  (see);
        \draw [arrow,rounded corners,  thick] (s6) -|  (see);
        \draw [arrow,rounded corners,  thick] (s14) -|  (see);
        \draw [arrow,rounded corners,  thick] (s5) --  (s6);
        \draw [arrow,rounded corners,  thick] (s7) --  (s012);
        \draw [arrow,rounded corners,  thick] (s012) --  (s9);
        \draw [arrow,rounded corners,  thick] (s13) --  (s14);
        \draw [arrow,rounded corners,  thick] (s2) --  (see);
        
        \draw [arrow,rounded corners,  thick] (s0) |-  (-7.2,-5.1);
        \draw [arrow,rounded corners,  thick] (s7) |-  (-5.7,-5.6);
        \draw [arrow,rounded corners,  thick] (s7) |-  (-5.7,-6.2);
        \draw [arrow,rounded corners,  thick] (s7) |-  (-4.2,-6.4);
        \draw [arrow,rounded corners,  thick] (s7) |-  (-4.2,-7.0);
        \draw [arrow,rounded corners,  thick] (s0) |-  (-4.2,-2.9);
        \draw [arrow,rounded corners,  thick] (s0) |-  (-4.2,-3.3);
        \draw [arrow,rounded corners,  thick] (s0) |-  (-7.2,-4.7);
        \draw [arrow,rounded corners,  thick] (s0) |-  (-7.2,-11.1);
        \draw [arrow,rounded corners,  thick] (s0) |-  (-7.2,-11.5);
        \draw [arrow,rounded corners,  thick] (s0) |-  (-5.7,-10.5);
        \draw [arrow,rounded corners,  thick] (s0) |-  (-5.7,-10.1);

        \draw [arrow,rounded corners,  thick] (-7.7,-7.3) --  (-2.7,-7.3);
        \draw [arrow,rounded corners,  thick] (-7.7,-7.7) --  (-2.7,-7.7);
        \draw [arrow,rounded corners,  thick] (s07) -|  (-5.5,-9.9);
        \draw [arrow,rounded corners,  thick] (s07) -|  (-4.9,-9.9);

        \coordinate[label = left:{\fontsize{7}{7}\selectfont $\mathcal{t}_{01}$}](fa) at(-5.2,-1.5);
        \coordinate[label = left:{\fontsize{7}{7}\selectfont $\mathcal{t}_{02}$}](fa) at(-0.9,-3.5);
        \coordinate[label=left:{\fontsize{7}{7}\selectfont$\mathcal{t}_{16}$}](fa)at(-7.5,-3.1);
\coordinate[label=left:{\fontsize{7}{7}\selectfont$\mathcal{t}_{05}$}](fa)at(-7.5,-2.7);
\coordinate[label=left:{\fontsize{7}{7}\selectfont$\mathcal{t}_{07}$}](fa)at(-7.5,-2.1);
\coordinate[label=left:{\fontsize{7}{7}\selectfont$\mathcal{t}_{06}$}](fa)at(-2.4,-3.0);
        \coordinate[label=left:{\fontsize{7}{7}\selectfont$\mathcal{t}_{33}$}](fa)at(-4.0,-4.1);
\coordinate[label=left:{\fontsize{7}{7}\selectfont$\mathcal{t}_{08}$}](fa)at(-4.0,-2.1);
\coordinate[label=left:{\fontsize{7}{7}\selectfont$\mathcal{t}_{32}$}](fa)at(-7.5,-3.7);
\coordinate[label=left:{\fontsize{7}{7}\selectfont$\mathcal{t}_{30}$}](fa)at(-5.7,-4.1);
\coordinate[label=left:{\fontsize{7}{7}\selectfont$\mathcal{t}_{12}$}](fa)at(-7.5,-4.5);
\coordinate[label=left:{\fontsize{7}{7}\selectfont$\mathcal{t}_{31}$}](fa)at(-7.5,-4.9);
\coordinate[label=left:{\fontsize{7}{7}\selectfont$\mathcal{t}_{18}$}](ta)at(-5.7,-6.1);
\coordinate[label=left:{\fontsize{7}{7}\selectfont$\mathcal{t}_{20}$}](fa)at(-5.7,-5.8);
\coordinate[label=left:{\fontsize{7}{7}\selectfont$\mathcal{t}_{21}$}](fa)at(-5.7,-5.4);
\coordinate[label=left:{\fontsize{7}{7}\selectfont$\mathcal{t}_{26}$}](fa)at(-5.7,-4.7);
\coordinate[label=left:{\fontsize{7}{7}\selectfont$\mathcal{t}_{27}$}](fa)at(-4.4,-5.4);
\coordinate[label=left:{\fontsize{7}{7}\selectfont$\mathcal{t}_{19}$}](fa)at(-2.4,-5.7);
\coordinate[label=left:{\fontsize{7}{7}\selectfont$\mathcal{t}_{13}$}](fa)at(-4.1,-6.2);
\coordinate[label=left:{\fontsize{7}{7}\selectfont$\mathcal{t}_{16}$}](fa)at(-5.2,-6.5);
\coordinate[label=left:{\fontsize{7}{7}\selectfont$\mathcal{t}_{25}$}](fa)at(-5.2,-6.8);
\coordinate[label=left:{\fontsize{7}{7}\selectfont$\mathcal{t}_{22}$}](fa)at(-2.4,-6.5);
\coordinate[label=left:{\fontsize{7}{7}\selectfont$\mathcal{t}_{31}$}](f31)at(-4.9,-9.7);
\coordinate[label=left:{\fontsize{7}{7}\selectfont$\mathcal{t}_{40}$}](f36)at(-4.3,-9.7);
\coordinate[label=left:{\fontsize{7}{7}\selectfont$\mathcal{t}_{03}$}](f3)at(-5.2,-7.1);
\coordinate[label=left:{\fontsize{7}{7}\selectfont$\mathcal{t}_{09}$}](f3)at(-5.2,-7.5);
\coordinate[label=left:{\fontsize{7}{7}\selectfont$\mathcal{t}_{04}$}](f4)at(-1.1,-7.3);
\coordinate[label=left:{\fontsize{7}{7}\selectfont$\mathcal{t}_{10}$}](f8)at(-5.9,-8.1);
\coordinate[label=left:{\fontsize{7}{7}\selectfont$\mathcal{t}_{11}$}](f9)at(-2.4,-8.1);
\coordinate[label=left:{\fontsize{7}{7}\selectfont$\mathcal{t}_{12}$}](f14)at(-7.5,-9.3);
\coordinate[label=left:{\fontsize{7}{7}\selectfont$\mathcal{t}_{23}$}](f11)at(-3.2,-10.1);
\coordinate[label=left:{\fontsize{7}{7}\selectfont$\mathcal{t}_{28}$}](f28)at(-5.9,-8.7);
\coordinate[label=left:{\fontsize{7}{7}\selectfont$\mathcal{t}_{29}$}](f29)at(-2.4,-8.7);
\coordinate[label=left:{\fontsize{7}{7}\selectfont$\mathcal{t}_{37}$}](f34)at(-4.7,-11.1);
\coordinate[label=left:{\fontsize{7}{7}\selectfont$\mathcal{t}_{34}$}](f35)at(-3.2,-9.3);
\coordinate[label=left:{\fontsize{7}{7}\selectfont$\mathcal{t}_{38}$}](f39)at(-0.9,-11.1);
\coordinate[label=left:{\fontsize{7}{7}\selectfont$\mathcal{t}_{24}$}](f12)at(-0.9,-10.1);
\coordinate[label=left:{\fontsize{7}{7}\selectfont$\mathcal{t}_{35}$}](f32)at(-7.5,-10.9);
\coordinate[label=left:{\fontsize{7}{7}\selectfont$\mathcal{t}_{36}$}](f37)at(-7.5,-11.3);
\coordinate[label=left:{\fontsize{7}{7}\selectfont$\mathcal{t}_{42}$}](f13)at(-7.5,-9.9);
\coordinate[label=left:{\fontsize{7}{7}\selectfont$\mathcal{t}_{41}$}](f13)at(-7.5,-10.3);

    \node[state4,right of = A,yshift = -5.5cm,xshift=-0.2cm](s0){$\mathcal{S}_{00}$};
        \node[state4,right of = s0,xshift=5.0cm](s2){$\mathcal{S}_{02}$};
        \node[state4,right of = s2,xshift=0.5cm](see){$\mathcal{S}_{os}$};
        \node[state4,right of = s0,xshift=3.5cm,yshift=2.8cm](s3){$\mathcal{S}_{03}$};
        \node[state4,right of = s0,xshift=0.5cm,yshift=3.4cm](s04){$\mathcal{S}_{04}$};
        \node[state4,right of = s0,xshift=3.5cm,yshift=0.4cm](s8){$\mathcal{S}_{07}$};
        \node[state4,right of = s0,xshift=0.5cm,yshift=0.9cm](s7){$\mathcal{S}_{06}$};
        \node[state4,right of = s7,xshift=0.5cm](s12){$\mathcal{S}_{08}$};
        \node[state4,right of = s7,xshift=0.5cm,yshift=1.4cm](s9){$\mathcal{S}_{13}$};
        \node[state4,right of = s0,xshift=5cm,yshift=3.8cm](s1){$\mathcal{S}_{01}$};
        \node[state4,right of = s0,xshift=3.5cm,yshift=-0.6cm](s4){$\mathcal{S}_{05}$};
        \node[state4,right of = s0,xshift=0.5cm,yshift=-1.4cm](s07){$\mathcal{S}_{06}$};
        \node[state4,right of = s0,xshift=2cm,yshift=-2cm](s5){$\mathcal{S}_{09}$};
        \node[state4,right of = s5,xshift=2cm](s6){$\mathcal{S}_{10}$};
        \node[state4,right of = s0,xshift=0.5cm,yshift=-2.6cm](s13){$\mathcal{S}_{14}$};
        \node[state4,right of = s13,xshift=3.5cm](s14){$\mathcal{S}_{15}$};
        \node[state4,right of = s07,xshift=0.5cm,yshift=0.4cm](s11){$\mathcal{S}_{12}$};
        \node[state4,right of = s7,xshift=0.5cm,yshift=0.7cm](s012){$\mathcal{S}_{11}$};
       \draw [arrow,rounded corners,  thick] (s0) |-  (s1);
        \draw [arrow,rounded corners,  thick] (s1) -|  (see);
        \draw [arrow,rounded corners,  thick] (s0) |-  (s04);
        \draw [arrow,rounded corners,  thick] (s0) |-  (s7);
        \draw [arrow,rounded corners,  thick] (s7) |-  (s9);
        \draw [arrow,rounded corners,  thick] (s7) |-  (s8);
        \draw [arrow,rounded corners,  thick] (s7) |-  (s012);
        \draw [arrow,rounded corners,  thick] (s0) |-  (s4);
        \draw [arrow,rounded corners,  thick] (s0) |-  (s9);
        \draw [arrow,rounded corners,  thick] (s0) |-  (s07);
        \draw [arrow,rounded corners,  thick] (s07) |-  (s11);
        \draw [arrow,rounded corners,  thick] (s0) |-  (s5);
        \draw [arrow,rounded corners,  thick] (s0) |-  (s13);
        \draw [arrow,rounded corners,  thick] (s3) -|  (s1);
        \draw [arrow,rounded corners,  thick] (s9) -|  (s3);
        \draw [arrow,rounded corners,  thick] (s12) -|  (s3);
        \draw [arrow,rounded corners,  thick] (s12) -|  (s1);
        \draw [arrow,rounded corners,  thick] (s8) -|  (s1);
        \draw [arrow,rounded corners,  thick] (s4) -|  (s2);
        \draw [arrow,rounded corners,  thick] (s11) -|  (s2);
        \draw [arrow,rounded corners,  thick] (s07) -|  (s2);
        \draw [arrow,rounded corners,  thick] (s6) -|  (see);
        \draw [arrow,rounded corners,  thick] (s14) -|  (see);
        \draw [arrow,rounded corners,  thick] (s0) --  (s2);
        \draw [arrow,rounded corners,  thick] (s7) --  (s12);
        \draw [arrow,rounded corners,  thick] (s012) --  (s12);
        \draw [arrow,rounded corners,  thick] (s5) --  (s6);
        \draw [arrow,rounded corners,  thick] (s13) --  (s14);
        \draw [arrow,rounded corners,  thick] (s04) -|  (s3);
        \draw [arrow,rounded corners,  thick] (s0) |-  (s3);
        \draw [arrow,rounded corners,  thick] (s2) --  (see);
        \draw [arrow,rounded corners,  thick] (s07) -|  (s5);

        \draw [arrow,rounded corners, line width=8pt] (4.4,-8.3) --  (4.4,-9.3);

        \node[state5,right of = A,yshift = -9.8cm,xshift=0.4cm](B){5};
       
       \node[state2,right of = A,yshift = -9.8cm,xshift=3.9cm](B){Graphs of Four Approaches};

  
       \node[right of = A,yshift = -10.9cm,xshift=1cm](s00){};
       \node[state6,below of =s00,xshift= -1.2cm,yshift=-1cm](s0){$\mathcal{S}_{00}$};
        \node[state6,right of = s0,xshift=6.5cm](see){$\mathcal{S}_{os}$};
        \node[state6,right of = s0,xshift=3.5cm,yshift=1.2cm](s3){$\mathcal{S}_{03}$};
        \node[state6,right of = s0,xshift=0.5cm,yshift=1.8cm](s4){$\mathcal{S}_{04}$};
        \node[state6,right of = s0,xshift=2cm,yshift=-1.5cm](s8){$\mathcal{S}_{07}$};
        \node[state6,right of = s0,xshift=2cm,yshift=-0.8cm](s012){$\mathcal{S}_{11}$};
        \node[state6,right of = s0,xshift=0.5cm](s7){$\mathcal{S}_{06}$};
        \node[state6,right of = s7,xshift=0.5cm](s12){$\mathcal{S}_{08}$};
        \node[state6,right of = s7,xshift=0.5cm,yshift=0.6cm](s9){$\mathcal{S}_{13}$};
        \node[state6,right of = s0,xshift=5cm,yshift=2.2cm](s1){$\mathcal{S}_{01}$};
        \draw [arrow,rounded corners,  thick] (s0) |-  (s1);
        \draw [arrow,rounded corners,  thick] (s0) |-  (s4);
        \draw [arrow,rounded corners,  thick] (s7) |-  (s9);
        \draw [arrow,rounded corners,  thick] (s7) |-  (s012);
        \draw [arrow,rounded corners,  thick] (s0) |-  (s9);
        \draw [arrow,rounded corners,  thick] (s7) |-  (s8);
        \draw [arrow,rounded corners,  thick] (s0) |-  (s3);
        \draw [arrow,rounded corners,  thick] (s0) --  (s7);
        \draw [arrow,rounded corners,  thick] (s7) --  (s12);
        \draw [arrow,rounded corners,  thick] (s012) --  (s12);
        \draw [arrow,rounded corners,  thick] (s12) -|  (s1);
        \draw [arrow,rounded corners,  thick] (s4) -|  (s3);
        \draw [arrow,rounded corners,  thick] (s1) -|  (see);
        \draw [arrow,rounded corners,  thick] (s3) -|  (s1);
        \draw [arrow,rounded corners,  thick] (s9) -|  (s3);
        \draw [arrow,rounded corners,  thick] (s12) -|  (s3);
        \draw [arrow,rounded corners,  thick] (s8) -|  (s1);
        \coordinate[label = center:{\fontsize{8}{8}\selectfont $\Omega_{1}$: Math Expression Approach}](S) at(2,-10.5); 

       \node[state6,below of =s00,xshift= -1.2cm,yshift=-4cm](s0){$\mathcal{S}_{00}$};
        \node[state6,right of = s0,xshift=5.0cm](s2){$\mathcal{S}_{02}$};
        \node[state6,right of = s2,xshift=0.5cm](see){$\mathcal{S}_{os}$};
        \node[state6,right of = s0,xshift=2cm,yshift=0.6cm](s4){$\mathcal{S}_{05}$};
        \node[state6,right of = s0,xshift=2cm,yshift=-0.6cm](s11){$\mathcal{S}_{12}$};
        \node[state6,right of = s0,xshift=0.5cm,yshift=-1.2cm](s7){$\mathcal{S}_{06}$};
        \draw [arrow,rounded corners,  thick] (s0) |-  (s4);
        \draw [arrow,rounded corners,  thick] (s0) |-  (s7);
        \draw [arrow,rounded corners,  thick] (s7) |-  (s11);
        \draw [arrow,rounded corners,  thick] (s0) --  (s2);
        \draw [arrow,rounded corners,  thick] (s2) --  (see);
        \draw [arrow,rounded corners,  thick] (s4) -|  (s2);
        \draw [arrow,rounded corners,  thick] (s11) -|  (s2);
        \draw [arrow,rounded corners,  thick] (s7) -|  (s2);
        \coordinate[label = center:{\fontsize{8}{8}\selectfont $\Omega_{2}$: Linear Equations Approach}](S) at(2,-14.9);

        \node[state6,below of =s00,yshift = -6.7cm,xshift= -1.2cm](s30){$\mathcal{S}_{00}$};  
        \node[state6,right of = s30,yshift =0.0cm,xshift=2cm](s35){$\mathcal{S}_{09}$};
        \node[state6,right of = s35,yshift =0.0cm,xshift=2cm](s36){$\mathcal{S}_{10}$};
        \node[state6,right of = s30,yshift =0.5cm,xshift=0.5cm](s37){$\mathcal{S}_{06}$};
        \node[state6,right of = s36,xshift=0.5cm](s39){$\mathcal{S}_{os}$}; 
  
        \draw [arrow,rounded corners,  thick] (s30) --  (s35);
        \draw [arrow,rounded corners,  thick] (s30) |-  (s37);
        \draw [arrow,rounded corners,  thick] (s37) -|  (s35);
        \draw [arrow,rounded corners,  thick] (s35) --  (s36);
        \draw [arrow,rounded corners,  thick] (s36) --  (s39); 
        \coordinate[label = center:{\fontsize{8}{8}\selectfont $\Omega_{3}$: Relation-Centric Approach}](S) at(2,-17.7);

       
       \node[state6,below of =s00,yshift = -7.7cm,xshift= -1.2cm](s50){$\mathcal{S}_{00}$};  
        \node[state6,right of = s50,yshift =0.0cm,xshift=1.5cm](s513){$\mathcal{S}_{14}$};
        \node[state6,right of = s513,xshift=1.5cm](s514){$\mathcal{S}_{15}$};
        \node[state6,right of = s514,xshift=1.5cm](s59){$\mathcal{S}_{os}$}; 
  
        \draw [arrow,rounded corners,  thick] (s50) --  (s513);
        \draw [arrow,rounded corners,  thick] (s513) --  (s514); 
        \draw [arrow,rounded corners,  thick] (s514) --  (s59); 
        \coordinate[label = center:{\fontsize{8}{8}\selectfont $\Omega_{4}$: xRelation-Centric Approach}](S) at(2,-19.2);

       \node[state2,right of = A,yshift = -19.8cm,xshift=-5.5cm](B){Theory Base: State Transform Theory};
       \node[state8,right of = A,yshift = -19.8cm,xshift=-9.2cm](B){1};

        \node[state7,align=left,yshift = -21cm](Q){1.All the algorithms for solving algebra problems can be depicted by State Transform Graphs.\\2.All the representative algorithms for solving algebra problems developed since 2014 are decomposable.\\3.All the representative algorithms for solving algebra problems developed since 2014 share a set of states and transforms.};
        
        \node[state9,align=center,yshift = -17.9cm,xshift=-4cm](Q){(Transform,Algorithm,Approach)\\$\times$
        \\(six types of problems)};

        \node[state2,right of = A,yshift = -16.6cm,xshift=-4.9cm](B){Perspective Confusion Comparison};
       \node[state5,right of = A,yshift = -16.6cm,xshift=-8.4cm](B){6};

       \node[state10,align=left,yshift = -14.4cm,xshift=-4.8cm](Q){algorithms: $\mathcal{A}$ = \{$\mathcal{A}_{01}$, $\mathcal{A}_{02}$,..., $\mathcal{A}_{30}$\}\\states: $\Delta$ = \{$\mathcal{S}_{00}$, $\mathcal{S}_{01}$,..., $\mathcal{S}_{15}$\}
        \\transforms: $\Pi$ = \{$\mathcal{t}_{01}$, $\mathcal{t}_{02}$,..., $\mathcal{t}_{42}$\}};

        \node[state2,right of = A,yshift = -13.2cm,xshift=-5.8cm](B){Analysis of Individual Algorithms};
       \node[state5,right of = A,yshift = -13.2cm,xshift=-9.3cm](B){2};
       
       \draw [arrow, line width=8pt] (-8.3,-19) --  (-8.3,-14);
       \draw [arrow, line width=8pt] (-5,-12.7) --  (-5,-11.6);
       \draw [arrow, line width=8pt] (-4,-15.2) --  (-4,-16.2);
       \draw [arrow,line width=8pt] (0.5,-16.6) --  (-0.7,-16.6);
     \end{tikzpicture}
     \caption{}
     \label{example_A11}
\end{figure*}

\begin{highlights}
\item[] This article conducts a theoretical review on solving algebra problems. It makes the following main contributions:  

\item  Developed a theoretical framework for reviewing algorithms for solving algebra problems (APs).
\item  Built the State Transform Theory (STT) as the theoretical basis of the developed review framework. 
\item  Proposed a State Transform Analysis method of analyzing individual and collective algorithms based on the STT. 
\item Designed a Perspective Confusion Comparison for evaluating transform, algorithm, and approach on different types of APs.
\item Proposed future directions based on insights gained from the State Transform Analysis (STA) and Perspective Confusion Comparison (PCC).


That this article meets the submission criteria of the journal is stated as follows.

Professor Xinguo YU has been leading a research team focused on developing solving algorithms for basic education for over 10 years. The team has published 10 SCI-indexed papers in this field, introducing innovative techniques, algorithms, and approaches. Of these publications, three were featured in JCR Q1 journals, three in JCR Q2, and four in JCR Q4. The full list of publications is provided below:

Three JCR Q1 papers: 
\begin{enumerate}
    \item X. Lyu Xiaopan, Xinguo YU, Peng Rao. Vector relation acquisition and scene knowledge for solving arithmetic word problems. Journal of King Saud University - Computer and Information Sciences, 2023, 35(8):101673. (JCR Q1)
    \item Xinguo YU, X. Lyu, R. Peng, J. Shen. Solving arithmetic word problems by synergizing syntax-semantics extractor for explicit relations and neural network miner for implicit relations. Complex Intell. Syst. (2022).  (JCR Q1)
    \item Xinguo YU, H. Sun, C. Sun. A relation-centric algorithm for solving text-diagram function problems, Journal of King Saud University - Computer and Information Sciences, 2022.  (JCR Q1)
\end{enumerate}  

\vspace{20em}

Three JCR Q2 papers: 
\begin{enumerate}
     \item L. Huang, Xinguo YU*, L. Niu, Z. Feng. Solving algebraic problems with geometry diagrams using syntax-semantics diagram understanding, Computers, Materials and Continua, 2023. (JCR Q2)
    \item H. Meng, Xinguo YU*, B. He, et al. Solving arithmetic word problems of entailing deep implicit relations by qualia syntax-semantic model, Computers, Materials and Continua, 2023.  (JCR Q2)
    \item B. He, Xinguo Yu*, P. Jian, T.g Zhang. Solving direct current circuit problems, Applied Intelligence, 2020/03/03. (JCR Q2)
\end{enumerate}

Four JCR Q4 papers: 
\begin{enumerate}
      \item Xinguo Yu, M. Wang, W. Gan, B. He and N. Ye.  
      A framework for solving explicit arithmetic word problems and proving plane geometry theorems. 
      International Journal of Pattern Recognition and Artificial Intelligence, 33(7), 1940005 (2019) 
      \item W. Gan, Xinguo Yu and M. Wang. 
      Automatic understanding and formalization of plane geometry proving problems in natural language: a supervised approach,
      International Journal on Artificial Intelligence Tools, Vol. 28, No. 4, 1940003, 2019.   
    \item P. Jian, C. Sun, Xinguo Yu,  B. He and M. Xia. An end-to-end algorithm for solving circuit problems, International Journal of Pattern Recognition and Artificial Intelligence, 19400044, 2019.
   \item  W. Gan, Xinguo Yu, T. Zhang and M. Wang. Automatic proving plane geometry theorems stated by text and diagram. International Journal of Pattern Recognition and Artificial Intelligence, 33(7), 19400032 (2019).  
\end{enumerate}

As per the above description, Xinguo YU (the first author of this submission) is the first author of two JCR Q1 papers, the corresponding and second author of one JCR Q1 paper and three JCR Q2 papers. He is, also, the author of another four SCI papers. All other first authors (except B. He) are PhD students of Xinguo YU.  
\end{highlights}

\vspace{80em}

\end{frontmatter}

\pagebreak

\section{Introduction}
\label{}

As interest in solving algebra problems (APs) is continuing to gain momentum, the absence of theoretical justifications for new works is more keenly felt than ever. It is thus imperative and timely that a theoretical perspective is applied to the AP-solving algorithms because the relevant conditions are present. These include the fact that significant numbers of algorithms for solving APs have appeared in the literature over the past six decades as well as the existence of several important review papers that provided substantial background knowledge for this  review \citep{Carifio1994, Mukherjee2008,Kotwal2019,Mandal2019,zhang2020gap,Faldu2021,Lu2023}. 
For example, several papers developed the methods and algorithms for solving text-diagram problems\citep{Jian2019,He2020,YuSun2022}, which extended the scope of problems from Algebra Word Problems into Algebra Problems (mainly from mathematics, physics, and chemistry in basic education). 

The papers referred to these overviews, however, demonstrate that current research is still unsatisfactory in terms of inclusions, time span, and theoretical underpinnings despite attempts justifying AP-solving algorithms at model-level of which the state action frameworks developed by \citep{Xia2021,Yu2023} are significant examples. As many others have done, the authors stress the importance of contributions made by action techniques but pay scant attention to ``states'', which, in fact, play an essential role in determining the properties of solving algorithms, providing further justification for our work.

The development of a theory-based review of algorithms for solving APs is, thus, justified by the research communities’ need, significant number of models and paradigms that have appeared, creating the relevant conditions for the development of a theory:(a) a large body of algorithms exists, (b) a range of review papers have laid the groundwork \citep{Carifio1994,Mukherjee2008,Kotwal2019,Mandal2019,zhang2020gap,Faldu2021,Lu2023}, and (c) the work of \citep{Xia2021,Yu2023}  has provided specific paradigms that have the capacity to be extended to a theory.

As none of the review papers theoretically justified their work, this creates an urgent need for a theory that offers a basis of the analysis of AP-solving algorithms, explains their processes, provides a uniform basis for comparison, and provides theoretical guidance for the further development of AP-solving algorithms. To achieve this, we draw on the work on state transition paradigms in \citep{Xia2021,Yu2023} which provides a starting point for the development of a theory for reviewing AP-solving algorithms.

In the first step, to construct the State Transform Theory, this paper draws from the state action paradigm proposed in \citep{Yu2023}. This construct considers problem-solving in three phases, with the first and most significant consisting of states and actions. This first phase provides a logical basis for progression to the subsequent phases of problem-solving represented by “transforms” from the state to the algorithmic output. We posit that the presence of a “transform” in the work structure of an algorithm indicates its consistency with a theoretical model.

Some researchers have tried to use the state transition paradigm to explain algorithms in abstract level, through a mathematical model that describes the dynamic process of a finite state machine \citep{Gill1960}. The state transition paradigm is used to reveal the structure and properties of algorithms \citep{Luan2001}, to design the algorithm \citep{Yu2023,Zhou2020}, or to analyze algorithms \citep{Luan2001}. The state transition paradigm (borrowed from finite state machine) is used as a theoretical model to describe some constructs of algorithms. Both its states and transitions are the abstract concepts whereas states and transitions in "finite state machine" are concrete physical or mathematical contents.  States express the abstract common properties of the intermediate outcomes of algorithms; transitions express the theoretical methods of fulfilling state transition.  To rephrase, for algorithm analysis the state transition paradigm is a mental model for abstract phenomena but not a concrete physical configuration.

\textcolor{black}{
Hence, we propose the State Transform Theory (STT), which uses "state" and "transform" as key constructs to build State Transform Graphs (STGs). These graphs visualize the structure of an algorithm and confirms its decomposability. Building on the STT, we introduce the State Transform Analysis method, a graphical approach to analyze algorithms in three steps. First, we list all representative algorithms since 2014, identifying their respective states and key transforms. Second, we represent each algorithm with its individual STG, followed by an aggregated graph visualizing all collected algorithms. This aggregated graph is abstracted to create a state graph. Additionally, the four approaches are uniformly defined on the aggregated graph, and a taxonomy tree of the algorithms is generated. Both the aggregated graph and the state graph enable us to draw global conclusions about the algorithms. To further understand the solving algorithms, we examine the progression of states and transforms for each of the four approaches. For comparative analysis of perspective confusion, we classify algebra problems into different types, allowing us to evaluate the distribution of transforms, algorithms, and approaches across these types. This provides a performance landscape for solving algebraic problems (APs). However, this paper presents only sample evaluations rather than a complete analysis, due to the focus of this study.
}

\subsection{Contributions}

This paper integrates and develops several existing paradigms into a State Transform Theory by generalizing the paradigms to all representative algorithms and making statements about properties of individual algorithms and collective algorithms. Using this theory as a basis, a review of research into solving algebra problems is conducted whereby algorithms are reviewed at three level: transform, algorithm, and approach. Through conducting this novel review, this paper makes four contributions as follows:
 
\begin{itemize}

\item It develops the State Transform Theory (STT), serving as the theoretical basis for reviewing AP-solving algorithms. 

\item It proposes the State Transform Analysis method to depict and analyze both individual and collective algorithms based on the STT.

\item It proposes the review paradigm to define and review approaches that analyze the progression of approaches by reviewing the evolution to their states and transforms.
  
\item It proposes the Perspective Confusion Comparison to critically compare transforms, algorithms, and approaches, a method superior to the ablation experiment.

\end{itemize}

\subsection{Organization}

The organization of the remainder of the paper is as follows. The related work is reviewed in Section \ref{related_work}. The STT is built in Section \ref{sec-theory} while section \ref{GraphicalAlgorithmAnalysis} presents the State Transform Analysis method of reviewing algorithms in which State Transform Graphs are used to depict individual and collective algorithms based on the STT. Section \ref{sec-approach-analysis} reviews the approaches of AP-solving algorithms through narrating the evolution of states and transforms. Section \ref{sec-confusion} presents the Perspective Confusion Comparison on transforms, algorithms, and approaches. Section \ref{sec-conclusion} concludes the paper and discusses future research directions.

\section{Related Work} \label{related_work}

Solving APs is longstanding research problem which continues to be a challenge. Despite dating back to the 1960s, the interest in solving APs has intensified only recently with the advance of Artificial Intelligence (AI) and the demand for more intelligent education. During this long research history, seven survey papers have reviewed problem-solving algorithms with the aim of identifying the difficulties, as well as mining and generally useful techniques to build a body of knowledge \citep{Carifio1994, Mukherjee2008,Kotwal2019,Mandal2019,zhang2020gap,Faldu2021,Lu2023} which will be reviewed in this section. Different from the survey papers, several researchers-initiated work of developing  a model to  explain  their algorithms but not historical algorithms \citep{Xia2021,Yu2023}. These papers inspire this work to build a theory to not only explain individual algorithms but also analyze collective algorithms, further guide algorithm development and algorithm evaluation. 

In computer science, several algorithm evaluation paradigms have previously been proposed to evaluate types of algorithms. Generally, the researchers adopted a benchmark paradigm to evaluate AI algorithms \citep{Hooker1995,Mendling2021}. As these attempts that were based on ‘post’ and competitive evaluation and comparison, they were unable to reveal the inner structure and techniques of algorithms, nor could they identify the theoretical relations among them. This research gap inspires us to develop a perspective confusion comparison in this paper.


\subsection{Accumulation of Review Knowledge for AP Solving}

A significant body of knowledge has been accumulated over the past six decades on AP-solving found in seven review papers \citep{Carifio1994, Mukherjee2008,Kotwal2019,Mandal2019,zhang2020gap,Faldu2021,Lu2023} each of which has contributed today’s understanding of these issues within the research community.

\cite{Carifio1994} published the first review paper on solving APs wherein they proposed a two-stage review scheme, which divides the process of solving APs into representation and solution stages. They placed emphasis on identifying direct and inferred problem representation to obtain answers at the theoretical level. 

Almost fifteen years after \cite{Carifio1994}, \cite{Mukherjee2008} took the issue up again with their review. Again, the authors followed a two-stage review scheme although this time analyzing the workflow. This paper offered the observation that understanding natural language is the core step to solving mathematical problems (including algebra, geometry, physics, mechanics, etc.), consequently reviewing methods employed for understanding descriptions of problems. The authors also included an interesting discussion of the progress of rule-based methods published before 2008.

This attempt at categorization was advanced by  \cite{MandalNaskar2019} who proposed two schemes for reviewing literature on AP-solving of which the first one was a stage scheme that divided the history of the research on these specific problem-solving algorithms into two periods: Early Stage (1964-2006) and New Era (2007-2017). They concluded that in the Early Stage, researchers viewed solving APs either from a point of cognitive science, psychology, or education. Researchers developed multiple systems and algorithms for solving limited numbers of specific types of simple problems aimed at simulation of the processes by which the human students solve problems. In consequence, the proposed methods are mainly rule-based targeting APs with specific processes.
The algorithms developed in New Era were mainly published between 2007 and 2017 \citep{MandalNaskar2019}. 

During this period, machine learning based methods came into being and were applied to the development of problem-solving algorithms. This scheme classifies the algorithms into the following four categories based on the characteristics of the technique used to build them: 1) equation template-based; 2) equation tree-based; 3) entity and state transition-based and 4) tag/meaning based. The gained insights from this work are of significance for this paper as the first survey identified that solvers have two phases of problem understanding and symbolic solver (symbolic solver originally called solution stage). These three reviews agreed that algorithms are largely defined by their construction techniques, and these have a strong timestamp. They also agreed that all solving algorithms have two phases: problem understanding and symbolic solving. Whilst some of these insights are not precisely new, they re-enforce our current understanding and are, therefore, useful for our work.
 
In the same year as \cite{MandalNaskar2019},  \cite{Kotwal2019} used a quantity scheme to classify problems into categories according to the number of unknowns in each problem, subsequently applying a stage scheme to divide the algorithms into first generation and data driven. This research generated the idea to divide the AP corpus into groups according to their difficulty based on the belief that the number of unknowns in each problem is an indicator of the difficulty. This problem division inspires this  paper to divide APs into more categories according to more indicators.

A year later, \cite{zhang2020gap} re-adopted a cognitive approach by developing a systematic ``semantic gap'' review. 
The authors think that the difficulty in solving APs lies in the semantic gap that exists between human readable words and machine understandable logic. They, again, used a stage scheme, this time to classify the algorithms into, three generations according to the techniques underpinning them: the first generation (1960-2010) was characterized by rule-based matching, the second (2011-2017) by semantic parsing, feature engineering and statistical learning, and the third generation (2017-2019) by Deep Learning (DL) and Reinforcement Learning (RL). This paper also used the unknown quantity scheme to divide the solvers into two categories, arithmetic word solvers (AWS) capable of solving problems with one unknown, and equation set solvers (more than one unknown). The work further provided a confusion performance comparison of two different types of algorithms on different datasets and a discussion of advantages and disadvantages  The researchers used, in addition, a confusion comparison to discover the performances that solvers work on the various datasets. The take-away from this paper was that chronology schemes have repeatedly been used in surveys, indicating that. 
\begin{itemize}
    \item the solvers have a strong timestamp.
    \item the scheme aids classification of stages and categorization of AP-solving.
    \item the confusion comparison is a useful way to identify the actions that solvers carry out on datasets.
\end{itemize}

Also reviewing based on methodology were \cite{Faldu2021} who revisited APs, identifying the use of neural networks as an indicator of a quality solver. The paper further inspected non-neural and neural methods narrated in natural language applied as categorizer (1) direct answer generator, (2) expression tree generator for inferring answers, and (3) template retriever for answer computation. This work  showed that neural methods can output two types of results: direct answer and expression tree. This indicates that some neural networks can transform the vector sequence into a text-based math expression or expression tree with one unknown.

\cite{Lu2023} conducted a survey on mathematical reasoning tasks and neural networks for mathematical reasoning  which identified the four mathematical reasoning tasks: math word problem solving, theorem proving, geometry problem solving, and math question answering. It identified four types of neural networks for mathematical reasoning: seq2seq-based neural networks, graph-based networks, attention-based networks, and other neural networks.

Clearly, these works have made significant contributions in one or more areas; however, they would have benefited from   
\begin{itemize}
    \item a theoretical justification of their review schemes, absence of which prevents justification of fairness, completeness, and use of the work for guidance.
    \item a more comprehensive review which could have included relation-centric algorithms and algorithms for solving APs narrated through text and diagrams.
    \item identification and exploration of intermediate representations of the problems.
    \item a technique-based  scaffold would have been of benefit.
    \item a perspective-based understanding of AP-solving areas, superior to the single-level confusion comparison.
     
 \end{itemize}

Motivated by need, this paper conducts a theoretical review, based on the state transform theory (STT), which includes the identification of the intermediate representations, multiple level of techniques, and perspective confusion comparison for representative algorithms designed to solve APs.

\subsection{Theory Development in AP Solving}

Creating general model to analyze algorithms is a fascinating research problems due to the algorithm pseudo codes and structures are not built on the sound theory or uniform terms. Based on this cognition, \cite{Luan2001} proposed a formal specification based on first-order language to describe the internal state transition of algorithms. Although promising, this work remains in its early stage without further development or application. This state transition is a certain similarity with finite state machine. A finite state machine is a computational model that can be implemented with hardware or software and used to simulate logic structure and certain computer programs \citep{Gill1960}.  Finite state machines can model problems in many fields including mathematics, artificial intelligence, games, and linguistics. There is an understanding that representations and techniques appearing in algorithms have a certain similarity with states and transitions in finite state machines  \citep{Gill1960}. Based on this cognition, some researchers used state transition model to build some algorithms since they realized that it may be too ambitions to create a state transition model for analyzing general algorithms. The following is one of examples. \cite{Zhou2020} proposed a state transition algorithm,  which was proposed on the concepts of state, state transition and state space representation in modern control theory. Because of its excellent global search ability and fast convergence, the state transition algorithm has since been applied in various optimization problems, which inspires us a new way to improve solving algorithms in the future.

The rich facts in the literature in problem solving shed the light of that there is a model that can uniformly analyze most and main solving algorithms.  
Especially, All seven survey papers discussed above provided the substantial facts.  
They all focused on techniques that transform representations, indicating that reviewers uniformly agreed that representations and  techniques which transform representations are the main elements for constructing these algorithms, although implied rather than explicitly stated
\citep{Carifio1994, Mukherjee2008,Kotwal2019,Mandal2019,zhang2020gap,Faldu2021,Lu2023}. Several primary research papers provided more facts as follows.

\cite{Yu2019} first proposed an equivalent representation theory, which pointed out that math problem solving is to transform states from problem input to output. During this transition process, several intermediate states will be generated. In the sense of math problem solving, these states are equivalent,  that is, theoretically, the solution can be obtained from the input state through operations such as reasoning and calculation. The equivalent representation theory signifies that solving a problem is to transform one representation into its next, representing the seminal idea of algorithm structure of states and transforms. 

\cite{Xia2021} developed an equivalent representation theory developed in \cite{Yu2019} 
 into a state action framework to describe the process of solving APs by algorithms which was built around the development of solvers that solve APs similar to  students. For a solving algorithm, its input is the problem described by text and diagram while its output ideally is the solution to the given problem. The solving process aims to change one state to its next state through actions. The framework reveals the highly abstract nature of the cognitive model to explain the process of solving APs represented as text and diagram offering deep insights into the level of abstraction of algorithms. This makes it clear that  approaches are defined by determining instances of states, thus providing a formal way to define these.

\cite{Yu2023}  further developed the state action framework into a three-phase scheme to design algorithms from abstraction to concreteness. The first phase is to design a solving paradigm that proposes a state action paradigm for solving APs as the solving paradigm. The second phase is to determine the content of each state of the state action paradigm to produce a relation-centric approach to solving AWPs. The third phase is to determinate the actions of the relation-centric approach to instantiate an algorithm for solving AWPs.
Thus, this three-phase scheme works on the understanding that a solving algorithm can be abstracted into elements of states and actions. This represented a preliminary exploration of the work in this paper, and we intend to further develop and integrate the existing schemes into the STT.

The above-mentioned papers have contributed concepts, innovative ideas, and frameworks to building a theory capable of accounting for solving algorithms. By further developing these, this work builds an STT to explain the algorithms for solving APs, reveal the inner main states and transforms, form a new evaluation paradigm, and predict the trend in algorithm development.

 \subsection{Algorithm Evaluation Paradigms}

 Algorithm evaluation is an important topic because it is the basis for choosing the appropriate algorithms to solve the problems facing scientists; further, the algorithm is one of the key pillars in computer science. The first algorithm evaluation paradigm is to evaluate the algorithm through complexity analysis, being the important content in textbooks on algorithm analysis. A look at complexity analysis reveals that it is not applicable to most of the AI algorithms because their inputs are restricted to a narrow range of sizes, whereas complexity analysis requires that the input size ranges from finite to infinity. In consequence, a second algorithm evaluation paradigm was developed  \citep{Mendling2021}, which evaluates algorithms by their performance on benchmark datasets. The benchmark evaluation tests against performance indexes such as accuracy, recall, and running time on benchmark datasets. Such evaluations are seen as competitive tests inviting significant criticism. One of these is that the overemphasis on “state of the art” performance was “allowing researchers to publish dull papers that proposed small variations of existing algorithms and reported their small-but-significant incremental performance improvements in comparison studies”  \citep{Mendling2021}. The
second criticism is that it provides limited understanding of the reasons for the performance of algorithms. Therefore, in recent years, the ablation experiment has become a requirement for benchmark evaluation paradigms. This is an advance in algorithm evaluation since it offers some transparency in terms of the techniques that have contributed to the algorithm’s performance. However, it is still an “after algorithm” or “external algorithm” comparison that does not consider the constituents (i.e., technique) of algorithms.
This has prompted the development of other analysis schemes although they have not yet been fully developed. For example, a formal specification based on the first-order language was proposed to describe the internal state transition of algorithms \citep{Luan2001} which has so far not been adopted, to the best of our knowledge. Thus, the limitations of traditional review schemes are stimulating the search for new paradigms for reviewing algorithms. \cite{Xia2021}, for example, have developed an approach review scheme for algorithms that solve arithmetic word problems. This development demonstrated that the instances that instantiate abstract frameworks can be used to uniformly define the approaches instead of using heuristic indicators without technique  scaffold.
Based on the understanding gained from a thorough evaluation of existing review paradigms and the recognition that none of these would provide support for the development of theoretical work, this paper proposes the development of a new review paradigm capable of reviewing algorithms through identification of states and transforms, graphic analysis, and perspective confusion comparison.

\section{State Transform Theory}\label{sec-theory}

\textcolor{black}{
The review of the related work in the preceding section discloses that 
 the traditional review methods lack theory foundation. To fill this gap, this section develops the State Transform Theory (STT) as the theory base of the review method proposed in this paper.
 }

\subsection{Motivation and Justification}

In AP-solving, review papers concentrated on the new transforms (i.e., techniques), almost neglected the evaluation of state creation  and evolution \citep{Carifio1994, Mukherjee2008,Kotwal2019,Mandal2019,zhang2020gap,Faldu2021,Lu2023}.
This motivates to do a different review with  emphasis on the fact that it is capable of creation of state as well as technique. 

 Some other works with theoretical notions also attracted our attention such as those that developed basic concepts, abstract frameworks, theoretical explanations, theoretical processes for different purposes \citep{Luan2001,Xia2021,Yu2023}. 
Some researchers proposed an equivalent representation theory, which pointed out that mathematical problem solving is a state transition from problem input to output through intermediate states\citep{Yu2019}. This theory shows that several equivalent intermediate states will be generated during the entire transition process of problem solving. 
Then, the researchers further proposed a state action framework to explain the process of solving APs \citep{Xia2021}.
This framework consists of states and actions and it defines the approaches of solving algorithms by states. 
Recently \cite{Yu2023} created another usage of the state action framework. Namely, it proposed a three-phase scheme to develop algorithms for solving AWPss, which is a scheme to design algorithms from abstract to concreteness. In another view, this three-phase scheme reveals the relationship among framework, approach, algorithm, state, and action. These achievements motivate us to build a theory as the basis of review.

This section builds the State Transform Theory by further refining the theoretical concepts introduced in the previously discussed papers. The theory posits that the structural makeup of states dictates their core properties, and both these structures and the techniques applied to them influence performance. The theory will introduce several new assertions to highlight the essential characteristics of individual algorithms as well as groups of algorithms. First, all representative algorithms developed since 2014 can be modeled as graphs of states and transforms. Second, all such algorithms are decomposable. Third, these algorithms share a common set of states and transforms.

Leveraging the above stated theory accumulation, this paper develops a theoretical framework to review all representative algorithms, which reveals that the states represent the key progresses  whereas previous reviews used only transforms (techniques) to depict the progresses in AP solving. Based on this cognition, this paper mines all main states and then all transforms. This theoretical framework further uses crucial types of states to define the approaches of algorithms whereas existing review papers used the transforms (techniques) to define approaches. Our theoretical framework also creates a perspective confusion comparison which surpasses the ablation experiment in two aspects. First, it can reveal which transforms contribute to the performance of the whole algorithm. Second, it provides a perspective confusion comparison of transform, algorithm, and approach, deeper than step and algorithm comparison in the ablation experiment.



\subsection{Constructs of State Transform Theory}
\label{subsect_constructs}

The following concepts are the constructs of building state transform theory, in which state and transform are two pillar constructs. 
 
\begin{definition}[Corpus] 
 A set of algebra problems (APs), prepared by the research community, is called as a corpus of APs, denoted as $\mathcal{C}$. 
\end{definition}
 
 In this paper, Algebra Problems are Algebra Word Problems and  Algebra Text-Diagram Problems, which are mainly from mathematics, physics, and chemistry in basic education. 

\begin{definition}[Solving Algorithm]
\label{solving-algorithms}
Let $\mathcal{C}$ be a corpus of algebra problems. \AE~ is said to be a solving algorithm of $\mathcal{C}$ if \AE~  can correctly solve a significant percent of elements in $\mathcal{C}$.
\end{definition}


\begin{definition}[State] 
A state refers to an intermediate milestone representation generated by a solving algorithm while solving algebra problems, typically denoted a state as
$\mathcal{S}$ with subscription.
\end{definition}



\begin{definition}[Transform] 
 Let $\mathcal{S}_s$ and $\mathcal{S}_e$ be two states and $\mathcal{t}$ be a method or scheme that can be instantiated into a process of some solving algorithms that can transform instances in $\mathcal{S}_s$ into instances in $\mathcal{S}_e$ always for a corpus of problems. 
\end{definition}

 ``State'' and ``Transform'' are two pillar concepts of the theory.  ``State'' is the general name of various intermediate outcomes produced by solving algorithms and ``transform'' is an umbrella term of a range of processes. The concrete states and transforms are hidden in the algorithms in the literature. To identify them and to dissemble them from algorithms are two tasks in the next section. 
 
\textcolor{black}{
States are the representations of the results of the milestone intermediate goals. The understood form of problems is a sample intermediate goal. The results of such a goal could be a math expression, a set of linear equations, or a set of relations. 
The vector form of problems becomes another sample intermediate goal if machine learning is used to solve algebra problems. 
}

\textcolor{black}{
Here, two example states are on the understood form. The first example is Linear Equations, initially proposed by \cite{Kushman2014}. Linear equations are widely recognized as a representation of the intermediate results when solving arithmetic word problems \cite{Kushman2014, Zhou2015, Wang2017, Cao2021}.
So far the vector sequence is the only state for the vector form. 
The transforms are the methods that should produce the instances of states for algebra problems.  
}
\textcolor{black}{
The second example is Relation Set, as proposed by \cite{Yu2019}. This state is commonly used because it serves as an intermediate result for a broader range of math problems \cite{Yu2019, Gan2019a, He2020, Yu2023}.
}

\subsection {State Transform Graphs of Algorithms} \label{subsect_STGraph}
The state transform graph is defined to depict the concern algorithms. 


\begin{definition}[State Transform Graph] 
A directed graph, $\mathcal{G}$ = ($\mathcal{S}$, $\mathcal{T}$, $\phi$), is called as a state transform graph, where $\mathcal{S}$ is a set of state nodes and $\mathcal{T}$ is a set of transforms that directly link states, which appear in algorithms for solving algebra problems,  
$\phi: \{ \mathcal{T} \rightarrow {(x,y)} \hspace{0.11111em}\hspace{0.11111em} \| (x,y) \in \mathcal{S}X\mathcal{S}, x \neq y \}$
is an incidence function mapping every link to an ordered pair of nodes. 
\end{definition}


\begin{definition}[Decomposable Algorithm] \label{decomposable-alg}
An algorithm is called as a decomposable algorithm if its state transform graph has more than two states. 
\end{definition}

With these concepts, we have the following three statements.

\begin{statement}\label{State Transform-algorithm}
All the algorithms for solving algebra problems can be depicted by state transform graphs.  
\end{statement}
\begin{statement}\label{decomposable}
All the representative algorithms for solving algebra problems since 2014 in the literature are decomposable.
\end{statement}
\begin{statement}\label{finite-State Transform}
All the representative algorithms for solving algebra problems since 2014 in the literature share a set of states and transforms.
\end{statement}

The lists of algorithms, states, and transforms will be presented in Section \ref{subsec-StateTransform}. Three statements will be verified in Section \ref{subsect_STGraph}.  Statement \ref{State Transform-algorithm} and \ref{decomposable} are the core conclusions of the state transform theory because they show that this paper can use the state transform graph to analyze and compare the solving algorithms. Statement \ref{finite-State Transform} further tells us that the task of developing a solving algorithm can be broken into the state design, transform design, and state transform graph design. Thus, we can evaluate states, transforms, algorithms, and approaches in any form of combination.  

\section{State Transform Analysis}\label{GraphicalAlgorithmAnalysis}

\textcolor{black}{
With the built STT, this section presents a novel State Transform Analysis (STA) applied to algebra problem-solving. It begins with an overview of STA, followed by the real steps of the analysis. 
}

\subsection{Overview of STA}\label{subsec-STA-overview}

\textcolor{black}{
This section provides an explanation of State Transform Analysis (STA). To begin, the STA is defined as follows:
}

\begin{definition}[State Transform Analysis (STA)] \label{STA}
\textcolor{black}{
The State Transform Analysis is a method for analyzing algorithms based on State Transform Theory. It involves examining individual algorithms to generate state transform graphs that represent the transitions between different states during execution. These graphs are then aggregated to identify common properties and patterns across multiple algorithms. 
}
\end{definition}

The STA performs the following tasks: First, it identifies 47 representative solving algorithms from the literature since 2014. Next, it identifies 15 non-default states and 42 transforms. Following that, it constructs state transform graphs for all 47 algorithms and combines them into a single, aggregated state transform graph. Furthermore, it abstracts this aggregated graph into a state graph and defines four distinct approaches. 

The STA and the STT are highly related. The STT tells that solving algorithms have their own state transform graphs. This allows the algorithms to be uniformly compared and aggregated, highlighting their aggregate properties. In turn, the results from the STA validate the findings of the STT.


\subsection{States and Transforms}\label{subsec-StateTransform}
 Two criteria are set for selecting the reference papers. This paper selects papers with two criteria. The first one is that they published since 2014 (or after 2013). The other one is that they developed decent algorithms for solving algebra problems narrated by word and diagrams. In other words, we do not select the papers that just discuss some methods without presenting the algorithm. In addition, we select the final journal paper as our representative paper if one algorithm is developed from the immature  version in conference paper and the mature version in journal paper.  
 With the above criteria, this paper gains 47 papers, thus defined 47 algorithms $\mathcal{A}$ = \{$\mathcal{A}_{01}$, $\mathcal{A}_{02}$,...,$\mathcal{A}_{47}$\}, listed in Table \ref{table_all}. These 47 algorithms are divided into two categories of {\it original} and {\it incremental}.   
The algorithms in  {\it original} are the first 30 algorithms in $\mathcal{A}$, proposing new states and/or new transforms or made the significant progress, say that the developed algorithms are published in high rank journals; the algorithms in  {\it incremental} are the remaining 17 algorithms.

The 17 states are identified from $\mathcal{A}$ = \{$\mathcal{A}_{01}$, $\mathcal{A}_{02}$,...,$\mathcal{A}_{30}$\}, denoted as $\Delta$ = \{$\mathcal{S}_{00}$,$\mathcal{S}_{01}$,$\mathcal{S}_{02}$,...,$\mathcal{S}_{15}$, $\mathcal{S}_{os}$\}.
In $\Delta$, $\mathcal{S}_{00}$ and $\mathcal{S}_{os}$ are two default states because any solving algorithms must have them. Table \ref{table_all} lists the 15 non-default states and their basic facts. 
The items of the basic facts are reference paper (Within Table \ref{table_all}, we indicate reference papers by their serial numbers defined in the bottom of Table \ref{table_all}), development year, and occurrence number (the frequency with which the state is used in the 47 algorithms). We named each of the identified 17 states and described their original paper and features. Some further explanations are presented as follows.

\begin{table*}[!h]
\small
\centering
\caption{The lists and the basic facts of the identified 47 algorithms, 15 non-default states, and 42 transforms for solving algebra problems from the identified 47 papers since 2014}\label{table_all}%
\tabcolsep=0.65em
    \begin{subtable}[t]{\linewidth}
        \begin{tabular}{llllllllllllllll}
    
               \multicolumn{14}{l}{Table 1.1: The list of the identified 47 algorithms from the 47 identified papers and their basic facts}  \\    
     
                \toprule 
                 Notation & $\mathcal{A}_{01}$ & $\mathcal{A}_{02}$ & $\mathcal{A}_{03}$ & $\mathcal{A}_{04}$ & $\mathcal{A}_{05}$ & $\mathcal{A}_{06}$& $\mathcal{A}_{07}$ & $\mathcal{A}_{08}$ & $\mathcal{A}_{09}$ & $\mathcal{A}_{10}$ & $\mathcal{A}_{11}$ & $\mathcal{A}_{12}$ & $\mathcal{A}_{13}$  \\ 
                
                Year of Birth & 2014 & 2014 & 2015 & 2015 & 2015 & 2015 & 2017  &  2018  & 2018  & 2019  & 2019  & 2019  & 2019  \\ 
                
                SN of Paper & [2] & [1]& [5] & [6] & [3]  & [4]  & [12] &  [13]   &  [16] & [17]  & [19] &  [20]  & [21]    \\
                
               \# Created States &  1 & 1 & 1 & 1 & 0  & 1  & 2 & 0 & 0 & 1 &0 & 0  &0   \\
        
                \# Created Transforms &  2 &2 &2 &2 & 1  & 2  & 4 & 1  &  1 & 0  & 2 & 0  & 1    \\
          
              \midrule 
        
                Notation & $\mathcal{A}_{14}$ & $\mathcal{A}_{15}$  & $\mathcal{A}_{16}$ & $\mathcal{A}_{17}$ & $\mathcal{A}_{18}$ & $\mathcal{A}_{19}$ & $\mathcal{A}_{20}$ & $\mathcal{A}_{21}$& $\mathcal{A}_{22}$ & $\mathcal{A}_{23}$ & $\mathcal{A}_{24}$ & $\mathcal{A}_{25}$ & $\mathcal{A}_{26}$ &  \\ 
                
                Year of Birth & 2019   & 2020 & 2020 & 2020 & 2020  & 2020 & 2021 & 2021 & 2021 & 2021 & 2021 & 2022 & 2022  \\   
                  
                SN of Paper  &  [22] & [23] & [28] & [25] & [24] & [26] & [29] &  [30] & [34] & [31]  & [37]   & [41] & [42] \\ %
             
               \# Created States &  0 & 2  &  0 & 1 & 0& 0 & 1  &0  &1 &  0  &  0 &2  & 0   \\
        
                 \# Created Transforms &  1 & 3 &  0 &2 & 1 & 0 &2  &1  & 2 & 1  &  1& 4  & 1   \\
                
             \midrule           
                  Notation & $\mathcal{A}_{27}$ & $\mathcal{A}_{28}$ & $\mathcal{A}_{29}$ & $\mathcal{A}_{30}$ & $\mathcal{A}_{31}$  & $\mathcal{A}_{32}$  & $\mathcal{A}_{33}$  & $\mathcal{A}_{34}$  & $\mathcal{A}_{35}$  & $\mathcal{A}_{36}$  & $\mathcal{A}_{37}$  & $\mathcal{A}_{38}$  & $\mathcal{A}_{39}$ \\
                
                Year of Birth & 2023 &  2023 &  2023 &  2023 & 2016 & 2016  & 2016 & 2017 & 2017 & 2018 & 2018   & 2019  & 2020    \\ 
                         
                SN of Paper  & [47] & [45]   & [46]  & [43] & [7]  & [8]  & [9] & [10]  & [11]  & [14]  & [15] & [18]  & [27]  & \\   
                
               \# Created States &  0 & 0 & 0 & 0 & 0  & 0  & 0 &  0  &  0 & 0  & 0 &  0  & 0    \\
        
                \# Created Transforms & 1  & 0  &  1  &  1 &  0 & 0 & 0 & 0 & 0  & 0  & 0 &  0  &  0    \\  
                \midrule

                  Notation  & $\mathcal{A}_{40}$  & $\mathcal{A}_{41}$  & $\mathcal{A}_{42}$  & $\mathcal{A}_{43}$  & $\mathcal{A}_{44}$  & $\mathcal{A}_{45}$  & $\mathcal{A}_{46}$  & $\mathcal{A}_{47}$     &    &     &   &    &     \\
                
                Year of Birth &  2021 & 2021 &  2021 & 2021 & 2022 & 2022 & 2022 & 2023 &   &    &    &    &   &   &       \\
                              
                SN of Paper  & [40]  & [44] & [33] & [32] & [36] & [35] &  [38] & [39] &  &    &    &    &  
                  &   &   \\   
                
                \# Created States  & 0 & 0 & 0 & 0 & 0 & 0 & 0 & 0 & & & & &  \\
                \# Created Transforms & 0 & 0 & 0 & 0 & 0 & 0 & 0 & 0 & & & & &   \\
                
                \bottomrule  
        \end{tabular}
        \end{subtable}
    
        \begin{subtable}[t]{\linewidth}
        \begin{tabular}{lllllllllllllllll} 
        
                \multicolumn{16}{l}{}\\     
                \multicolumn{16}{l}{Table 1.2: The list of identified 15 non-default states from the first 30 identified algorithms and their basic facts}  \\    
                \toprule  
                
                Notation & $\mathcal{S}_{01}$ & $\mathcal{S}_{02}$ & $\mathcal{S}_{03}$ & $\mathcal{S}_{04}$ & $\mathcal{S}_{05}$ & $\mathcal{S}_{06}$ & $\mathcal{S}_{07}$& $\mathcal{S}_{08}$ & $\mathcal{S}_{09}$ & $\mathcal{S}_{10}$ & $\mathcal{S}_{11}$ & $\mathcal{S}_{12}$ & $\mathcal{S}_{13}$& $\mathcal{S}_{14}$ & $\mathcal{S}_{15}$ \\ 
                
                 Year of Birth & 2014 & 2014 &  2015 & 2015 &  2015 &  2017 & 2017 & 2018  &  2019 & 2019 & 2020  & 2021 & 2021 & 2022  & 2022  \\

                 Source Alg  & $\mathcal{A}_{01}$ &$\mathcal{A}_{02}$ &$\mathcal{A}_{03}$ & $\mathcal{A}_{06}$ & $\mathcal{A}_{04}$ & $\mathcal{A}_{07}$& $\mathcal{A}_{07}$ & $\mathcal{A}_{10}$ & $\mathcal{A}_{15}$ & $\mathcal{A}_{15}$& $\mathcal{A}_{17}$ &  $\mathcal{A}_{20}$ & $\mathcal{A}_{22}$ & $\mathcal{A}_{25}$ & $\mathcal{A}_{25}$ \\  
        
                 SN of Paper &[2] &  [1] &  [5] & [4] & [6]  & [12] & [12] & [17] & [23] & [23] & [25]  &  [29] & [34] & [41] & [41] \\ 
        
                \# Occurrence &  35 & 7 &  4 & 1 & 1  & 26 & 9  & 14 & 5 & 3 & 2  & 3  &   2 &   1  & 1    \\  
                   \bottomrule
        \end{tabular}
        \end{subtable}

        \begin{subtable}[t]{\linewidth}
        \begin{tabular}{lllllllllllllllll} 
                      \multicolumn{16}{l}{}  \\    
                      \multicolumn{16}{l}{Table 1.3: The list of the identified 42 transforms from the first 30 identified 
              algorithms and their basic facts}  \\    
                      \toprule 
                      
              Notation & $\mathcal{t}_{01}$ & $\mathcal{t}_{02}$ & $\mathcal{t}_{03}$ & $\mathcal{t}_{04}$ & $\mathcal{t}_{05}$ & $\mathcal{t}_{06}$ & $\mathcal{t}_{07}$ & $\mathcal{t}_{08}$ & $\mathcal{t}_{09}$ & $\mathcal{t}_{10}$ & $\mathcal{t}_{11}$ & $\mathcal{t}_{12}$ & $\mathcal{t}_{13}$ & $\mathcal{t}_{14}$ & $\mathcal{t}_{15}$  \\
          
                  Year of Birth & 2014  & 2014  & 2014  & 2014  & 2015 & 2015  & 2015  & 2015 & 2015  & 2015  & 2015  & 2017 & 2017 & 2017  & 2017  \\
              SN of Paper & [2] & [2] & [1] & [1] & [5] & [5] & [4] & [4] & [3] & [6]  & [6] & [12] & [12]  & [12] & [12]  \\
              Start State & $\mathcal{S}_{00}$ & $\mathcal{S}_{01}$ & $\mathcal{S}_{00}$ & $\mathcal{S}_{02}$ & $\mathcal{S}_{00}$ & $\mathcal{S}_{03}$ & $\mathcal{S}_{00}$ & $\mathcal{S}_{04}$ & $\mathcal{S}_{00}$ & $\mathcal{S}_{00}$ & $\mathcal{S}_{05}$   & $\mathcal{S}_{00}$ & $\mathcal{S}_{06}$ & $\mathcal{S}_{07}$ & $\mathcal{S}_{06}$ \\
              Target state & $\mathcal{S}_{01}$ & $\mathcal{S}_{os}$ & $\mathcal{S}_{02}$ & $\mathcal{S}_{os}$ & $\mathcal{S}_{03}$ & $\mathcal{S}_{01}$ & $\mathcal{S}_{04}$ &  $\mathcal{S}_{03}$ & $\mathcal{S}_{02}$ & $\mathcal{S}_{05}$ & $\mathcal{S}_{02}$  & $\mathcal{S}_{06}$ & $\mathcal{S}_{07}$ & $\mathcal{S}_{01}$ & $\mathcal{S}_{01}$  \\
              \# Occurrence & 3     & 32    & 3     & 7     & 3     & 6     & 1   &1  & 1     & 1     & 1  & 26   & 6  & 3   & 8     \\
              \midrule 
              
              Notation & $\mathcal{t}_{16}$ & $\mathcal{t}_{17}$ & $\mathcal{t}_{18}$ & $\mathcal{t}_{19}$ & $\mathcal{t}_{20}$ & $\mathcal{t}_{21}$ & $\mathcal{t}_{22}$ & $\mathcal{t}_{23}$ & $\mathcal{t}_{24}$ & $\mathcal{t}_{25}$ & $\mathcal{t}_{26}$ & $\mathcal{t}_{27}$ & $\mathcal{t}_{28}$ & $\mathcal{t}_{29}$ & $\mathcal{t}_{30}$ \\
              Year of Birth   & 2018 & 2018  & 2019    & 2019  & 2019  & 2019  & 2019  & 2019 & 2019 & 2020  & 2020  & 2020 & 2021 & 2021 & 2021   \\
              SN of Paper  & [13] & [16]  & [19] & [19]  & [21] & [22]  & [23] & [23] & [23] & [24] & [25] &[25]  & [29] & [29] & [30] \\
              Start State   & $\mathcal{S}_{06}$  & $\mathcal{S}_{06}$ & $\mathcal{S}_{06}$ &  $\mathcal{S}_{08}$ & $\mathcal{S}_{06}$ & $\mathcal{S}_{06}$ & {$\mathcal{S}_{06}$} & $\mathcal{S}_{09}$  & $\mathcal{S}_{10}$   & {$\mathcal{S}_{06}$} & $\mathcal{S}_{06}$  & $\mathcal{S}_{11}$  &  $\mathcal{S}_{06}$ & $\mathcal{S}_{12}$ & $\mathcal{S}_{06}$  \\
              Target state  & $\mathcal{S}_{07}$  & $\mathcal{S}_{08}$ &$\mathcal{S}_{08}$ &   $\mathcal{S}_{01}$ & $\mathcal{S}_{08}$ & $\mathcal{S}_{08}$ & $\mathcal{S}_{09}$  & $\mathcal{S}_{10}$  & $\mathcal{S}_{os}$  & $\mathcal{S}_{07}$ & $\mathcal{S}_{11}$  & $\mathcal{S}_{08}$ & $\mathcal{S}_{12}$ & $\mathcal{S}_{02}$ & $\mathcal{S}_{13}$  \\
              \# Occurrence   & 1  & 1 & 1  & 12 & 1 & 1 & 1 & 3 & 3 & 3 & 4 & 8 & 1 & 1 & 1     \\
              \midrule
          
              Notation & $\mathcal{t}_{31}$  & $\mathcal{t}_{32}$ & $\mathcal{t}_{33}$ & $\mathcal{t}_{34}$  & $\mathcal{t}_{35}$ & $\mathcal{t}_{36}$  &  $\mathcal{t}_{37}$ & $\mathcal{t}_{38}$ &   $\mathcal{t}_{39}$ & $\mathcal{t}_{40}$ &  $\mathcal{t}_{41}$ & $\mathcal{t}_{42}$ & & &       \\
              Year of Birth  & 2021 & 2021  & 2021 & 2021  & 2022  & 2022  & 2022  & 2022 & 2022 & 2023  &  2023 &  2023 &   &    &             \\
              SN of Paper  & [31] & [34] & [34] & [37] & [41] & [41] & [41] & [41] & [42] & [47]  & [46] & [43] & & & \\
              Start State  & $\mathcal{S}_{06}$ & $\mathcal{S}_{06}$ & $\mathcal{S}_{13}$ & $\mathcal{S}_{06}$ & $\mathcal{S}_{00}$ & $\mathcal{S}_{00}$ & $\mathcal{S}_{14}$ & $\mathcal{S}_{15}$ & $\mathcal{S}_{06}$ & $\mathcal{S}_{06}$ & $\mathcal{S}_{00}$ & $\mathcal{S}_{00}$ & &  &\\
              Target state  & $\mathcal{S}_{09}$ & $\mathcal{S}_{13}$ & $\mathcal{S}_{03}$ & $\mathcal{S}_{02}$ & $\mathcal{S}_{14}$ & $\mathcal{S}_{14}$ & $\mathcal{S}_{15}$ & $\mathcal{S}_{os}$  & $\mathcal{S}_{08}$  & $\mathcal{S}_{09}$ & $\mathcal{S}_{09}$ & $\mathcal{S}_{09}$ & & & \\
              \# Occurrence    & 3 & 1 & 1 & 1 & 1 & 1 & 1 & 1 & 1 & 3 & 1 & 1 & & &         \\
              
              \bottomrule 
              \end{tabular}
        \end{subtable}
    
        \begin{subtable}[t]{\linewidth}
        \scriptsize
                \begin{tabular}{@{}l@{}}
         The notations from [1] to [47] appeared in this table are defined as follows: 
                     {[}1{]}: \cite{Kushman2014} {[}2{]}: \cite{Hosseini2014} {[}3{]}: \cite{Zhou2015}\\ {[}4{]}: \cite{Koncel2015} 
                     {[}5{]}: \cite{Roy2015} {[}6{]}: \cite{Shi2015}
          {[}7{]}: \cite{Mitra2016} {[}8{]}: \cite{Roy2016a} {[}9{]}: \cite{Upadhyay2016} \\
          {[}10{]}: \cite{Huang2017} 
          {[}11{]}: \cite{Ling2017} {[}12{]}: \cite{Wang2017}
          {[}13{]}: \cite{Huang2018Neural} {[}14{]}: \cite{Huang2018Using} {[}15{]}: \cite{Roy2018} \\
          {[}16{]}: \cite{Wang2018MathDQN} 
          {[}17{]}: \cite{Wang2018Translating}{[}18{]}: \cite{Li2019}
          {[}19{]}: \cite{Liu2019} {[}20{]}: \cite{Rehman2019} {[}21{]}: \cite{Wang2019} {[}22{]}: \cite{Xie2019} \\
          {[}23{]}: \cite{Yu2019} 
          {[}24{]}: \cite{Kim2020}
          {[}25{]}: \cite{Li2020} {[}26{]}: \cite{Mandal2020} {[}27{]}: \cite{Wu2020} {[}28{]}: \cite{Zhang2020graph2tree} {[}29{]}: \cite{Cao2021} \\
          {[}30{]}: \cite{Lin2021}
          {[}31{]}: \cite{Lyu2021} {[}32{]}: \cite{Mandal2021} {[}33{]}: \cite{Qin2021}{[}34{]}: \cite{Tsai2021}{[}35{]}: \cite{Wu2021a} \\
          {[}36{]}: \cite{Wu2021b}
          {[}37{]}: \cite{Zaporojets2021} {[}38{]}: \cite{Jie2022} {[}39{]}: \cite{Li2022} {[}40{]}: \cite{Liang2022} {[}41{]}: \cite{YuSun2022} {[}42{]}: \cite{Zhang2022}\\
          {[}43{]}: \cite{Huang2023} {[}44{]}: \cite{Liang2023} {[}45{]}: \cite{Lin2023} {[}46{]}: \cite{Meng2023} {[}47{]}: \cite{Yu2023}  \\                                                                                                                                                     
          \bottomrule 
        \end{tabular}
        \end{subtable}

\end{table*}

\subsection{State Transform Graphs of Individual Algorithms} \label{subsec-indi-STG} 

With the identified 17 states and 42 transforms from the 47 algorithms in the preceding listed in Table \ref{table_all}, this section will show that all the 47 algorithms can be depicted by the State Transform Graphs. Figure \ref{example_A11} uses state transform graph to depict $\mathcal{A}_{07}$ proposed in \cite{Wang2017}, being an example to draw the state transform graph of an algorithm.

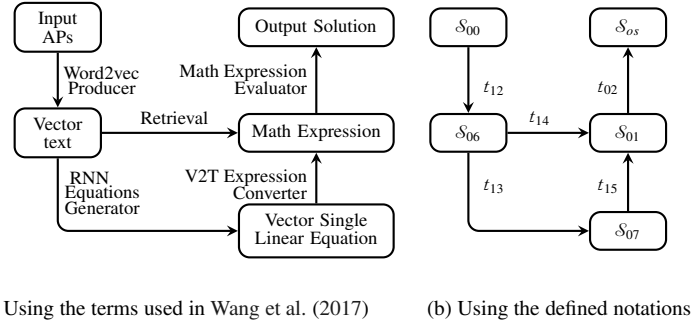
\begin{figure}[htp]
     \centering
    \thispagestyle{empty}
    \tikzstyle{state} = [rectangle, text width=1.8cm,thick,rounded corners, minimum width = 1cm, minimum height=0.5cm,text centered, draw = black,font=\fontsize{7}{7}\selectfont]
    \tikzstyle{state1} = [rectangle, text width=0.7cm,thick,rounded corners, minimum width = 1cm, minimum height=0.5cm,text centered, draw = black,font=\fontsize{7}{7}\selectfont]
    \tikzstyle{state2} = [rectangle, text width=0.9cm,thick,rounded corners, minimum width = 1cm, minimum height=0.5cm,text centered, draw = black,font=\fontsize{7}{7}\selectfont]
    \vspace{0.3cm}
    \tikzstyle{arrow} = [->,>=stealth]
    \begin{tikzpicture}[node distance=1cm]
        \selectcolormodel{gray}
        \node[state2](A){Input APs};
        \node[state2,right of = A,yshift = -1.4cm,xshift=-1cm](B){Vector text};
        \node[state,right of = B,yshift = -1.3cm,xshift=2.40cm](C){Vector Single Linear Equation};
        \node[state,right of = C,yshift = 1.3cm,xshift=-1.0cm](D){Math Expression};
        \node[state,right of = D,yshift = 1.4cm,xshift=-1.0cm](E){Output Solution};
        
        \node[state1,right of = E,xshift=1.0cm](s0){$\mathcal{S}_{00}$};
        \node[state1,right of = s0,yshift = -1.4cm,xshift=-1cm](s5){$\mathcal{S}_{06}$};
        \node[state1,right of = s5,yshift = -1.3cm,xshift=1.1cm](s6){$\mathcal{S}_{07}$};
        \node[state1,right of = s6,yshift = 1.3cm,xshift=-1cm](s3){$\mathcal{S}_{01}$};
        \node[state1,right of = s3,yshift = 1.4cm,xshift=-1cm](s100){$\mathcal{S}_{os}$};
        
        \draw [arrow,rounded corners,  thick] (A) -- (B);
        \draw [arrow,rounded corners,  thick] (B) |- (C);
        \draw [arrow,rounded corners,  thick] (B) -- (D);
        \draw [arrow,rounded corners,  thick] (C) -- (D);
        \draw [arrow,rounded corners,  thick] (D) -- (E);
        \draw [arrow,rounded corners,  thick] (s0) -- (s5);
        \draw [arrow,rounded corners,  thick] (s5) |- (s6);
        \draw [arrow,rounded corners,  thick] (s5) -- (s3);
        \draw [arrow,rounded corners,  thick] (s6) -- (s3);
        \draw [arrow,rounded corners,  thick] (s3) -- (s100);
        
        \coordinate[label = left:{\fontsize{7}{7}\selectfont Word2vec}](L1) at(1.2,-0.6);
        \coordinate[label = left:{\fontsize{7}{7}\selectfont Producer}](L1) at(1.15,-0.8);
        \coordinate[label = left:{\fontsize{7}{7}\selectfont RNN }](L2) at(0.8,-2);
        \coordinate[label = left:{\fontsize{7}{7}\selectfont Equations}](L2) at(1.2,-2.2);
        \coordinate[label = left:{\fontsize{7}{7}\selectfont Generator}](L2) at(1.2,-2.4);
        \coordinate[label = left:{\fontsize{7}{7}\selectfont Retrieval}](L3) at(2.1,-1.2);
        \coordinate[label = left:{\fontsize{7}{7}\selectfont Math Expression}](L5) at(3.4,-0.6);
        \coordinate[label = left:{\fontsize{7}{7}\selectfont  Evaluator}](L5b) at(3.4,-0.8);
        \coordinate[label = left:{\fontsize{7}{7}\selectfont V2T Expression}](L6) at(3.4,-2);
        \coordinate[label = left:{\fontsize{7}{7}\selectfont Converter}](L6b) at(3.4,-2.2);
        
        \coordinate[label = left:{\fontsize{7}{7}\selectfont $\mathcal{t}_{12}$}](L6) at(6.0,-0.8);
        \coordinate[label = left:{\fontsize{7}{7}\selectfont $\mathcal{t}_{13}$}](L7) at(6.0,-2.1);
        \coordinate[label = left:{\fontsize{7}{7}\selectfont $\mathcal{t}_{14}$}](L8) at(6.6,-1.2);
        \coordinate[label = left:{\fontsize{7}{7}\selectfont $\mathcal{t}_{15}$}](L9) at(7.5,-2.1);
        \coordinate[label = left:{\fontsize{7}{7}\selectfont $\mathcal{t}_{02}$}](L10) at(7.5,-0.8);

        \coordinate[label = below:{\fontsize{8}{8}\selectfont (a) Using the terms used in \cite{Wang2017}}](L13) at(1.5,-3.5);
        \coordinate[label = below:{\fontsize{8}{8}\selectfont (b) Using the defined notations}](L13) at(6.6,-3.5);

     \end{tikzpicture}
     \caption{Two representations of state transform graphs for $\mathcal{A}_{07}$: (a) using the terminology from the original paper, and (b) adopting the notations introduced in this paper.}
     \label{example_A11}
\end{figure}

\cite{Wang2017} proposed two methods, namely a seq2seq model and a similarity-based retrieval model to obtain expressions from Input APs. Thus, the algorithm in \cite{Wang2017} has two paths to obtain the expressions. 
Figure \ref{example_A11}(a) and \ref{example_A11}(b) are the two versions of the state transform graph of $\mathcal{A}_{07}$, where \ref{example_A11}(a) is the the first version, which uses the terminologies in the original paper \citep{Wang2017} and  \ref{example_A11}(b) is the second version, which uses the notations defined in this paper.


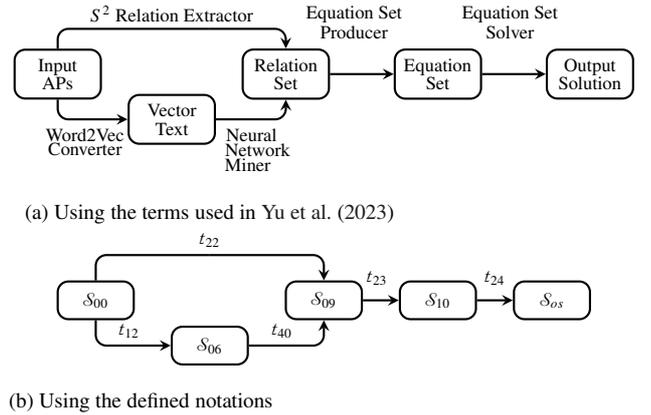
\begin{figure}[htp]
     \centering
    \thispagestyle{empty}
    \tikzstyle{state} = [rectangle, text width=1.8cm,thick,rounded corners, minimum width = 1cm, minimum height=0.5cm,text centered, draw = black,font=\fontsize{7}{7}\selectfont]
    \tikzstyle{state1} = [rectangle, text width=0.7cm,thick,rounded corners, minimum width = 1cm, minimum height=0.5cm,text centered, draw = black,font=\fontsize{7}{7}\selectfont]
    \tikzstyle{state2} = [rectangle, text width=0.9cm,thick,rounded corners, minimum width = 1cm, minimum height=0.5cm,text centered, draw = black,font=\fontsize{7}{7}\selectfont]
    \tikzstyle{E}=[]
    \vspace{0.3cm}
    \tikzstyle{arrow} = [->,>=stealth]
    \begin{tikzpicture}[node distance=1cm]
        \selectcolormodel{gray}

        \node[state2,yshift =-3.2cm,xshift=1cm](a96){Input APs};  
        \node[state2,right of = a96,yshift =-0.6cm,xshift=0.5cm](a97){Vector Text};
        \node[state2,right of = a96,xshift=2cm](a98){Relation Set};
        \node[state2,right of = a98,xshift=1cm](a99){Equation Set};
        \node[state2,right of = a99,xshift=1cm](a100){Output Solution}; 
  
        \draw [arrow,rounded corners,  thick] (a96) |-  (a97);
        \draw [arrow,rounded corners,  thick] (a97) -|  (a98); 
        \draw [arrow,rounded corners,  thick] (a98) --  (a99); 
        \draw [arrow,rounded corners,  thick] (a99) --  (a100); 
        \draw [rounded corners,  thick] (a96) |-  (3,-2.6); 
        \draw [arrow,rounded corners,  thick] (3,-2.6) -|  (a98); 
  
        \coordinate[label = center:{\fontsize{7}{7}\selectfont $S^2$ Relation Extractor}](S) at(2.5,-2.4);
        \coordinate[label = center:{\fontsize{7}{7}\selectfont Word2Vec}](S) at(1.36,-4);
         \coordinate[label = center:{\fontsize{7}{7}\selectfont Converter}](S) at(1.36,-4.2);
        
        \coordinate[label = center:{\fontsize{7}{7}\selectfont Neural}](S) at(3.55,-4);
        \coordinate[label = center:{\fontsize{7}{7}\selectfont Network}](S) at(3.63,-4.2);
        \coordinate[label = center:{\fontsize{7}{7}\selectfont Miner}](S) at(3.5,-4.4);
        \coordinate[label = center:{\fontsize{7}{7}\selectfont Equation Set}](S) at(4.9,-2.4);
        \coordinate[label = center:{\fontsize{7}{7}\selectfont Producer}](S) at(4.9,-2.65);
        \coordinate[label = center:{\fontsize{7}{7}\selectfont Equation Set}](S) at(6.95,-2.4);
        \coordinate[label = center:{\fontsize{7}{7}\selectfont Solver}](S) at(6.95,-2.65);

        \node[state1,right of = a96,yshift =-3cm,xshift=-0.5cm](96){$\mathcal{S}_{00}$};  
        \node[state1,right of = 96,yshift =-0.6cm,xshift=0.5cm](97){$\mathcal{S}_{06}$};
        \node[state1,right of = 96,xshift=2cm](98){$\mathcal{S}_{09}$};
        \node[state1,right of = 98,xshift=0.5cm](99){$\mathcal{S}_{10}$};
        \node[state1,right of = 99,xshift=0.5cm](100){$\mathcal{S}_{os}$}; 
  
        \draw [arrow,rounded corners,  thick] (96) |-  (97);
        \draw [arrow,rounded corners,  thick] (97) -|  (98); 
        \draw [arrow,rounded corners,  thick] (98) --  (99); 
        \draw [arrow,rounded corners,  thick] (99) --  (100); 
        \draw [rounded corners,  thick] (96) |-  (3,-5.6); 
        \draw [arrow,rounded corners,  thick] (3,-5.6) -|  (98); 
  
        \coordinate[label = center:{\fontsize{7}{7}\selectfont $\mathcal{t}_{22}$}](S) at(3,-5.4);
        \coordinate[label = center:{\fontsize{7}{7}\selectfont $\mathcal{t}_{12}$}](S) at(1.94,-6.6);
        \coordinate[label = center:{\fontsize{7}{7}\selectfont $\mathcal{t}_{40}$}](S) at(3.95,-6.6);
        \coordinate[label = center:{\fontsize{7}{7}\selectfont $\mathcal{t}_{23}$}](S) at(5.2,-5.9);
        \coordinate[label = center:{\fontsize{7}{7}\selectfont $\mathcal{t}_{24}$}](S) at(6.75,-5.9);

        \coordinate[label = below:{\fontsize{8}{8}\selectfont (a) Using the terms used in \cite{Yu2023}}](L13) at(3,-4.8);
        \coordinate[label = below:{\fontsize{8}{8}\selectfont (b) Using the defined notations}](L13) at(2.1,-7.3);
         
     \end{tikzpicture}
     \caption{Two representations of state transform graphs for $\mathcal{A}_{27}$: (a) using the terminology from the original paper, and (b) adopting the notations introduced in this paper.}
     \label{example_A27}
\end{figure}

Figure \ref{example_A27} uses state transform graphs to depict $\mathcal{A}_{27}$ proposed in \cite{Yu2023}, being another example to draw the state transform graph of an algorithm. \cite{Yu2023} proposed a relation-centric algorithm for solving arithmetic word problems, in which obtaining a set of relations as the understood state of the algorithm is the key task.  
The algorithm obtains the explicit and implicit relations separately.  Hence, the algorithm in \cite{Yu2023} has two paths to obtain the relations. 
Figure \ref{example_A27}(a) and \ref{example_A27}(b) are the two versions of the state transform graph of $\mathcal{A}_{27}$, where \ref{example_A27}(a) is the the first version, which uses the terminologies in the original paper \citep{Yu2023} and  \ref{example_A27}(b) is the second version, which uses the notations defined in this paper.

From both the above examples, we can know that the state transform graphs for depicting individual algorithms can be a true graph, though most of such graphs may degrade into simple paths and that both algorithms can be depicted by state transform graphs, though they use the very different techniques.

\subsection{Aggregated State Transform Graph}  
\label{subsec_AggreSTG}

Figure \ref{example_A11} shows that the algorithm proposed in \cite{Wang2017} is decomposable and can be abstracted into a graph with states as its nodes and transforms as its links. From the above analysis, we know that all the algorithms share the 17 states and 42 transforms. Based on these facts, the graphs of depicting a collection of algorithms can merge into an aggregated graph. Thus, we have a definition as follows.

\begin{definition}[Aggregated Graph]
Let $\mathcal{A}$ be a collection of algorithms for solving algebra problems.  A graph $\mathcal{H}$ is called  as an aggregated graph of $\mathcal{A}$ if it is formed by integrating all the graphs of depicting all individual algorithms in $\mathcal{A}$.  
\end{definition}

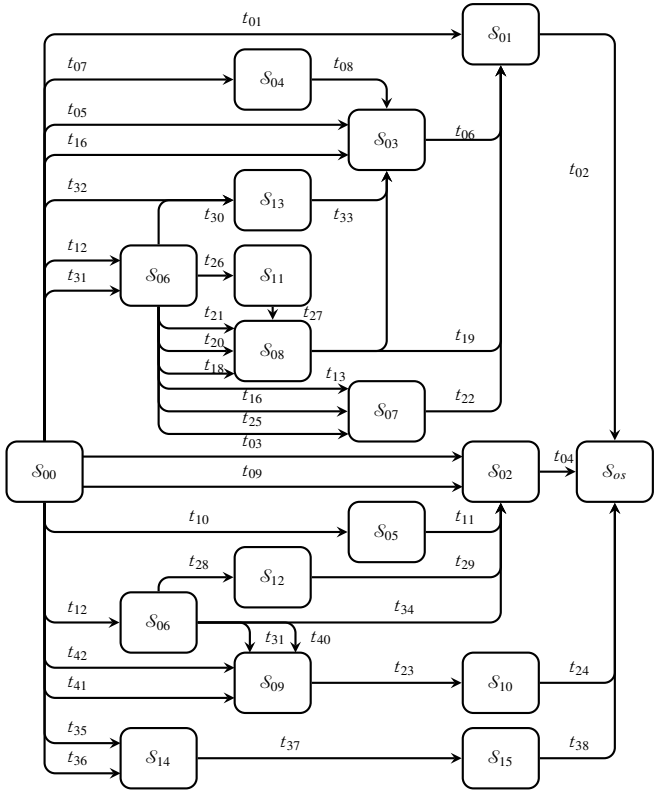
\begin{figure}[htp]
      \centering
    \thispagestyle{empty}
    \tikzstyle{state} = [rectangle, thick,rounded corners, minimum width = 1cm, minimum height=0.8cm,text centered, draw = black,font=\fontsize{7}{7}\selectfont]
    
    \vspace{0.3cm}
    \tikzstyle{arrow} = [->,>=stealth]
    \begin{tikzpicture}[node distance=1cm]
    \node[state](s0){$\mathcal{S}_{00}$};
        \node[state,right of = s0,xshift=5.0cm](s2){$\mathcal{S}_{02}$};
        \node[state,right of = s2,xshift=0.5cm](see){$\mathcal{S}_{os}$};
        \node[state,right of = s0,xshift=3.5cm,yshift=0.8cm](s8){$\mathcal{S}_{07}$};
        \node[state,right of = s0,xshift=0.5cm,yshift=2.6cm](s7){$\mathcal{S}_{06}$};
        \node[state,right of = s7,xshift=0.5cm,yshift=1cm](s12){$\mathcal{S}_{13}$};
        \node[state,right of = s12,xshift=0.5cm,yshift=0.8cm](s3){$\mathcal{S}_{03}$};
        \node[state,right of = s3,xshift=-2.5cm,yshift=0.8cm](s04){$\mathcal{S}_{04}$};
        \node[state,right of = s7,xshift=0.5cm,yshift=-1cm](s9){$\mathcal{S}_{08}$};
        \node[state,right of = s04,xshift=2cm,yshift=0.6cm](s1){$\mathcal{S}_{01}$};
        \node[state,right of = s0,xshift=3.5cm,yshift=-0.8cm](s4){$\mathcal{S}_{05}$};
        \node[state,right of = s0,xshift=0.5cm,yshift=-2cm](s07){$\mathcal{S}_{06}$};
        \node[state,right of = s0,xshift=2cm,yshift=-2.8cm](s5){$\mathcal{S}_{09}$};
        \node[state,right of = s5,xshift=2cm](s6){$\mathcal{S}_{10}$};
        \node[state,right of = s0,xshift=0.5cm,yshift=-3.8cm](s13){$\mathcal{S}_{14}$};
        \node[state,right of = s13,xshift=3.5cm](s14){$\mathcal{S}_{15}$};
        \node[state,right of = s07,xshift=0.5cm,yshift=0.6cm](s11){$\mathcal{S}_{12}$};
        \node[state,right of = s7,xshift=0.5cm](s012){$\mathcal{S}_{11}$};
        \draw [arrow,rounded corners,  thick] (s0) |-  (1,2.4);
        \draw [arrow,rounded corners,  thick] (s0) |-  (s12);
        \draw [arrow,rounded corners,  thick] (s0) |-  (s04);
        \draw [arrow,rounded corners,  thick] (s0) |-  (s1);
        \draw [arrow,rounded corners,  thick] (s7) |-  (s12);
        \draw [arrow,rounded corners,  thick] (s7) |-  (s9);
        \draw [arrow,rounded corners,  thick] (s7) |-  (2.5,1.9);
        \draw [arrow,rounded corners,  thick] (s7) |-  (2.5,1.3);
        \draw [arrow,rounded corners,  thick] (s7) |-  (s8);
        \draw [arrow,rounded corners,  thick] (s7) |-  (4,1.1);
        \draw [arrow,rounded corners,  thick] (s7) |-  (4,0.5);
        \draw [arrow,rounded corners,  thick] (s0) |-  (s4);
        \draw [arrow,rounded corners,  thick] (s0) |-  (4,4.6);
        \draw [arrow,rounded corners,  thick] (s0) |-  (4,4.2);
        \draw [arrow,rounded corners,  thick] (s0) |-  (s07);
        \draw [arrow,rounded corners,  thick] (s0) |-  (1,2.8);
        \draw [arrow,rounded corners,  thick] (s07) |-  (s11);

        \draw [arrow,rounded corners,  thick] (s0) |-  (2.5,-2.6);
        \draw [arrow,rounded corners,  thick] (s0) |-  (2.5,-3);
        
        \draw [arrow,rounded corners,  thick] (s0) |-  (1,-3.6);
        \draw [arrow,rounded corners,  thick] (s0) |-  (1,-4.0);
        \draw [arrow,rounded corners,  thick] (s04) -|  (s3);
        \draw [arrow,rounded corners,  thick] (s3) -|  (s1);
        \draw [arrow,rounded corners,  thick] (s12) -|  (s3);
        \draw [arrow,rounded corners,  thick] (s9) -|  (s3);
        \draw [arrow,rounded corners,  thick] (s9) -|  (s1);
        \draw [arrow,rounded corners,  thick] (s8) -|  (s1);
        \draw [arrow,rounded corners,  thick] (s4) -|  (s2);
        \draw [arrow,rounded corners,  thick] (s11) -|  (s2);
        \draw [arrow,rounded corners,  thick] (s07) -|  (s2);
        \draw [arrow,rounded corners,  thick] (s1) -|  (see);
        \draw [arrow,rounded corners,  thick] (s6) -|  (see);
        \draw [arrow,rounded corners,  thick] (s14) -|  (see);
        \draw [arrow,rounded corners,  thick] (0.5,0.2) --  (5.5,0.2);
        \draw [arrow,rounded corners,  thick] (0.5,-0.2) --  (5.5,-0.2);
        \draw [arrow,rounded corners,  thick] (s5) --  (s6);
        \draw [arrow,rounded corners,  thick] (s7) --  (s012);
        \draw [arrow,rounded corners,  thick] (s012) --  (s9);
        \draw [arrow,rounded corners,  thick] (s13) --  (s14);
        \draw [arrow,rounded corners,  thick] (s2) --  (see);
        \draw [arrow,rounded corners,  thick] (s07) -|  (2.7,-2.4);
        \draw [arrow,rounded corners,  thick] (s07) -|  (3.3,-2.4);
        \coordinate[label = left:{\fontsize{7}{7}\selectfont $\mathcal{t}_{01}$}](fa) at(3,6);
        \coordinate[label = left:{\fontsize{7}{7}\selectfont $\mathcal{t}_{02}$}](fa) at(7.3,4);
        \coordinate[label = left:{\fontsize{7}{7}\selectfont $\mathcal{t}_{16}$}](fa) at(0.7,4.4);
        \coordinate[label = left:{\fontsize{7}{7}\selectfont $\mathcal{t}_{05}$}](fa) at(0.7,4.8);
        \coordinate[label = left:{\fontsize{7}{7}\selectfont $\mathcal{t}_{07}$}](fa) at(0.7,5.4);
        \coordinate[label = left:{\fontsize{7}{7}\selectfont $\mathcal{t}_{06}$}](fa) at(5.8,4.5);
        \coordinate[label = left:{\fontsize{7}{7}\selectfont $\mathcal{t}_{33}$}](fa) at(4.2,3.4);
        \coordinate[label = left:{\fontsize{7}{7}\selectfont $\mathcal{t}_{08}$}](fa) at(4.2,5.4);
        \coordinate[label = left:{\fontsize{7}{7}\selectfont $\mathcal{t}_{32}$}](fa) at(0.7,3.8);
        \coordinate[label = left:{\fontsize{7}{7}\selectfont $\mathcal{t}_{30}$}](fa) at(2.5,3.4);
        \coordinate[label = left:{\fontsize{7}{7}\selectfont $\mathcal{t}_{12}$}](fa) at(0.7,3);
        \coordinate[label = left:{\fontsize{7}{7}\selectfont $\mathcal{t}_{31}$}](fa) at(0.7,2.6);
        \coordinate[label = left:{\fontsize{7}{7}\selectfont $\mathcal{t}_{18}$}](ta) at(2.5,1.4);
        \coordinate[label = left:{\fontsize{7}{7}\selectfont $\mathcal{t}_{20}$}](fa) at(2.5,1.74);
        \coordinate[label = left:{\fontsize{7}{7}\selectfont $\mathcal{t}_{21}$}](fa) at(2.5,2.1);
        \coordinate[label = left:{\fontsize{7}{7}\selectfont $\mathcal{t}_{26}$}](fa) at(2.5,2.8);
        \coordinate[label = left:{\fontsize{7}{7}\selectfont $\mathcal{t}_{27}$}](fa) at(3.8,2.1);
        \coordinate[label = left:{\fontsize{7}{7}\selectfont $\mathcal{t}_{19}$}](fa) at(5.8,1.8);
        \coordinate[label = left:{\fontsize{7}{7}\selectfont $\mathcal{t}_{13}$}](fa) at(4.1,1.3);
        \coordinate[label = left:{\fontsize{7}{7}\selectfont $\mathcal{t}_{16}$}](fa) at(3,1);
        \coordinate[label = left:{\fontsize{7}{7}\selectfont $\mathcal{t}_{25}$}](fa) at(3,0.7);
        \coordinate[label = left:{\fontsize{7}{7}\selectfont $\mathcal{t}_{22}$}](fa) at(5.8,1);
        \coordinate[label = left:{\fontsize{7}{7}\selectfont $\mathcal{t}_{31}$}](f31) at(3.3,-2.2);
        \coordinate[label = left:{\fontsize{7}{7}\selectfont $\mathcal{t}_{40}$}](f36) at(3.9,-2.2);
        \coordinate[label = left:{\fontsize{7}{7}\selectfont $\mathcal{t}_{03}$}](f3) at(3,0.4);
        \coordinate[label = left:{\fontsize{7}{7}\selectfont $\mathcal{t}_{09}$}](f3) at(3,0);
        \coordinate[label = left:{\fontsize{7}{7}\selectfont $\mathcal{t}_{04}$}](f4) at(7.1,0.2);
        \coordinate[label = left:{\fontsize{7}{7}\selectfont $\mathcal{t}_{10}$}](f8) at(2.3,-0.6);
        \coordinate[label = left:{\fontsize{7}{7}\selectfont $\mathcal{t}_{11}$}](f9) at(5.8,-0.6);
        \coordinate[label = left:{\fontsize{7}{7}\selectfont $\mathcal{t}_{12}$}](f14) at(0.7,-1.8);
        \coordinate[label = left:{\fontsize{7}{7}\selectfont $\mathcal{t}_{23}$}](f11) at(5,-2.6);
        \coordinate[label = left:{\fontsize{7}{7}\selectfont $\mathcal{t}_{28}$}](f28) at(2.3,-1.2);
        \coordinate[label = left:{\fontsize{7}{7}\selectfont $\mathcal{t}_{29}$}](f29) at(5.8,-1.2);
        \coordinate[label = left:{\fontsize{7}{7}\selectfont $\mathcal{t}_{37}$}](f34) at(3.5,-3.6);
        \coordinate[label = left:{\fontsize{7}{7}\selectfont $\mathcal{t}_{34}$}](f35) at(5,-1.8);
        \coordinate[label = left:{\fontsize{7}{7}\selectfont $\mathcal{t}_{38}$}](f39) at(7.3,-3.6);
        \coordinate[label = left:{\fontsize{7}{7}\selectfont $\mathcal{t}_{24}$}](f12) at(7.3,-2.6);
        \coordinate[label = left:{\fontsize{7}{7}\selectfont $\mathcal{t}_{35}$}](f32) at(0.7,-3.4);
        \coordinate[label = left:{\fontsize{7}{7}\selectfont $\mathcal{t}_{36}$}](f37) at(0.7,-3.8);
        \coordinate[label = left:{\fontsize{7}{7}\selectfont $\mathcal{t}_{42}$}](f13) at(0.7,-2.4);
        \coordinate[label = left:{\fontsize{7}{7}\selectfont $\mathcal{t}_{41}$}](f13) at(0.7,-2.8);

    \end{tikzpicture}
     \caption{The aggregated graph of the identified 47 algorithms.}
     \label{fig-AggregateGraph}
\end{figure}

Figure \ref{fig-AggregateGraph} is the aggregated graph of the algorithm set $\mathcal{A}=\{\mathcal{A_{01}},\mathcal{A_{02}},\dots,\mathcal{A_{47}}\}$. 
Figure \ref{fig-AggregateGraph} includes all the states, all the transforms, and their links of all 47 algorithms for solving algebra problems in a graph. 
Hence, Figure \ref{fig-AggregateGraph} reflects the overall picture of all the algorithms. 

{\bf Notice:}When some transforms can generate a linear equation set, of course they can generate a linear equation. However, 
Figure \ref{fig-AggregateGraph} does not include the links that they can generate a linear equation for emphasizing their highest capabilities only.  

\subsection{State Graph and Approach} \label{subsect_STnA}

Since the aggregated graph in Figure  \ref{fig-AggregateGraph}  is for the collection of 47 algorithms, Figure \ref{fig-AggregateGraph} can reflect the advance in developing algorithms for AP-solving. Hence, the conclusions about the global progress can be made by analyzing this graph. For example, we can focus on the reachability from one state to another without concerning the concrete transforms to conduct. For this objective, we can have the following state graph. 

\begin{definition}[State Graph]
Let $\mathcal{H}$ be an aggregated graph. $\mathcal{H}^{'}$ is called as the state graph of $\mathcal{H}$ if it is formed by converting the multiple links of $\mathcal{H}$ that link the same pair of states into single link.   
\end{definition}

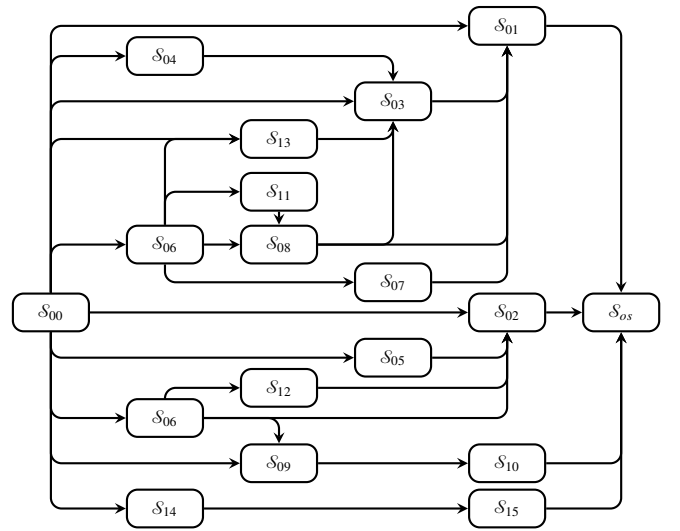
\begin{figure}[htp]
      \centering
    \thispagestyle{empty}
    \tikzstyle{state} = [rectangle, thick,rounded corners, minimum width = 1cm, minimum height=0.5cm,text centered, draw = black,font=\fontsize{7}{7}\selectfont]
    
    \vspace{0.3cm}
    \tikzstyle{arrow} = [->,>=stealth]
    \begin{tikzpicture}[node distance=1cm]
    \node[state](s0){$\mathcal{S}_{00}$};
        \node[state,right of = s0,xshift=5.0cm](s2){$\mathcal{S}_{02}$};
        \node[state,right of = s2,xshift=0.5cm](see){$\mathcal{S}_{os}$};
        \node[state,right of = s0,xshift=3.5cm,yshift=2.8cm](s3){$\mathcal{S}_{03}$};
        \node[state,right of = s0,xshift=0.5cm,yshift=3.4cm](s04){$\mathcal{S}_{04}$};
        \node[state,right of = s0,xshift=3.5cm,yshift=0.4cm](s8){$\mathcal{S}_{07}$};
        \node[state,right of = s0,xshift=0.5cm,yshift=0.9cm](s7){$\mathcal{S}_{06}$};
        \node[state,right of = s7,xshift=0.5cm](s12){$\mathcal{S}_{08}$};
        \node[state,right of = s7,xshift=0.5cm,yshift=1.4cm](s9){$\mathcal{S}_{13}$};
        \node[state,right of = s0,xshift=5cm,yshift=3.8cm](s1){$\mathcal{S}_{01}$};
        \node[state,right of = s0,xshift=3.5cm,yshift=-0.6cm](s4){$\mathcal{S}_{05}$};
        \node[state,right of = s0,xshift=0.5cm,yshift=-1.4cm](s07){$\mathcal{S}_{06}$};
        \node[state,right of = s0,xshift=2cm,yshift=-2cm](s5){$\mathcal{S}_{09}$};
        \node[state,right of = s5,xshift=2cm](s6){$\mathcal{S}_{10}$};
        \node[state,right of = s0,xshift=0.5cm,yshift=-2.6cm](s13){$\mathcal{S}_{14}$};
        \node[state,right of = s13,xshift=3.5cm](s14){$\mathcal{S}_{15}$};
        \node[state,right of = s07,xshift=0.5cm,yshift=0.4cm](s11){$\mathcal{S}_{12}$};
        \node[state,right of = s7,xshift=0.5cm,yshift=0.7cm](s012){$\mathcal{S}_{11}$};
       \draw [arrow,rounded corners,  thick] (s0) |-  (s1);
        \draw [arrow,rounded corners,  thick] (s1) -|  (see);
        \draw [arrow,rounded corners,  thick] (s0) |-  (s04);
        \draw [arrow,rounded corners,  thick] (s0) |-  (s7);
        \draw [arrow,rounded corners,  thick] (s7) |-  (s9);
        \draw [arrow,rounded corners,  thick] (s7) |-  (s8);
        \draw [arrow,rounded corners,  thick] (s7) |-  (s012);
        \draw [arrow,rounded corners,  thick] (s0) |-  (s4);
        \draw [arrow,rounded corners,  thick] (s0) |-  (s9);
        \draw [arrow,rounded corners,  thick] (s0) |-  (s07);
        \draw [arrow,rounded corners,  thick] (s07) |-  (s11);
        \draw [arrow,rounded corners,  thick] (s0) |-  (s5);
        \draw [arrow,rounded corners,  thick] (s0) |-  (s13);
        \draw [arrow,rounded corners,  thick] (s3) -|  (s1);
        \draw [arrow,rounded corners,  thick] (s9) -|  (s3);
        \draw [arrow,rounded corners,  thick] (s12) -|  (s3);
        \draw [arrow,rounded corners,  thick] (s12) -|  (s1);
        \draw [arrow,rounded corners,  thick] (s8) -|  (s1);
        \draw [arrow,rounded corners,  thick] (s4) -|  (s2);
        \draw [arrow,rounded corners,  thick] (s11) -|  (s2);
        \draw [arrow,rounded corners,  thick] (s07) -|  (s2);
        \draw [arrow,rounded corners,  thick] (s6) -|  (see);
        \draw [arrow,rounded corners,  thick] (s14) -|  (see);
        \draw [arrow,rounded corners,  thick] (s0) --  (s2);
        \draw [arrow,rounded corners,  thick] (s7) --  (s12);
        \draw [arrow,rounded corners,  thick] (s012) --  (s12);
        \draw [arrow,rounded corners,  thick] (s5) --  (s6);
        \draw [arrow,rounded corners,  thick] (s13) --  (s14);
        \draw [arrow,rounded corners,  thick] (s04) -|  (s3);
        \draw [arrow,rounded corners,  thick] (s0) |-  (s3);
        \draw [arrow,rounded corners,  thick] (s2) --  (see);
        \draw [arrow,rounded corners,  thick] (s07) -|  (s5);
    \end{tikzpicture}
     \caption{The state graph of the algorithm set $\mathcal{A}=\{\mathcal{A_{01}},\mathcal{A_{02}},\dots,\mathcal{A_{47}}\}$.}
     \label{fig-state graph}
\end{figure}

From the above definition, state graph concerns that whether there is a link to link a pair of states. In other word, it pays attention to the reachability of states.
The state graph in Figure \ref{fig-state graph} is formed from Figure \ref{fig-AggregateGraph}. From Figure \ref{fig-state graph}, approach can be defined as follows.

\begin{definition}[\textbf{Understood State}]
\label{defn:Understood-State}
A state is an understood state if a symbolic solver can produce the solution from the instance of the state of any AP.  
\end{definition} 

From Figure \ref{fig-state graph}, four understood states are identified; they are math expression, linear equations, general equations, and xEquations.

\begin{definition}[Approach]\label{def-approach}
An approach of solving algebra problems is the partial graph of the state graph that includes all the simple paths of passing the concern understood state.
\end{definition}

\begin{figure}[!h]
     \centering
    \thispagestyle{empty}
    \tikzstyle{state} = [rectangle, thick,rounded corners, minimum width = 1cm, minimum height=0.5cm,text centered, draw = black,font=\fontsize{7}{7}\selectfont]
    
    \vspace{0.3cm}
    \tikzstyle{arrow} = [->,>=stealth]
    \begin{tikzpicture}[node distance=1cm]
 %
       \node[](s00){};
       \node[state,below of =s00,xshift= -1.2cm,yshift=-1cm](s0){$\mathcal{S}_{00}$};
        \node[state,right of = s0,xshift=6.5cm](see){$\mathcal{S}_{os}$};
        \node[state,right of = s0,xshift=3.5cm,yshift=1.2cm](s3){$\mathcal{S}_{03}$};
        \node[state,right of = s0,xshift=0.5cm,yshift=1.8cm](s4){$\mathcal{S}_{04}$};
        \node[state,right of = s0,xshift=2cm,yshift=-1.5cm](s8){$\mathcal{S}_{07}$};
        \node[state,right of = s0,xshift=2cm,yshift=-0.8cm](s012){$\mathcal{S}_{11}$};
        \node[state,right of = s0,xshift=0.5cm](s7){$\mathcal{S}_{06}$};
        \node[state,right of = s7,xshift=0.5cm](s12){$\mathcal{S}_{08}$};
        \node[state,right of = s7,xshift=0.5cm,yshift=0.6cm](s9){$\mathcal{S}_{13}$};
        \node[state,right of = s0,xshift=5cm,yshift=2.2cm](s1){$\mathcal{S}_{01}$};
        \draw [arrow,rounded corners,  thick] (s0) |-  (s1);
        \draw [arrow,rounded corners,  thick] (s0) |-  (s4);
        \draw [arrow,rounded corners,  thick] (s7) |-  (s9);
        \draw [arrow,rounded corners,  thick] (s7) |-  (s012);
        \draw [arrow,rounded corners,  thick] (s0) |-  (s9);
        \draw [arrow,rounded corners,  thick] (s7) |-  (s8);
        \draw [arrow,rounded corners,  thick] (s0) |-  (s3);
        \draw [arrow,rounded corners,  thick] (s0) --  (s7);
        \draw [arrow,rounded corners,  thick] (s7) --  (s12);
        \draw [arrow,rounded corners,  thick] (s012) --  (s12);
        \draw [arrow,rounded corners,  thick] (s12) -|  (s1);
        \draw [arrow,rounded corners,  thick] (s4) -|  (s3);
        \draw [arrow,rounded corners,  thick] (s1) -|  (see);
        \draw [arrow,rounded corners,  thick] (s3) -|  (s1);
        \draw [arrow,rounded corners,  thick] (s9) -|  (s3);
        \draw [arrow,rounded corners,  thick] (s12) -|  (s3);
        \draw [arrow,rounded corners,  thick] (s8) -|  (s1);
        \coordinate[label = center:{\fontsize{8}{8}\selectfont $\Omega_{1}$: Math Expression Approach}](S) at(0,0.5); 

       \node[state,below of =s00,xshift= -1.2cm,yshift=-4.2cm](s0){$\mathcal{S}_{00}$};
        \node[state,right of = s0,xshift=5.0cm](s2){$\mathcal{S}_{02}$};
        \node[state,right of = s2,xshift=0.5cm](see){$\mathcal{S}_{os}$};
        \node[state,right of = s0,xshift=2cm,yshift=0.6cm](s4){$\mathcal{S}_{05}$};
        \node[state,right of = s0,xshift=2cm,yshift=-0.6cm](s11){$\mathcal{S}_{12}$};
        \node[state,right of = s0,xshift=0.5cm,yshift=-1.2cm](s7){$\mathcal{S}_{06}$};
        \draw [arrow,rounded corners,  thick] (s0) |-  (s4);
        \draw [arrow,rounded corners,  thick] (s0) |-  (s7);
        \draw [arrow,rounded corners,  thick] (s7) |-  (s11);
        \draw [arrow,rounded corners,  thick] (s0) --  (s2);
        \draw [arrow,rounded corners,  thick] (s2) --  (see);
        \draw [arrow,rounded corners,  thick] (s4) -|  (s2);
        \draw [arrow,rounded corners,  thick] (s11) -|  (s2);
        \draw [arrow,rounded corners,  thick] (s7) -|  (s2);
        \coordinate[label = center:{\fontsize{8}{8}\selectfont $\Omega_{2}$: Linear Equations Approach}](S) at(0,-4.1);

        \node[state,below of =s00,yshift = -7.2cm,xshift= -1.2cm](s30){$\mathcal{S}_{00}$};  
        \node[state,right of = s30,yshift =0.0cm,xshift=2cm](s35){$\mathcal{S}_{09}$};
        \node[state,right of = s35,yshift =0.0cm,xshift=2cm](s36){$\mathcal{S}_{10}$};
        \node[state,right of = s30,yshift =0.5cm,xshift=0.5cm](s37){$\mathcal{S}_{06}$};
        \node[state,right of = s36,xshift=0.5cm](s39){$\mathcal{S}_{os}$}; 
  
        \draw [arrow,rounded corners,  thick] (s30) --  (s35);
        \draw [arrow,rounded corners,  thick] (s30) |-  (s37);
        \draw [arrow,rounded corners,  thick] (s37) -|  (s35);
        \draw [arrow,rounded corners,  thick] (s35) --  (s36);
        \draw [arrow,rounded corners,  thick] (s36) --  (s39); 
        \coordinate[label = center:{\fontsize{8}{8}\selectfont $\Omega_{3}$: Relation-Centric Approach}](S) at(0,-7.1);

       
       \node[state,below of =s00,yshift = -8.5cm,xshift= -1.2cm](s50){$\mathcal{S}_{00}$};  
        \node[state,right of = s50,yshift =0.0cm,xshift=1.5cm](s513){$\mathcal{S}_{14}$};
        \node[state,right of = s513,xshift=1.5cm](s514){$\mathcal{S}_{15}$};
        \node[state,right of = s514,xshift=1.5cm](s59){$\mathcal{S}_{os}$}; 
  
        \draw [arrow,rounded corners,  thick] (s50) --  (s513);
        \draw [arrow,rounded corners,  thick] (s513) --  (s514); 
        \draw [arrow,rounded corners,  thick] (s514) --  (s59); 
        \coordinate[label = center:{\fontsize{8}{8}\selectfont $\Omega_{4}$: xRelation-Centric Approach}](S) at(0,-8.9);

        \end{tikzpicture}
     \caption{Four graphs of depicting the identified four approaches from Figure \ref{fig-state graph}.}
     \label{fig-separate-five-approach}
\end{figure}
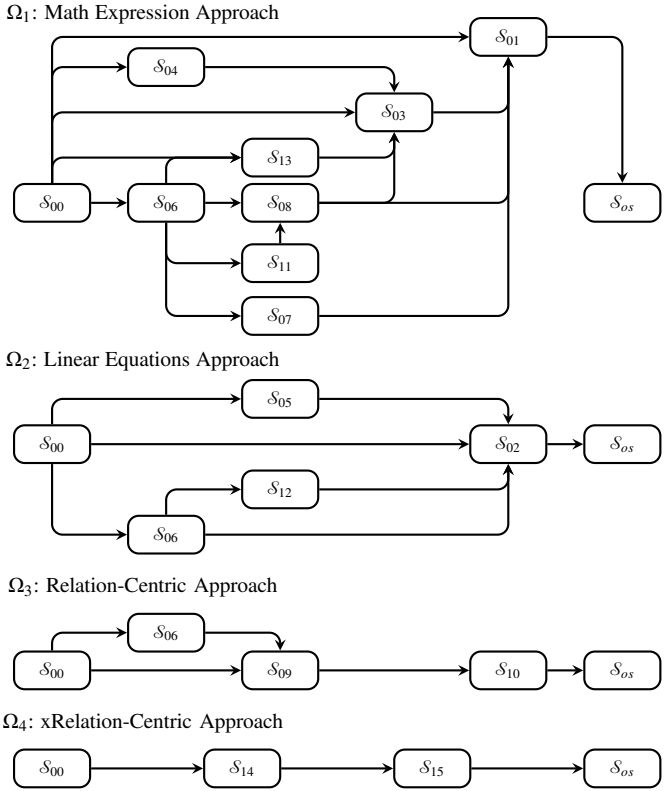

From the above definition, an approach is an abstract description of the algorithms of sharing the same understood state, and an algorithm is an instance of an approach. In addition, the factors that can reflect the difference of algorithms are its states, followed by its transforms.
 According to Definition \ref{def-approach}, four approaches are identified and listed in Figure \ref{fig-separate-five-approach}. They are $\Omega_{1}$=Math Expression, $\Omega_{2}$=Linear Equations, $\Omega_{3}$=Relation-Centric, $\Omega_{4}$=xRelation-Centric.

\vspace{-1mm}

\subsection{Taxonomy of Algorithms} \label{subsec_taxonomy}

\textcolor{black}{
Appendix B presents the state transform graphs for each of the 47 individual algorithms. In Section \ref{subsec_AggreSTG}, these individual graphs are combined into an aggregate state transform graph, which is further abstracted into a state graph, and four distinct approaches are defined. These analyses and results offer deeper insights into the problem-solving algorithms.
}

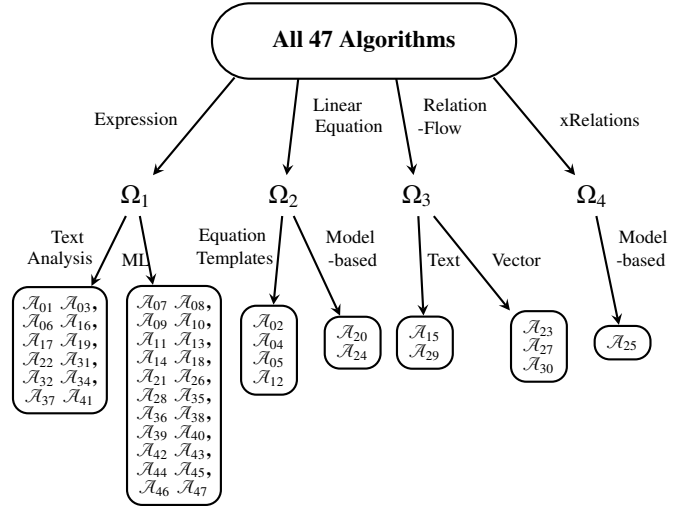
\begin{figure}[t]
     \centering

    \tikzstyle{S0} = [rectangle, thick, rounded corners=5mm,minimum width = 4cm, minimum height=1cm,text centered, draw = black,font=\fontsize{9}{10}\selectfont\bfseries]
    \tikzstyle{D} = [rectangle,thick, minimum width = 0.2cm,text width=0.5cm,rounded corners=2mm, text centered, draw = black,font=\fontsize{7}{7}\selectfont\bfseries]
    \tikzstyle{D1} = [rectangle,thick, minimum width = 0.2cm,text width=1cm,rounded corners=2mm, text centered, draw = black,font=\fontsize{7}{7}\selectfont\bfseries]
    \tikzstyle{E} = []
    \tikzstyle{F} = [font=\fontsize{7}{7}\selectfont]

    \tikzstyle{A} = [->,>=stealth,rounded corners, thick]
    \tikzstyle{AA} = [->,>=stealth,rounded corners, thick,blue]
    \begin{tikzpicture}[node distance=1cm]

        \node[S0](Start){All 47 Algorithms};
        \node[E,right of = Start,yshift = -2.05cm,xshift=-4cm](E1){$\Omega_{1}$};
        \node[E,right of = Start,yshift = -2.05cm,xshift=-2.05cm](E2){$\Omega_{2}$};
        \node[E,right of = Start,yshift = -2.05cm,xshift=-0.3cm](E3){$\Omega_{3}$};
        \node[E,right of = Start,yshift = -2.05cm,xshift=2cm](E4){$\Omega_{4}$};
        \node[E,right of = Start,yshift = -0.36cm,xshift=-2.6cm](P1){};
        \node[E,right of = Start,yshift = -0.37cm,xshift=-1.85cm](P2){};
        \node[E,right of = Start,yshift = -0.37cm,xshift=-0.6cm](P3){};
        \node[E,right of = Start,yshift = -0.36cm,xshift=0.6cm](P4){};
        
        \node[F,right of = Start,yshift = -1.05cm,xshift=-4cm](T1){Expression};
        \node[F,right of = Start,yshift = -0.85cm,xshift=-1.35cm](T2){Linear};
        \node[F,right of = Start,yshift = -1.15cm,xshift=-1.2cm](T22){Equation};
        \node[F,right of = Start,yshift = -0.85cm,xshift=0.2cm](T3){Relation};
        \node[F,right of = Start,yshift = -1.15cm,xshift=0cm](T33){-Flow};
        \node[F,right of = Start,yshift = -1.05cm,xshift=2.1cm](T4){xRelations};

        \node[D1,right of = E1,yshift = -2.05cm,xshift=-2cm](F1){$\mathcal{A}_{01}$ $\mathcal{A}_{03}$, $\mathcal{A}_{06}$ $\mathcal{A}_{16}$, $\mathcal{A}_{17}$ $\mathcal{A}_{19}$, $\mathcal{A}_{22}$ $\mathcal{A}_{31}$, $\mathcal{A}_{32}$ $\mathcal{A}_{34}$, $\mathcal{A}_{37}$ $\mathcal{A}_{41}$};
        \node[D1,right of = E1,yshift = -2.65cm,xshift=-0.5cm](F2){$\mathcal{A}_{07}$ $\mathcal{A}_{08}$, $\mathcal{A}_{09}$ $\mathcal{A}_{10}$, $\mathcal{A}_{11}$ $\mathcal{A}_{13}$, $\mathcal{A}_{14}$ $\mathcal{A}_{18}$, $\mathcal{A}_{21}$ $\mathcal{A}_{26}$, $\mathcal{A}_{28}$ $\mathcal{A}_{35}$,
        $\mathcal{A}_{36}$ $\mathcal{A}_{38}$, $\mathcal{A}_{39}$ $\mathcal{A}_{40}$, $\mathcal{A}_{42}$ $\mathcal{A}_{43}$, $\mathcal{A}_{44}$ $\mathcal{A}_{45}$, $\mathcal{A}_{46}$ $\mathcal{A}_{47}$};
        \node[D,right of = E2,yshift = -2.05cm,xshift=-1.2cm](F3){$\mathcal{A}_{02}$ $\mathcal{A}_{04}$ $\mathcal{A}_{05}$ $\mathcal{A}_{12}$};
        \node[D,right of = E2,yshift = -1.95cm,xshift=-0.1cm](F4){$\mathcal{A}_{20}$ $\mathcal{A}_{24}$};

        \node[D,right of = E3,yshift = -1.95cm,xshift=-0.9cm](F5){$\mathcal{A}_{15}$ $\mathcal{A}_{29}$};
        \node[D,right of = E3,yshift = -2.cm,xshift=0.6cm](F6){$\mathcal{A}_{23}$ $\mathcal{A}_{27}$ $\mathcal{A}_{30}$};

         \node[D,right of = E4,yshift = -1.95cm,xshift=-0.6cm](F7){$\mathcal{A}_{25}$};

         \node[F,right of = E1,yshift = -0.55cm,xshift=-1.9cm](t1){Text};
         \node[F,right of = E1,yshift = -0.85cm,xshift=-2cm](t11){Analysis};
         \node[F,right of = E1,yshift = -0.85cm,xshift=-1cm](t2){ML};

         \node[F,right of = E2,yshift = -0.55cm,xshift=-1.68cm](t3){Equation};
         \node[F,right of = E2,yshift = -0.85cm,xshift=-1.65cm](t33){Templates};

         \node[F,right of = E2,yshift = -0.55cm,xshift=-0.13cm](t4){Model};
         \node[F,right of = E2,yshift = -0.85cm,xshift=-0.1cm](t44){-based};

         \node[F,right of = E3,yshift = -0.85cm,xshift=-0.65cm](t5){Text};
         \node[F,right of = E3,yshift = -0.85cm,xshift=0.3cm](t6){Vector};

         \node[F,right of = E4,yshift = -0.55cm,xshift=-0.33cm](t7){Model};
         \node[F,right of = E4,yshift = -0.85cm,xshift=-0.36cm](t77){-based};

        \draw [A] (P1) -- (E1);
        \draw [A] (P2) -- (E2);
        \draw [A] (P3) -- (E3);
        \draw [A] (P4) -- (E4);

        \draw [A] (E1) -- (F1);
        \draw [A] (E1) -- (F2);
        \draw [A] (E2) -- (F3);
        \draw [A] (E2) -- (F4);

        \draw [A] (E3) -- (F5);
        \draw [A] (E3) -- (F6);
        
        \draw [A] (E4) -- (F7);
   
     \end{tikzpicture}
     \caption{Two-level taxonomy of the 47 identified algorithms for solving algebraic problems.}
     \label{fig-taxonomy}
\end{figure}

\textcolor{black}{
Building on this improved understanding, a new taxonomy for the algorithms is introduced, organized as a two-level hierarchy. At the first level, the algorithms are grouped into four overarching approaches. At the second level, they are further categorized based on the primary techniques they employ to solve problems. This taxonomy is illustrated in Figure \ref{fig-taxonomy}.
}

\section{Evolution of Four Approaches}\label{sec-approach-analysis}
For the four approaches defined earlier, this section will delve into the detailed establishment and progression of each approach. The focus will be on examining the development of these approaches over time, including both the evolution of their states and their transforms.

\subsection{Math Expression Approach}
The first approach denoted as $\Omega_{1}$ and its characteristics is to acquire a math expression that can produce the value of the unknown of the problem by evaluating it. Hence, the algorithms belong this approach can solve only the problems with single unknown. 

\subsubsection{$\Omega_{1}$ State Evolution}
 The essential task of $\Omega_{1}$ is to accomplish the transform from Input AP ($\mathcal{S}_{00}$) to Math Expression ($\mathcal{S}_{01}$) for the given problem. Since this task is hard to be done by a single transform, researchers proposed some bridge states.  

\cite{Roy2015} proposed Expression Tree ($\mathcal{S}_{03}$) as the first bridge state.
\cite{Koncel2015} continued the idea of using Expression Trees ($\mathcal{S}_{04}$) to bridge Input AP ($\mathcal{S}_{00}$) and Expression Tree ($\mathcal{S}_{03}$). Expression Trees($\mathcal{S}_{04}$) are the set of the possible expression trees, which is prepared to identify the wanted expression tree. 

Inspired by the success of seq2seq translation, \cite{Wang2017} first brought seq2seq method into solving APs. Hence, two states, vector text ($\mathcal{S}_{06}$) and vector single linear equation ($\mathcal{S}_{07}$), were proposed as bridge states. Since then, various deep neural networks were proposed to translate $\mathcal{S}_{06}$ into $\mathcal{S}_{07}$ \citep{Wang2019,Mandal2020}. 
\cite{Wang2018Translating} used vector computing to translate from $\mathcal{S}_{06}$ into vector expression tree $\mathcal{S}_{08}$ instead of $\mathcal{S}_{07}$. Since vector expression tree can keep the structure information, more  transforms were developed to transform from  $\mathcal{S}_{06}$ to  $\mathcal{S}_{08}$ \citep{Liu2019,Wang2019,Xie2019,Lin2021,Zaporojets2021,Wu2021a,Wu2021b,Zhang2022,Jie2022}.

Inspired by the better performance by adding structural information in expressions\citep{Wang2019,Liu2019,Xie2019}, \cite{Zhang2020graph2tree} firstly proposed Quantity Cell Graph and Quantity Comparison Graph to enrich the quantity representations by capturing the relation  and order structure between quantities. \cite{Li2020} combined the word sequence with the corresponding dependency or constituency tree to enrich input information. These expressions, which combine graphs and problem text, are named as TextGraph ($\mathcal{S}_{11}$) and is the extended state of $\mathcal{S}_{06}$.


\subsubsection{$\Omega_{1}$ Transform Evolution}
In the early years, the researchers developed the multiple transforms from $\mathcal{S}_{00}$ to $\mathcal{S}_{01}$.  
With the observation that verbs may determine the operands in the expression of a given AP, \cite{Hosseini2014} developed a transform, being denoted as $\mathcal{t}_{01}$. They presented a verb categorization based method, namely ARIS to solve simple addition and subtraction APs. ARIS first analyzes each of the sentences in the problem to identify the relevant variables and their values, and then mapped this information into a predefined logic template, namely ``container'' so as to capture the relations between variables. A verb in a sentence is classified into one of seven categories to determine the operator between two quantities. 
\cite{Mitra2016} adopted a similar idea, first using predefined logical templates to capture the types of relations between variables, thereby generating expressions. 
Although these methods perform well on some datasets, they are considered unable to be popularized on large datasets due to the high cost of data annotation.

Taking Expression Tree ($\mathcal{S}_{03}$) as a bridge state, \cite{Roy2015} formalized the AP-solving problem into multiple classification problems, and generated the expression by judging whether the number in the expression participated in the composition of the expression and the operator between the two candidate numbers. More specifically, they proposed $\mathcal{t}_{05}$ to generate Expression Tree ($\mathcal{S}_{03}$) from Input APs ($\mathcal{S}_{00}$), and $\mathcal{t}_{06}$ to generate Math Expression ($\mathcal{S}_{01}$) from Expression Tree ($\mathcal{S}_{03}$). 
\cite{Koncel2015} took the Expression Trees ($\mathcal{S}_{04}$) as a bridge state between $\mathcal{S}_{01}$ and $\mathcal{S}_{03}$ and proposed $\mathcal{t}_{07}$, namely integer linear programming to generate candidate expression trees, and $\mathcal{t}_{08}$ to identify a wanted expression tree by scoring the candidate expression trees through local and global discriminant models and selecting the one with the highest score.

Compared with the algorithm in \cite{Hosseini2014}, the ones in \cite{Roy2015} and \cite{Koncel2015}  have two progresses. First, they are machine learning based so that they are compact without requiring tedious processes such as annotation, classification. Second, they have a higher accuracy. However, due to the limited expression ability of the state $\mathcal{S}_{03}$, they can solve only problems with single unknown.

Inspired by the recent advancements in NLP (Natural Language Processing), the seq2x approach is developed (seq2x is the umbrella name of seq2seq, seq2tree, seq2graph, etc). This approach targets to obtain the expression that can produce the answer by  various neural networks. It first transforms the text into the vector text by an encoder and then decodes vector text into vector expression, or vector expression tree, or vector graph.  
 
DNS \citep{Wang2017} is the first work to use seq2seq model for solving APs, which aims to directly generate Vector Single Linear Equation ($\mathcal{S}_{07}$) from Vector Text ($\mathcal{S}_{06}$), namely the input and the output are the sequence of Vector Text ($\mathcal{S}_{06}$) and  the sequence of Vector Single Linear Equation ($\mathcal{S}_{07}$).  
It adopted the gated recurrent units (GRU) to convert $\mathcal{S}_{06}$ to $\mathcal{S}_{07}$, and the long short-memory (LSTM) to convert $\mathcal{S}_{07}$ to $\mathcal{S}_{02}$. 
\cite{Wang2018Translating} followed the idea of DNS \citep{Wang2017}, and proposed equation normalization to solve the problem of equation duplication in DNS \citep{Wang2017} by generated an expression tree.
\cite{Huang2018Neural} adjusted the output-sequence generation process by copy and alignment mechanism (namely CASS) to the seq2seq model, and solved the problem of the false numbers in the wrong position in the seq2seq model. In \cite{Huang2018Neural}, they proposed $\mathcal{t}_{16}$ to generate Vector Single Linear Equation ($\mathcal{S}_{07}$) from Vector Text ($\mathcal{S}_{06}$) using the reinforcement learning (RL).

\cite{Wang2018Translating,Wang2018MathDQN} showed the important contribution of structured information in equations in improving the performance of solvers. A new state of Vector Expression Tree ($\mathcal{S}_{08}$) is proposed.
With the bridge states of Vector Text ($\mathcal{S}_{06}$) and Vector Expression Tree ($\mathcal{S}_{08}$), some works \citep{Wang2019,Liu2019,Xie2019} used the structural information of the expression to improve the effect of the seq2seq models. 
\cite{Liu2019} proposed $\mathcal{t}_{18}$ to generate $\mathcal{S}_{08}$ from $\mathcal{S}_{06}$, and $\mathcal{t}_{19}$ to transform $\mathcal{S}_{06}$ to $\mathcal{S}_{01}$.
\cite{Wang2019} proposed $\mathcal{t}_{20}$ to generate $\mathcal{S}_{08}$ from $\mathcal{S}_{06}$.
\cite{Xie2019} proposed $\mathcal{t}_{21}$ to explicitly generate $\mathcal{S}_{08}$ from $\mathcal{S}_{06}$ in a human-like goal-driven way. So it could be seen as an expansion of the work of \cite{Wang2018Translating}, and is the first tree-structured neural model for APs. 

The problem text also contain some structure information, which could contribute to improve the performance of the solving algorithms \citep{Wang2018Translating,Huang2018Using}. 
Inspired by this, a new bridge state of TextGraph ($\mathcal{S}_{11}$) was proposed. With this bridge state, \cite{Zhang2020graph2tree} firstly adopted the graph2tree model for AP-solving which is a novel deep learning architecture that combines the merits of the graph-based encoder and tree-based decoder to generate better solution expressions.
\cite{Li2020} proposed a novel Graph2Tree Neural Network, consisting of a graph encoder and a hierarchical tree decoder, that encodes an augmented graph-structured input and decodes a tree-structured output.
Both of them claimed that their models outperform the state of-the-art baselines on benchmark datasets, proving the superiority of $\mathcal{S}_{11}$ and graph2Tree models in AP-solving.

\subsection{Linear Equations Approach}

This approach is denoted as $\Omega_{2}$, targeting to acquire a set of linear equations to represent the given problem. Hence, this approach can solve the problems with multiple unknowns. 
The goal of this approach is to generate Linear Equations ($\mathcal{S}_{02}$) from the Input APs ($\mathcal{S}_{00}$). 
 
\subsubsection{ $\Omega_{2}$ State Evolution}
\cite{Kushman2014} first built this approach. The algorithm in \cite{Kushman2014} used a state of Linear Equations ($\mathcal{S}_{02}$) to represent the given problem. It used a template based method to obtain the system of linear equations. The algorithm has a set of templates of abstract linear equations; it used a machine learning based  procedure to select a template and to instantiate  with the quantity entities from the given problem.   

 \cite{Shi2015} first proposed ``Formal Language'' ($\mathcal{S}_{05}$) as the first bridge state. ``Formal Language'' appeared in a form of tree, whose each internal node is a function, and each leaf node can be a constant, a class, or a zero-argument function. 

The previous seq2tree models have proved the importance of structural information. However, seq2tree cannot handle multiple unknown problems. Because the equations in the equation set problem are unordered, but for the seq2tree and seq2seq models, the generation of equations will involve the order problems, which will add unnecessary burden to the model. To solve this problem, \cite{Cao2021} firstly proposed $\mathcal{S}_{12}$ (DAG = Direct Acyclic Graph) as the new bridge state, where leaf nodes of DAG are the quantities, and other (internal and root) nodes are arithmetic or comparison operators. It integrates multiple expression trees of an equation set problem into the same solving scenario, so as to obtain a DAG. In this way, it not only preserves the structural information of the expression, but also extends the scope to the equation set problems. 

\subsubsection{ $\Omega_{2}$ Transform Evolution}
\cite{Kushman2014} proposed a template-based transform ($\mathcal{t}_{03}$), which identifies a template from the corpus of predefined templates and then fills the number slots and unknown slots with the quantity entities extracted from the text to produce the instances of $\mathcal{S}_{02}$. It also proposed to use $\mathcal{t}_{04}$ to solve $\mathcal{S}_{02}$ to get $\mathcal{S}_{os}$.
In order to reduce the hypothesis space, \cite{Zhou2015} improved the work of \cite{Kushman2014} by only considering assigning the number slots. They developed another transform ($\mathcal{t}_{09}$), which is a log-linear model, to directly transform $\mathcal{S}_{00}$ into $\mathcal{S}_{02}$. 

With the bridge state of ``Formal Language'' ($\mathcal{S}_{05}$), \cite{Shi2015} firstly proposed a context-free grammar parser ($\mathcal{t}_{10}$) to translate Input APs ($\mathcal{S}_{00}$) into $\mathcal{S}_{05}$, and a reasoning module ($\mathcal{t}_{02}$) to derive Linear Equation Set ($\mathcal{S}_{02}$) from $\mathcal{S}_{05}$. This method uses  the symbolic reasoning to obtain equations, making enhancement in evaluation and parsing.

With the bridge state of $\mathcal{S}_{12}$ (DAG), \cite{Cao2021} proposed a Seq2DAG model for AP-solving.  This paper proposed DAG Extraction Decoder ($\mathcal{t}_{28}$) to transform Vector Text of Input AP directly into a DAG structure, and $\mathcal{t}_{29}$ to generate Linear Equation Sets from DAG. These transforms together extended the solving scope by including the problems with multiple unknowns because the system of linear equations can find the values of multiple unknowns. 
 
 
\subsection{Relation-Centric Approach}

This approach $\Omega_{3}$ was first proposed by \cite{Yu2019}. This approach is called as relation-centric approach because its main actions are to acquire Relation Set ($\mathcal{S}_{09}$) and to transform $\mathcal{S}_{09}$ to obtain Equation Set ($\mathcal{S}_{10}$), and then to find the Output Solution ($\mathcal{S}_{os}$), in which relations are atomic operands of transforms. The relation-centric approach is comparable with Math Expression approach ($\Omega_{1}$) and Linear Equations approach ($\Omega_{2}$), but it is different from them in two aspects. First, the equations converted from the relations can be high-order equations. Such high-order equation set can represent the larger scope of algebra problems than the set of linear equations can do. Second,  acquiring relations is easier than acquiring equations, although the processes of acquiring relations and acquiring equations have just little difference in appearance. On the other hand, it is relatively easy to convert relations into equations, but it is hard to reverse  equations into relations.  

\subsubsection{$\Omega_{3}$ State Evolution}
On relation-centric approach ($\Omega_{3}$), the initial algorithms have two exclusive states $\mathcal{S}_{09}$ to $\mathcal{S}_{10}$ in \cite{Yu2019}. 
Inspired by the success of vector computing in NLP, \cite{Lyu2023} introduced vector computing technology into relation-centric approach. 
Thus, Vector Text ($\mathcal{S}_{06}$) was brought into this approach, first proposed in \cite{Wang2017} in 2017.
  
\subsubsection{$\Omega_{3}$ Transform Evolution}


\cite{Yu2019} proposed a relation-centric approach and constructed an algorithm with $\mathcal{t}_{22}$ ($S^2$ Relation Extractor), $\mathcal{t}_{23}$ (Equation Set Producer), and $\mathcal{t}_{24}$(Equation Set Solver).  
In 2023, \cite{Lyu2023} evolved $\mathcal{t}_{22}$ into $\mathcal{t}_{31}$, a new transform that uses the vector $S^2$ method to extract $\mathcal{S}_{09}$  from $\mathcal{S}_{06}$ of APs.  \cite{Lyu2023} is the updated version of \cite{Lyu2021}. 

Acquiring implicit relations from algebra problems is an indispensable  problem because solving problems requires to use implicit relations \citep{Yu2019}. 
On the other hand, acquiring implicit relations from algebra problems is a challenge problem because students use various ways to acquire them when they do homework. 
\cite{Yu2023} developed $\mathcal{t}_{40}$ (Neural Network Miner), a neural network based transform, mining implicit relations from $\mathcal{S}_{06}$ (Vector Text). 
\cite{Yu2023} developed a high-performance algorithm for solving arithmetic word problems by synergizing syntax-semantics extractor for explicit relations and neural network miner for implicit relations. 
\cite{Lyu2023} enhanced the algorithm developed in \cite{Yu2023} in multiple aspects by using vector syntax-semantics extractor. 
\cite{Meng2023} developed $\mathcal{t}_{42}$ (DI Relation Extractor, which extracts Deep Implicit relations), a new relation extractor that can extract DI relations from problems text. 
\cite{Huang2023} developed $\mathcal{t}_{41}$ (GD Relation Extractor, which extracts relations from diagrams), a new relation extractor that can extract GD relations from Geometry Diagrams. Both$\mathcal{t}_{41}$ and  $\mathcal{t}_{42}$ extend the scope of solving algorithms.


\subsubsection{$\Omega_{3}$ Application in Solving Other Types of Problems}

$\Omega_{3}$ has also applied to solve other types of problems. \cite{Gan2019a, Gan2019b} have successfully used $S^2$ relation extractor in $\Omega_{3}$
to understand plane geometry theorems. \cite{Jian2019} and \cite{He2020} have successfully used $S^2$ relation extractor in $\Omega_{3}$
to understand linear circuit problems.

\subsection{xRelation-Centric Approach}

\subsubsection{$\Omega_{4}$ State Evolution}
\cite{YuSun2022} developed a new approach $\Omega_{4}$ (xRelation-Centric Approach). $\Omega_{4}$ is the first approach that has mechanism to handle the compound object of functions. The algorithm in \cite{YuSun2022} is the first and only algorithm in this approach so far. The algorithm has two new states: xRelation Set ($\mathcal{S}_{14}$) and xEquation Set ($\mathcal{S}_{14}$). 
xRelation set consists of universal relations and period relations; each clique of period relations potentially describes a function. xEquation Set consists of high-order equations and functions. 

\subsubsection{$\Omega_{4}$ Transform Evolution}
The algorithms in \cite{YuSun2022} has four new transforms. 
The first transform is $\mathcal{t}_{35}$ ($S^2P$ Relation Extractor), extracting period relations from text of function problems.
The second transform is  $\mathcal{t}_{36}$ ($L^2$ Relation Extractor), acquiring relations from diagrams of algebra problems by a line segment and label method, shorted as  $L^2$ method.  
The third transform is $\mathcal{t}_{37}$(xEquation Set Producer), converting $\mathcal{S}_{14}$ (xRelation Set) into $\mathcal{S}_{14}$ (xEquation Set).
The fourth transform is $\mathcal{t}_{38}$(xEquation Set Solver), obtaining the answers of function problems. This transform is required to interact between equations and functions to find the answers.

\section{Perspective Confusion Comparison}\label{sec-confusion}

The algorithms in the literature were developed to address various types of APs, each of which presents distinct challenges. To gain a performance overview of the existing algorithms, the types of problems are defined, and a comparative evaluation is performed to assess the transforms, algorithms, and approaches across these problem types. However, this section is not intended to provide an exhaustive comparative analysis. Instead, it aims to verify the feasibility of the new evaluation paradigm, given the specific focus of this paper.

\begin{definition}[Perspective Confusion Comparison]
The Perspective Confusion Comparison (PCC) is an evaluation paradigm for comparing a set of algorithms designed to solve a complex problem that its instances can be classified into different types. This paradigm assumes that algorithms can be represented as state transform graphs. The PCC involves conducting confusion experiments, where each transform, each algorithm, and each approach are evaluated across various problem types to assess their performance and applicability.
\end{definition}

\subsection{Types of Algebra Problems}\label{APs-type}
The types of problems are defined according to 3 features and 2 modalities, being defined below. 
Each problem will express some conditions so that solvers solve it by using these conditions. Problems may state some of these conditions explicitly; it also may tell some conditions implicitly. Hence, we have the following definition.

\begin{definition}[Explicit/Implicit Condition]
When solving an algebra problem, some conditions are needed. A condition is called as an explicit condition if the problem explicitly states it; otherwise, it is an implicit condition. Further, an implicit condition is called as shallow implicit condition if some expression of the problem can hint it; otherwise, it is a deep implicit condition.
\end{definition}  

The current indicator is  tentatively used to differ shallow implicit conditions from deep implicit conditions because this indicator is not perfect yet. 
According to which features the considered problem possesses, whether it has diagrams, and the non-double inclusion principle, the six types of algebra problems are defined.  The following non-double inclusion principle is introduced to avoid the occurrence that one problem is divided into two types. 

\textit{Non-double inclusion principle:}
Type $j$ do not contain $\mathcal{P}$ if Type $i$ has contained $\mathcal{P}$ and $i<j$.  

The algorithms will be analyzed on three datasets: M23K, TnD1K and DRAW-1K. M23K is from Math23K \citep{Wang2017} by discarding pure numerical problems and special symbol calculation, which contains 20910 algebra word problems with only one variable \citep{Yu2023}. TnD1K contains 1000 function problems \citep{YuSun2022}. DRAW-1K contains 747 algebra word problems with two unknowns and 255 algebra word problems with one unknown  \citep{Kushman2014}. 
Table \ref{types} shows: 1) all the three datasets do not include the problems of Type  5 and 6; 2) M23K and DRAW-1K, two of the representative popular datasets, do not include the problems of Type 4; 3) the distributions of these three datasets over Type 1 to 4 are presented. The next section  will analyze algorithms over four types of problems.

\begin{table}[h]
\small
    \begin{center}
            \caption{The definitions of 6 types of algebra problems and the distribution of datasets M23K, TnD1K, DRAW-1K,  DIR-AWPs, and GCP-PS over Type 1 to 5}\label{types}%
            \begin{tabular}{@{}llllll@{}}
                \toprule
                \multirow{2}{*}{Table}&\multicolumn{2}{l}{Modality} &\multicolumn{3}{l}{Three Features} \\
                \cmidrule{2-3} \cmidrule{4-6} 
                 & Text& Diagram&Explicit & Shallow & Deep \\ \midrule
                Type 1& \checkmark &&\checkmark& & \\ 
                Type 2&\checkmark&&\checkmark&\checkmark& \\ 
                Type 3&\checkmark&&\checkmark&\checkmark&\checkmark \\ 
                Type 4&\checkmark&\checkmark&\checkmark& & \\ 
                Type 5&\checkmark&\checkmark&\checkmark&\checkmark& \\ 
                Type 6&\checkmark&\checkmark&\checkmark&\checkmark&\checkmark \\ \midrule
                
                \multirow{2}{*}{Dataset } & \multicolumn{5}{l}{Distributions over Type 1 to 5}  \\      \cmidrule{2-6}
                & Type 1 & \multicolumn{1}{l}{Type 2}  & Type 3 & Type 4  &Type 5\\ \midrule
                M23K & 16568 & \multicolumn{1}{l}{3591}   & 741   & 0      & 0 \\ 
                TnD1K & 530   & \multicolumn{1}{l}{74}     & 0     & 395   & 0 \\ 
                DRAW-1K &  677  & \multicolumn{1}{l}{148 }     &   174   &  0   & 0 \\ 
                DIR-AWPs &  0  & \multicolumn{1}{l}{0 }     &   6030   &  0   & 0 \\ 
                GCP-PS &  0  & \multicolumn{1}{l}{0 }     &   0   &  0   & 217 \\ 
                \bottomrule
            \end{tabular}
  \begin{tablenotes}
        \footnotesize
        \item[1] Note: All the datasets in this table do not include the problems in Type 6. 
    \end{tablenotes}
    
    \end{center}
\end{table}


\subsection{Reachability and Efficiency of Transforms}
Some algorithms and transforms are selected to demonstrate the perspective confusion comparison. The selected algorithms are $\mathcal{A}_{07}$, $\mathcal{A}_{20}$, $\mathcal{A}_{21}$, $\mathcal{A}_{25}$, $\mathcal{A}_{27}$, $\mathcal{A}_{29}$, $\mathcal{A}_{30}$ and $\mathcal{A}_{35}$, 
 where  $\mathcal{A}_{07}$,  $\mathcal{A}_{21}$, and $\mathcal{A}_{35}$  belong to Math Expression ($\Omega_{3}$), $\mathcal{A}_{20}$ belongs to the Linear Equations, $\mathcal{A}_{27}$, $\mathcal{A}_{29}$, $\mathcal{A}_{30}$ belong to Relation Centric, and $\mathcal{A}_{25}$ belongs to the xRelation-Centric approach, respectively.  
The transform efficiency are define as follows.


\begin{definition}[Transform Efficiency]
Let $\mathcal{C}$ be a corpus and $\mathcal{t}(i,j)$ be a transform from $\mathcal{S}_i$ to $\mathcal{S}_{j}$ in solving algebra  problems. 
The transform efficiency is defined as the accuracy that $\mathcal{t}(i,j)$ transforms all $\mathcal{S}_i$ instances of $\mathcal{C}$ into $\mathcal{S}_{j}$.
\end{definition}


It is easy to define the transform efficiency, but it is hard to obtain it.  
The difficulty of obtaining the transform efficiency lies in that it is hard to know the correctness of instances of its start state and target state for some transforms. 
Especially for deep learning based algorithms, the instances in vector are difficult in knowing their correctness. 
Maybe, that is the reason that the research community requires only the ablation experiments.

Since it is hard to find the transform efficiency, we turn to find the efficiencies of the partial algorithms as an substitute. Let $\mathcal{E}_{1}$ be the efficiency of the partial graph of the state transform graph (STG) of $\mathcal{A}_{35}$ in Appendix B from $\mathcal{S}_{00}$ to $\mathcal{S}_{01}$ that includes $\mathcal{t}_{05}$  and $\mathcal{t}_{06}$ ; let $\mathcal{E}_{2}$ be the efficiency of the partial graph of the STG of $\mathcal{A}_{07}$ in Appendix B from $\mathcal{S}_{00}$ to $\mathcal{S}_{01}$ that includes $\mathcal{t}_{12}$, $\mathcal{t}_{13}$ and $\mathcal{t}_{15}$; 
let $\mathcal{E}_{3}$ be the efficiency of the partial graph of the STG of $\mathcal{A}_{21}$ in Appendix B from $\mathcal{S}_{00}$ to $\mathcal{S}_{01}$ that includes $\mathcal{t}_{12}$, $\mathcal{t}_{30}$ and $\mathcal{t}_{06}$; 
let $\mathcal{E}_{4}$ be the efficiency of the partial graph of the STG of $\mathcal{A}_{20}$ in Appendix B from $\mathcal{S}_{00}$ to $\mathcal{S}_{02}$ that includes $\mathcal{t}_{12}$,  $\mathcal{t}_{28}$ and $\mathcal{t}_{29}$.  
Because we can know the correctness of $\mathcal{S}_{02}$ and $\mathcal{S}_{01}$, we can calculate $\mathcal{E}_{1}$ and $\mathcal{E}_{2}$.  


Table \ref{confusion-appro} shows the perspective confusion comparison of 8 algorithms, 4 partial algorithms and 4 transforms from 4 approaches tested on 5-type distribution of 5 datasets of algebra problems. In Table \ref{confusion-appro} the data with superscript stars are from the corresponding published papers; the other data are produced by this paper. 

\begin{table*}[!h]
  \centering
  \small
  \caption{Confusion comparison results of 8 algorithms, 4 partial algorithms  and 4 transforms from 4 approaches over 5 types of problems of 5 datasets}\label{confusion-appro}
  \tabcolsep=0.3em
  \begin{threeparttable}
    \begin{tabular}{llllllllll|llllllll}
    \hline
    \multirow{2}{*}{$\Omega_{*}$} & \multirow{2}{*}{$\mathcal{A_{*}}$}  & \multirow{2}{*}{\makecell{Source \\ paper}} & \multirow{2}{*}{Dataset} & \multicolumn{6}{l|}{Algorithm accuracy on 5 types (\%)} & \multirow{2}{*}{Transform} & \multicolumn{5}{l}{Transform efficiency on 5 types (\%)} \\ \cline{5-10} \cline{12-16}
          &       &       &        & Type 1 & Type 2 & Type 3& Type 4  & Type 5 & Sum   &       & Type 1 & Type 2 & Type 3& Type 4 & Type 5 \\ \hline
          
     \multirow{3}{*}{$\Omega_{1}$} & $\mathcal{A}_{35}$   & \cite{Roy2018} & M23K  & 32.3  & 15.5  & 4.9   &     &   & 28.4  & $\mathcal{E}_{1}$  of $\mathcal{A}_{35}$  & 27.6 & 13.3    & 3.8  &  \\ 

     & $\mathcal{A}_{07}$    & \cite{Wang2017} & M23K  & 65.1  & 73.9  & 55.8  &   &     & 66.2  & $\mathcal{E}_{2}$  of $\mathcal{A}_{07}$  & 54.8  & 64.0    & 47.7  &  \\
     & $\mathcal{A}_{21}$ & \cite{Lin2021} & M23K  & 74.7  & 76.7  & 26.6 &     &   & 74.1  & $\mathcal{E}_{3}$  of $\mathcal{A}_{21}$  & 63.9  & 65.9  &20.8  &  \\ 
      \hline

    $\Omega_{2}$ & $\mathcal{A}_{20}$ & \cite{Cao2021} & DRAW-1K & 36.4 & 36.4 & 28.9&&  & {44.4*} &    $\mathcal{E}_{4}$ of $\mathcal{A}_{20}$   &   34.5    &   34.5    &  26.7     &  \\
    \hline
    \multirow{3}{*}{$\Omega_{3}$} & $\mathcal{A}_{27}$ & \cite{Yu2023} & M23K  & 74.6  & 89.8  & 65.1  &    &   & 77.7  &  $t_{40}$ of $\mathcal{A}_{25}$ & 94.5  & 81.6   & 97.3  &     \\ 
    
     & $\mathcal{A}_{29}$ & \cite{Meng2023} &DIR-AWPs  &  &     & 84.5* &&   & {84.5*}  & $t_{41}$  of $\mathcal{A}_{29}$  &   &   &84.5*  & &  \\
     
    & $\mathcal{A}_{30}$ & \cite{Huang2023} & GCP-PS  &   &      &   &&81.5*  & {81.5*}  & $t_{42}$ of $\mathcal{A}_{30}$  &   &   &  & &81.5* \\\hline
           
    $\Omega_{4}$& $\mathcal{A}_{25}$ & \cite{YuSun2022} & TnD1K & 83.0  & 26.2 &       & 88.1  & & 80.8  & $t_{38}$ of $\mathcal{A}_{25}$ & 87.3  & 32.1  &       & 90.5 \\ 
    
    \hline
    
    \end{tabular}%
    
    \begin{tablenotes}
        \footnotesize
        \item[1] $\Omega_{*}$ represents ``Approach'' and $\mathcal{A}_{*}$ represents ``Algorithm'' in order to save the space of the table.
        \item[2] ``Type 6'' does not appear in the table because no algorithm solve the problems in this type yet.
        \item[3] The numbers with superscript * are from the reference paper, while the other data are produced by this paper.
    \end{tablenotes}
      
    \end{threeparttable}

\end{table*}%

Table \ref{confusion-appro} reveals the following findings. First, 
 the algorithm $\mathcal{A}_{30}$ presented in \cite{Huang2023} is the only algorithm that can solve problems in Type 5 and it achieved an accuracy of 81.5\% on a dataset named as GCP-Ps. 
 Second, the xRelation-Centric approach can solve multiple types of problems. The algorithms $\mathcal{A}_{25}$ presented in \cite{YuSun2022} are the only algorithms for solving the problems with diagrams (namely Type 4) so far. Third, only the algorithm $\mathcal{A}_{25}$ presented in \cite{Meng2023} has achieved an accuracy of 84.5\% on a dataset named as DIR-AWPs, while all other existing algorithms have relatively low accuracy in solving problems in Type 3. 
Thus, it is the potential research problem to develop methods to acquire the shallow implicit relations to improve the algorithm accuracy for solving APs.  
Fourth, $\mathcal{A}_{31}$ also reports the accuracy of only 26.2\% over Type 2, though it can solve more types of APs.  It shows that $\mathcal{A}_{31}$ does not adopt the best techniques in the existing algorithms.

Compared with one-level confusion comparison and ablation experiment, the perspective confusion comparison (PCC) has the following features:  

1) The PCC can evaluate not only the external indicators of algorithms but also their internal indicators such as the transform efficiency over the various types of problems. Thus, it can reveal the advantages and disadvantages of transforms, algorithms and approaches on the various types of problems. 
    
2) Compared with the ablation experiment and one-level confusion evaluation, the PCC can reveal more research gaps. The results of the PCC are three dimension space covering the results of the competitive comparison as its partial space. Hence, the PCC can show more research gaps because any point of the three dimension space has potential to have research gaps. 
    

\section{Conclusions and Research Directions}\label{sec-conclusion}

This paper has done a theoretical review on solving algebra problems, which has built the state transform theory (STT), performed the state transform analysis (STA) of individual algorithms and collective algorithms, and proposed the PCC for solving algebra problems. Through this review, we have lead the five conclusions presented in section \ref{section-conclusion} and we can propose the six research directions presented in section \ref{section-direction}.  


\subsection{Conclusions and Discussions} \label{section-conclusion}
 This paper has developed a theoretical review framework for reviewing the algorithms for solving algebra problems, which are published since 2014 (or after 2013).  This framework consists of State Transform theory, State Transform Analysis method for structure analysis of individual and collective algorithms, state/transform approach review method and perspective confusion comparison scheme. The main conclusions on the above mentioned five aspects are presented as follows.
 
\subsubsection{Theoretical Review Framework} 
This paper has developed the theoretical review framework for reviewing algorithms for solving algebra problems since 2014. This framework consists of STT, STA, and PCC. This framework differs from the traditional review way in three aspects. First, it reviews algorithms based on the STT. Second, it uses the various state transform graphs to depict individual algorithms and collective algorithms; it can uniformly define approaches on the state graph, instead of the traditional way, i.e. defining approaches based on the hand-craft indicators. Third, it extends the traditional algorithm comparison scope into a perspective confusion comparison for transform, algorithm, and approach. Thus, it reveals not only the competitive performance among algorithms but also the internal contributions of algorithms (transform contributions) and the performance of approaches.  
 
\subsubsection{State Transform Theory} 
This paper introduces a State Transform Theory (STT), which highlights the crucial roles of states and transforms in solving algorithms. The significance of this theory lies in its distinction between two capacities: the theoretical capacity, determined by the structure of state transforms, and the technical capacity, governed by the specific transforms used in algorithms. According to this theory, the evolution of both states and transforms defines the progress of solving algorithms, whereas traditional surveys and reviews have primarily focused on the progression of transforms (techniques).
The core assertions of the STT are as follows: 1) Most algorithms for solving algebraic problems can be represented as a state transform graph;
2) Representative algorithms developed since 2014 are decomposable;
3) All representative algorithms for solving algebraic problems developed since 2014 are constructed using 17 states and 42 transforms.

\textcolor{black}{
Earlier review frameworks also support the STT. The first review paper on solving algorithms already marks the emergence of the STT. In that paper, the two-stage review framework is considered the beginning of the STT. This framework is based on the cognition that solving algorithms involve two stages: problem understanding and symbolic solver. The outcome of the first stage is essentially the understood representation within the STT. Notably, most review papers adopt this two-stage framework, even if implicitly. These papers focus on the advancement of techniques, which are, in essence, the implementation methods for the transforms that are more abstract than the techniques themselves.  
}

\subsubsection{State Transform Analysis} 
This paper has developed the State Transform Analysis (STA)  based on the STT. In this method, each algorithm can be depicted by a state transform graph, a collection of algorithms can be depicted by an aggregated graph, and the state structures of a collection of algorithms can be depicted by a state graph. Furthermore, the approaches can be uniformly defined on the state graph according to the understood states. Such analysis identified 17 states, 4 understood states and 42 transforms for the solving algorithms proposed since 2014. The 4 understood states further defined 4 approaches. 
The traditional review or survey papers emphasized on reviewing the techniques of algorithms whereas this paper emphasized on both  states and transforms, which is the umbrella terms of the techniques.   
 
\subsubsection{State Transform Review Scheme} This paper has built a state/transform review scheme. This scheme reviews an approach based on its state progress and transform progress. The traditional review papers emphasized on the transforms of approaches whereas this paper emphasized on both states and transforms. 
  
\subsubsection{Perspective Confusion Comparison} 

This paper introduces the Perspective Confusion Comparison (PCC) framework based on state transform theory. The proposed scheme extends traditional one-level confusion comparison into a three-level structure, encompassing transforms, algorithms, and approaches. The PCC systematically compares these elements across various problem types, offering a more comprehensive understanding of the knowledge framework underlying solving algorithms. Unlike ablation experiments, the PCC not only evaluates performance across multiple levels but also provides a more granular analysis of specific problem types.

The PCC offers several advantages over ablation experiments when evaluating solving algorithms. 
First, the PCC captures all experimental results obtained through ablation. In the context of solving algorithms, each ablation test can be viewed as evaluating an algorithm with a new transform, as removing a method or feature alters the original structure. Consequently, the results of ablation experiments are inherently encompassed within PCC outcomes. 
Second, the PCC generates additional insights that ablation experiments cannot. Specifically, it analyzes the effectiveness of individual transforms on different problem types, offering a level of analysis that ablation methods are not designed to provide. Ablation experiments are generally limited to deep learning algorithms, which cannot always be decomposed into distinct components, thus restricting their scope.
The PCC is explicitly designed for the systematic and detailed evaluation of approaches, algorithms, and transforms, establishing a new evaluation paradigm. It allows for a deeper and more nuanced analysis compared to traditional methods. However, it is important to note that the PCC is only applicable to algorithms that can be broken down into component parts.
 
\subsection{Future Research Directions}\label{section-direction}
The theoretical review on solving algebra problems reveals six research directions. 

First, analyzing the solving capacities of approaches is a valuable theory problem. The state graph of an approach decides the theoretical ability that the approach can solve APs; the transforms tell the technique ability of the approach. The prior survey papers \citep{ZhangLee2021} mainly emphasized on summarizing the transforms (techniques) of understanding problems, while they cannot reveal the solving capacity of an approach. This paper defined the state structure of each approach. Thus, we can study the theoretical capacity of each approach by analyzing the representation ability of states of constructing the approach. Along this thinking, we can pose a question of what kind of states can represent what types of the algebra problems well.  

Second, developing solving algorithms that can meet the requirements of the end users is another urgent research direction. Many algorithms emphasize on the accuracy only. As you know, the accuracy is not all the requirements that end users want from solving algorithms. Hence, it is high time to explore that the need of end users and that end users want which features of the solving algorithms. Thus, it is a valuable research problem how we can achieve these features. We should develop the algorithms with the required properties in order to realize the wide application.  One of main objectives of developing solving algorithms is to use the algorithms as the core technology for developing intelligent tutoring systems. For this objective, we should develop the algorithms that can generate humanoid solutions. Unfortunately, many algorithms developed in the recent years  did not address this requirement at all. In the future work, researchers are expected to strengthen the efforts on this direction. 
 
Third, another research direction is to widen the application of perspective confusion comparison. Most of algorithms compare with the baseline algorithms on some representative datasets. Such comparisons do not reveal that the algorithm can solve what types of problems in what performance. This paper just gave some samples of  perspective confusion comparison, not a comprehensive comparison due to that it is not the goal of this paper. Hence, we plan to accomplish an comprehensive perspective confusion comparison for all the representative transforms, algorithms and approaches on more datasets in the near future. 

Fourth, extending the solvable scope is a research direction. Among the selected 47 algorithms, so far only one paper addressed solving the algebra problems with compound object and diagram. Hence, we should emphasize on developing more methods to solve the algebra problems with compound objects and diagrams. Currently two papers proposed methods to extract relations from diagrams. One paper extracted relations from diagrams of function problems and to build functions, being a kind of compound objects. Another paper extracted relations from geometry diagrams.  Hence, we expect that more methods will be proposed to acquire the knowledge from various types of diagrams so that we can solve more types of problems with diagrams. 

Fifth, building a solving engine not just an algorithm. This paper has showed that some algorithms have the better performances on some specific types of problems. More importantly, this paper has built the aggregated graph of batch of algorithms. On the other hand, \cite{Zhou2020} proposed a state transition algorithm,  which was applied in various optimization problems. The above-mentioned three facts inspire us a new way to develop  solving engine in the future. Concretely, we borrow the method in  \cite{Zhou2020} to find the optimal portion of the aggregated graph to form a solving engine, which integrates the high performance methods from multiple algorithms. 

Sixth, a new direction is applying LLMs into AP-solving. In the recent years, LLMs (large language models), such as GPT-4 (Generative Pretrained Transformer)\citep{openai2023gpt4} or PaLM-2 (Pretrained Autoregressive Language Model)\citep{Bi2020_PALM} have been applying into building algorithms for solving math word problems \citep{Zhu2023,Zhou2023_gpt, Zong2023}. Once BERT, the first LLM, was published in 2018, researchers applied it into building solving algorithms and they even built MWP-BERT, which is special for solving math word problems\citep{Liang2022}. They mainly used BERT (or specialized BERT) as tool of converting problems text into vector sequence of problem text to replace the previous word2vec \citep{Liang2022, Yu2023, Lyu2023}. 
With the success of ChatGPT, some researchers proposed to use LLMs as QA (Question and Answer) tool. Following this idea, Zong and Krishnamachari published a paper that tested the performances of GPT-3 answers some questions related to the problems solving \citep{Zong2023}. 
\cite{Zhou2023_gpt} uses LLM as the main base of  solving math problems, which is another way of applying LLMs. This paper explores the effect of code on enhancing LLMs’ reasoning capability by introducing different constraints on the Code Usage Frequency of GPT-4 Code Interpreter. It found that its success can be largely attributed to its powerful skills in generating and executing code, evaluating the output of code execution, and rectifying its solution when receiving unreasonable outputs. Based on this cognition, they proposed a new prompting method, explicit code-based self-verification (CSV), to further boost the mathematical reasoning potential of GPT-4 Code Interpreter.  
\cite{LuVista2023} conducts a comprehensive evaluation of
11 prominent open-source and proprietary foundation models (LLMs, LLMs augmented
with tools, and LMMs), and early experiments with GPT-4V. The evaluation finds that the best performing model, Multimodal Bard, achieves only 58\% of human performance(34.8\% vs 60.3\%), indicating ample room for further improvement.  
These papers have shown that LLMs can improve the performance of solving math word problems, but LLMs' competency in solving math word problems is not known yet. These papers show us a new research  direction, which is how to make use LLMs' competency and the traditional methods together to build the high-performance solving algorithms. 

\section*{Declaration of competing interest}
The authors declare that they have no known competing financial interests or personal relationships that could have appeared to influence the work reported in this paper.

\section*{Data availability}
The datasets used in this paper are available on our web page.

\section*{Acknowledgements}
This work is partially supported by the general Project  of the National Natural Science Foundation of China (Grant No: 62277022).

\bibliographystyle{elsarticle-harv}
\bibliography{main}

\section*{Appendix A: Descriptions of the 17 Identified States and 42 Identified Transforms}\label{appendix1}
This section presents the state transform graphs (STG) for the 47 representative algorithms.

$\mathcal{S}_{00}$=:{\bf Input APs:}  {\it Input APs} is the state that algebra problems are input into algorithms, including text and diagrams.

$\mathcal{S}_{01}$=:{\bf Math Expression:} It first appeared in \cite{Hosseini2014} in 2014, being the verbal math  expression to express the answer key in numbers and four operators. 

$\mathcal{S}_{02}$=:{\bf Linear Equation Set:}  It first appeared in \cite{Kushman2014} in 2014. This becomes a popular state quickly because the available math tools can solve a set of linear equations. 

$\mathcal{S}_{03}$=:{\bf Expression Tree:} It first appeared in \cite{Roy2015} in 2015. This state is close to math expression because they can convert each other by simple procedures. 

$\mathcal{S}_{04}$=:{\bf Expression Trees:} It first appeared in \cite{Koncel2015} in 2015, being a state of multiple math trees, which can be used to solve the algebra problems with multiple unknowns. 

$\mathcal{S}_{05}$=:{\bf Formal Language:}  It was proposed in \cite{Shi2015} in 2015. DOL is the formal language used in \cite{Shi2015}, being a structured semantic representation of natural language text. 

$\mathcal{S}_{06}$=:{\bf Vector Text:}  It first appeared in \cite{Wang2017} in 2017, being the vector sequence of the text. The word2vec (or word embedding) converts text into vector sequence. The most important advantage is that the vector computing technology can be applied into solving algebra problems after problem description can be converted into vector representation. 
  
$\mathcal{S}_{07}$=:{\bf Vector Single Linear Equation:} It first appeared in \cite{Wang2017} in 2017, which is the output of an RNN-based seq2seq model. It directly transforms vector text into a math equation template, being a single equation containing one unknown variable. More algorithms use various transforms to generate linear equations \citep{Liu2019,Kim2020}.  

$\mathcal{S}_{08}$=:{\bf Vector Expression Tree:}  It is an expression tree generated by a seq2seq neural network, which is equivalent to a math expression, first appeared in \cite{Wang2018Translating}, more papers used expression tree \citep{Li2020}.  
 
$\mathcal{S}_{09}$=:{\bf Relation Set:} It was first proposed in \cite{Yu2019} in 2019. It is similar to equation set but it is superior to it in two aspects, as \cite{Yu2019,Yu2023} pointed out, which are: 1) compared with equation set, it is easier to obtain; 2) it is more readable than equation set especially by pupils. 

$\mathcal{S}_{10}$=:{\bf Equation Set:} It is first proposed in \cite{Yu2019} in 2019, being a set of equations, in which equations can be high-order equations. 

$\mathcal{S}_{11}$=: {\bf Vector TextGraph:} It is a sequence of vectors, first appeared in \cite{Li2020}, encoded from text and graph derived from text, where graph derived from text aims to represent the structural relations of entities of the problem. 

$\mathcal{S}_{12}$=:{\bf Vector Equation Graph:} It is a graph from which multiple equations can be deciphered.  {\it Direct acyclic graph} is first appeared in \cite{Cao2021} in 2021. \cite{Cao2021} proposed to use seq2DAG to produce DAG from vector text.   

$\mathcal{S}_{13}$=:{\bf Vector Operation Tree:} It is a tree in which the nodes can be formulas with arbitrary number of arguments, proposed in \cite{Tsai2021}. The objective is to incorporate domain knowledge into problem solving and produce readable solutions via using vector operation tree.  

$\mathcal{S}_{14}$=:{\bf xRelation Set:} {\it xRelation set} was first proposed in \cite{YuSun2022}; an xRelation set is a set of relations and period relations. xRelation set differs from relation set in two aspects. The first one is that it has period relations. The second one is that these period relations form some cliques of relations and each clique forms one function, being a compound object. 

$\mathcal{S}_{15}$=:{\bf xEquation Set:} It is first proposed in \cite{YuSun2022}, being a set of equations, period equations, and function. The equations in the set can be high-order equations; period equations are equations with period constraints; function is a compound object that can interact with equations. 

$\mathcal{S}_{os}$=:{\bf Output Solution:} It is the state appeared in all the solving algorithms. 

We identified 42 transforms from the 28 algorithms, $\Pi$ = \{$\mathcal{t}_{01}$,$\mathcal{t}_{02}$,$\mathcal{t}_{03}$,..., $\mathcal{t}_{40}$\}.
Table \ref{table_all} lists all the 42 transforms and their basic facts. We explain these transforms briefly for concise. The items of the basic facts are the original sources of the transforms, borne year, the occurrence number (the number that the transform used in the 47 algorithms).  We describe the 42 transforms as follows: 

$\mathcal{t}_{01}$=:{\bf Math Expression Generator:}  It generates math expression from problem text, namely transforming from $\mathcal{S}_{00}$ to $\mathcal{S}_{01}$, first appeared in \cite{Hosseini2014} in 2014. 

$\mathcal{t}_{02}$=:{\bf Math Expression Evaluator:}  It evaluates math expression to produce the answer of the given problem, namely transforming from $\mathcal{S}_{01}$ to $\mathcal{S}_{os}$, first appeared in \cite{Hosseini2014} in 2014. 

$\mathcal{t}_{03}$=:{\bf Linear Equations Generator:}  It generates a set of linear equations from problem text, namely transforming from $\mathcal{S}_{00}$ to $\mathcal{S}_{02}$, first appeared in \cite{Kushman2014} in 2014. 
 
$\mathcal{t}_{04}$=:{\bf Solver of Linear Equations:}  It solves a set of linear equations to produce the answer of the given problem, being an application of Gaussian eliminator, namely transforming from $\mathcal{S}_{02}$ to $\mathcal{S}_{os}$, first appeared in \cite{Kushman2014} in 2014.  

$\mathcal{t}_{05}$=:{\bf Expression Tree Producer:}  It is a producer that produces an expression tree, being equivalent to an linear equation, namely transforming from $\mathcal{S}_{00}$ to $\mathcal{S}_{03}$, proposed in \cite{Roy2015} in 2015. 

$\mathcal{t}_{06}$=:{\bf Expression Tree Traversal:}  It produces a math expression by traversing expression tree and replacing vectors with numbers and operands, namely transforming from $\mathcal{S}_{03}$ to $\mathcal{S}_{01}$, appeared in \cite{Roy2015} in 2015.

$\mathcal{t}_{07}$=:{\bf Expression Trees Producer:}  It produces multiple expression trees, namely transforming from $\mathcal{S}_{00}$ to $\mathcal{S}_{04}$, proposed in \cite{Koncel2015} in 2015. This transform enumerates all the possible equation trees by integer linear programming. 

$\mathcal{t}_{08}$=:{\bf Expression Tree Identifier:}  It identifies the wanted expression tree by scoring the likelihood of equation trees by learning discriminative models trained from a small number of word problems and their solutions and selecting the tree with the highest score, namely transforming from $\mathcal{S}_{04}$ to $\mathcal{S}_{03}$, proposed in \cite{Koncel2015} in 2015. 

$\mathcal{t}_{09}$=:{\bf QP Equations Generator:}  It generates a set of linear equations from problem text, namely transforming from $\mathcal{S}_{00}$ to $\mathcal{S}_{02}$, proposed in \cite{Zhou2015} in 2015. This transform improves ${t}_{03}$ by using a QP (quadratic programming) to instantiate the selected template.  

$\mathcal{t}_{10}$=:{\bf DOL Tree Producer:} It converts problem text into DOL trees, namely transforming from $\mathcal{S}_{00}$ to $\mathcal{S}_{05}$ defined in \cite{Shi2015} in 2015. This transform uses a set of grammar rules to build DOL trees from text. 

$\mathcal{t}_{11}$=:{\bf DOL2Expression Generator:}  It generates math expression from DOL trees, namely transforming from $\mathcal{S}_{05}$ to $\mathcal{S}_{02}$, first appeared in \cite{Shi2015} in 2015. 
 
$\mathcal{t}_{12}$=:{\bf Word2Vec Converter:}  It  converts text into vector sequence by a Word2Vec procedure, namely transforming from $\mathcal{S}_{00}$ to $\mathcal{S}_{06}$, proposed in \cite{Wang2017} in 2017. The vector text becomes popular soon in the various types of machine learning based algorithms.  

$\mathcal{t}_{13}$=:{\bf RNN Equations Generator:}  It generates equations from vector text, namely transforming from $\mathcal{S}_{06}$ to $\mathcal{S}_{07}$ by RNN (Recurrent Neural Network), proposed in \cite{Wang2017} in 2017. The gated recurrent units (GRU) and long short-memory (LSTM) cells are used for encoding and decoding, respectively. 

$\mathcal{t}_{14}$=:{\bf Equations Retrieval:}  It generates equations from vector text through similarity matching, namely transforming from $\mathcal{S}_{00}$ to $\mathcal{S}_{01}$, proposed in \cite{Wang2017} in 2017.  

$\mathcal{t}_{15}$=:{\bf V2T Expression Converter:}  It converts vector expression to math expression, namely transforming from $\mathcal{S}_{06}$ to $\mathcal{S}_{01}$, used in \cite{Wang2017} in 2017. 



$\mathcal{t}_{16}$=:{\bf RL Vector Expression Generator:}  It generates vector expression from vector text, namely transforming from $\mathcal{S}_{06}$ to $\mathcal{S}_{07}$ by RL (Reinforcement Learning), proposed in \cite{Huang2018Neural} in 2018. This transform improves ${t}_{17}$ by incorporating copy and alignment mechanism. 

$\mathcal{t}_{17}$=:{\bf DQN Vector Expression Generator:} It generates vector expression from vector text, namely transforming from $\mathcal{S}_{06}$ to $\mathcal{S}_{08}$ by DQN (Deep Q-Network), first appeared in \cite{Wang2018MathDQN} in 2018. 

$\mathcal{t}_{18}$=:{\bf Seq2Tree Generator:} It generates vector tree from vector text, namely transforming from $\mathcal{S}_{06}$ to $\mathcal{S}_{08}$, first appeared in \cite{Liu2019} in 2019. 

$\mathcal{t}_{19}$=:{\bf V2E Tree Converter:}  It converts vector tree to expression tree, namely transforming from $\mathcal{S}_{08}$ to $\mathcal{S}_{01}$, first appeared in \cite{Liu2019} in 2019. 

$\mathcal{t}_{20}$=:{\bf tRNN Seq2Tree Generator:}  It generates vector expression tree from vector text, namely transforming from $\mathcal{S}_{06}$ to $\mathcal{S}_{08}$ by tRNN (template-based Recursive Neural Network), proposed in \cite{Wang2019} in 2019. 

$\mathcal{t}_{21}$=:{\bf Goal-Driven Seq2Tree Producer:}  It produces vector tree from vector text by using goal-driven tree-structured neural model, namely transforming from $\mathcal{S}_{06}$ to $\mathcal{S}_{08}$, proposed in \cite{Xie2019} in 2019. 

$\mathcal{t}_{22}$=:{\bf $S^2$ Relation Extractor:}  It acquires explicit relations from annotated problem text, namely transforming from $\mathcal{S}_{00}$ to $\mathcal{S}_{09}$, proposed in \cite{Yu2019} in 2019. $S^2$ extractor uses a pool of $S^2$ models ($S^2$ stands for syntax-semantics) to extract explicit relations one by one from the annotated text of describing a problem. 

$\mathcal{t}_{23}$=:{\bf Equation Set Producer:}  It produces an equation set, namely transforming from $\mathcal{S}_{09}$ to $\mathcal{S}_{10}$, first appeared in \cite{Yu2019} in 2019, converting a relation set into an equation set by replacing entities with defined variables. 

$\mathcal{t}_{24}$=:{\bf Equation Set Solver:}  It solves equation set to produce the answer of the given problem, namely transforming from $\mathcal{S}_{10}$ to $\mathcal{S}_{os}$, first appeared in \cite{Yu2019} in 2019. 

$\mathcal{t}_{25}$=:{\bf EPT Vector L-Equation Generator:}  It generates vector linear equation from vector text, namely transforming from $\mathcal{S}_{06}$ to $\mathcal{S}_{07}$ by using EPT (Expression-Pointer Transformer), first appeared in \cite{Kim2020} in 2020. This transform uses ``expression'' token and operand-context pointers when generating vector linear equation and adopts a pretrained language model as the encoder.



$\mathcal{t}_{26}$=:{\bf TextGraph Encoder:} It first acquires the graph from text and then encodes text and graph into a vector sequence, namely transforming from $\mathcal{S}_{06}$ to $\mathcal{S}_{11}$, proposed in \cite{Li2020}. 

$\mathcal{t}_{27}$=:{\bf Graph2tree Decoder:} It generates vector tree from the vector sequence of TextGraph, namely transforming from $\mathcal{S}_{11}$ to $\mathcal{S}_{08}$, such as equation tree or expression tree in AP-solving tasks, proposed in \cite{Li2020}. 

$\mathcal{t}_{28}$=:{\bf Seq2DAG Generator:}  It generates direct acyclic graph from vector text, namely transforming from $\mathcal{S}_{06}$ to $\mathcal{S}_{12}$, first appeared in \cite{Cao2021} in 2021. 

$\mathcal{t}_{29}$=:{\bf DAG Traversal:}  It produces a set of equations by traversing direct acyclic graph where equations may be high-order in appearance but they are linear essentially, namely transforming from $\mathcal{S}_{12}$ to $\mathcal{S}_{02}$, first appeared in \cite{Cao2021} in 2021.  

$\mathcal{t}_{30}$=:{\bf HMS Operation Tree Generator:}  It generates seq2seq operation tree from vector text, namely transforming from $\mathcal{S}_{06}$ to $\mathcal{S}_{13}$ by HMS (Hierarchical Math
Solver), first appeared in \cite{Lin2021} in 2021. 


$\mathcal{t}_{31}$=:{\bf V$\circ$S$^2$ Relation Extractor:}  It uses V$\circ$S$^2$ method to extract relations from vector text, namely transforming from $\mathcal{S}_{06}$ to $\mathcal{S}_{09}$ by using vector syntax semantics method, first appeared in \cite{Lyu2021} in 2021 and further developed in \cite{Lyu2023} in 2023. V$\circ$S$^2$ extractor is the improved version of $S^2$ extractor. 

$\mathcal{t}_{32}$=:{\bf Vector Operation Tree Producer:}  It is a seq2seq neural network that produces operation tree in \cite{Tsai2021}, namely transforming from $\mathcal{S}_{06}$ to $\mathcal{S}_{13}$.  

$\mathcal{t}_{33}$=:{\bf Operation Tree Evaluator:}  It is a procedure that converts an operation tree into an expression tree, namely transforming from $\mathcal{S}_{13}$ to $\mathcal{S}_{03}$. The algorithm in \cite{Tsai2021} should have this procedure, but it did not present it.  
 
$\mathcal{t}_{34}$=:{\bf Tree-RNN Generator:} It is a neural network to produce an equation from vector text, namely transforming from $\mathcal{S}_{06}$ to $\mathcal{S}_{02}$, first appeared in \cite{Zaporojets2021} in 2021. This transform scores the generated candidate equations using Tree-RNN (Tree-structured Recursive Neural Network).

$\mathcal{t}_{35}$=:{\bf $S^2P$ Relation Extractor:}  It extracts universal and period relations from text, namely transforming from $\mathcal{S}_{00}$ to $\mathcal{S}_{14}$, first appeared in \cite{YuSun2022} in 2022. 

$\mathcal{t}_{36}$=:{\bf $L^2$ Relation Extractor:}  It acquires period relation from problem diagrams, namely transforming from $\mathcal{S}_{00}$ to $\mathcal{S}_{14}$, first appeared in \cite{YuSun2022} in 2022,  which is designed to acquire the group of period relations from diagrams. This is the first transform that can acquire relations from diagrams for algebra problems. 

$\mathcal{t}_{37}$=:{\bf xEquation Set Producer:}  It is a producer that forms an xequation set, namely transforming from $\mathcal{S}_{14}$ to $\mathcal{S}_{15}$, first appeared in \cite{YuSun2022} in 2022. This transform does two main tasks. The first one is to identifies the appearances of each entity and assign variables to entities. The second one is to state the cliques of period relations and build a function for each clique.  

$\mathcal{t}_{38}$=:{\bf xEquation Set Solver:}  It solves xequation set to produce the answer of the given problem, namely transforming from $\mathcal{S}_{15}$ to $\mathcal{S}_{os}$, first appeared in \cite{YuSun2022} in 2022.  
 
$\mathcal{t}_{39}$=:{\bf HGEN Vector Expression Generator:}  It is a neural network to generate vector expression from vector text by HGEN (hierarchical heterogeneous graph encoding), namely transforming from $\mathcal{S}_{06}$ to $\mathcal{S}_{08}$, appeared in \cite{Zhang2022} in 2022. 

$\mathcal{t}_{40}$=:{\bf Neural Network Miner:}  It is a miner that mines implicit relation from vector text, namely transforming from $\mathcal{S}_{06}$ to $\mathcal{S}_{09}$, proposed in \cite{Yu2023} in 2023. 

$\mathcal{t}_{41}$=:{\bf DI Relation Extractor:}  It extracts DI (Deep Implicit) relations from problem text, namely transforming from $\mathcal{S}_{00}$ to $\mathcal{S}_{09}$, proposed in \cite{Meng2023} in 2023.  This transform is designed to acquire the deep implict relations by using qualia syntax-semantic model, extending the scope of problems that solving algorithms can solve.    

$\mathcal{t}_{42}$=:{\bf GD Relation Extractor:}  It extracts relations from geometry diagram, the partial description of the given problem, namely transforming from $\mathcal{S}_{00}$ to $\mathcal{S}_{09}$, proposed in \cite{Huang2023} in 2023.  This transform is designed to acquire the relations from geometry diagrams by using syntax-semantics diagram understanding, extending the scope of problems that solving algorithms can solve.  

\section*{Appendix B: State Transform Graphs for All 47 Algorithms}\label{appendix2}

\begin{figure}[!h]
    \thispagestyle{empty}
    \tikzstyle{state} = [rectangle, thick,rounded corners, minimum width = 0.74cm, minimum height=0.5cm,text centered, draw = black,font=\fontsize{6}{6}\selectfont]
    \tikzstyle{arrow} = [->,>=stealth]

    \label{a01}
\end{figure}
\vspace{-0.6cm}

\begin{figure}[!h]
    \thispagestyle{empty}
    \tikzstyle{state} = [rectangle, thick,rounded corners, minimum width = 0.74cm, minimum height=0.5cm,text centered, draw = black,font=\fontsize{6}{6}\selectfont]
    \tikzstyle{arrow} = [->,>=stealth]
    %
    \label{a02}
\end{figure}
\vspace{-0.6cm}

\begin{figure}[!h]
    \thispagestyle{empty}
    \tikzstyle{state} = [rectangle, thick,rounded corners, minimum width = 0.74cm, minimum height=0.5cm,text centered, draw = black,font=\fontsize{6}{6}\selectfont]
    \tikzstyle{arrow} = [->,>=stealth]
    %
    \label{a03}
\end{figure}
\vspace{-0.6cm}

        \begin{figure}[!h]
    \thispagestyle{empty}
    \tikzstyle{state} = [rectangle, thick,rounded corners, minimum width = 0.74cm, minimum height=0.5cm,text centered, draw = black,font=\fontsize{6}{6}\selectfont]
    \tikzstyle{arrow} = [->,>=stealth]
    %
    \label{a04}
\end{figure}
\vspace{-0.6cm}

\begin{figure}[!h]
    \thispagestyle{empty}
    \tikzstyle{state} = [rectangle, thick,rounded corners, minimum width = 0.74cm, minimum height=0.5cm,text centered, draw = black,font=\fontsize{6}{6}\selectfont]
    \tikzstyle{arrow} = [->,>=stealth]
    %
    \label{a05}
\end{figure}
\vspace{-0.6cm}

\begin{figure}[!h]
    \thispagestyle{empty}
    \tikzstyle{state} = [rectangle, thick,rounded corners, minimum width = 0.74cm, minimum height=0.5cm,text centered, draw = black,font=\fontsize{6}{6}\selectfont]
    \tikzstyle{arrow} = [->,>=stealth]
    %
    \label{a06}
\end{figure}
\vspace{-0.6cm}

\begin{figure}[!h]
    \thispagestyle{empty}
    \tikzstyle{state} = [rectangle, thick,rounded corners, minimum width = 0.74cm, minimum height=0.5cm,text centered, draw = black,font=\fontsize{6}{6}\selectfont]
    \tikzstyle{arrow} = [->,>=stealth]
    %
    \label{a07}
\end{figure}
\vspace{-0.6cm}

\begin{figure}[!h]
    \thispagestyle{empty}
    \tikzstyle{state} = [rectangle, thick,rounded corners, minimum width = 0.74cm, minimum height=0.5cm,text centered, draw = black,font=\fontsize{6}{6}\selectfont]
    \tikzstyle{arrow} = [->,>=stealth]
    %
    \label{a08}
\end{figure}
\vspace{-0.6cm}

    \begin{figure}[!h]
    \thispagestyle{empty}
    \tikzstyle{state} = [rectangle, thick,rounded corners, minimum width = 0.74cm, minimum height=0.5cm,text centered, draw = black,font=\fontsize{6}{6}\selectfont]
    \tikzstyle{arrow} = [->,>=stealth]
    %
    \label{a09}
\end{figure}
\vspace{-0.6cm}

\begin{figure}[!h]
    \thispagestyle{empty}
    \tikzstyle{state} = [rectangle, thick,rounded corners, minimum width = 0.74cm, minimum height=0.5cm,text centered, draw = black,font=\fontsize{6}{6}\selectfont]
    \tikzstyle{arrow} = [->,>=stealth]
    %
    \label{a10}
\end{figure}
\vspace{-0.6cm}

\begin{figure}[!h]
    \thispagestyle{empty}
    \tikzstyle{state} = [rectangle, thick,rounded corners, minimum width = 0.74cm, minimum height=0.5cm,text centered, draw = black,font=\fontsize{6}{6}\selectfont]
    \tikzstyle{arrow} = [->,>=stealth]
    %
    \label{a11}
\end{figure}
\vspace{-0.6cm}

\begin{figure}[!h]
    \thispagestyle{empty}
    \tikzstyle{state} = [rectangle, thick,rounded corners, minimum width = 0.74cm, minimum height=0.5cm,text centered, draw = black,font=\fontsize{6}{6}\selectfont]
    \tikzstyle{arrow} = [->,>=stealth]
    %
    \label{a12}
\end{figure}
\vspace{-0.6cm}

\begin{figure}[!h]
    \thispagestyle{empty}
    \tikzstyle{state} = [rectangle, thick,rounded corners, minimum width = 0.74cm, minimum height=0.5cm,text centered, draw = black,font=\fontsize{6}{6}\selectfont]
    \tikzstyle{arrow} = [->,>=stealth]
    %
    \label{a13}
\end{figure}
\vspace{-0.6cm}

\begin{figure}[!h]
    \thispagestyle{empty}
    \tikzstyle{state} = [rectangle, thick,rounded corners, minimum width = 0.74cm, minimum height=0.5cm,text centered, draw = black,font=\fontsize{6}{6}\selectfont]
    \tikzstyle{arrow} = [->,>=stealth]
    %
    \label{a14}
\end{figure}
\vspace{-0.6cm}

\begin{figure}[!h]
    \thispagestyle{empty}
    \tikzstyle{state} = [rectangle, thick,rounded corners, minimum width = 0.74cm, minimum height=0.5cm,text centered, draw = black,font=\fontsize{6}{6}\selectfont]
    \tikzstyle{arrow} = [->,>=stealth]
    %
    \label{a15}
\end{figure} 
\vspace{-0.6cm}

\begin{figure}[!h]
    \thispagestyle{empty}
    \tikzstyle{state} = [rectangle, thick,rounded corners, minimum width = 0.74cm, minimum height=0.5cm,text centered, draw = black,font=\fontsize{6}{6}\selectfont]
    \tikzstyle{arrow} = [->,>=stealth]
    %
    \label{a17}
\end{figure} 
\vspace{-0.6cm}
\begin{figure}[!h]
    \thispagestyle{empty}
    \tikzstyle{state} = [rectangle, thick,rounded corners, minimum width = 0.74cm, minimum height=0.5cm,text centered, draw = black,font=\fontsize{6}{6}\selectfont]
    \tikzstyle{arrow} = [->,>=stealth]
    %
    \label{a18}
\end{figure} 
\vspace{-0.6cm}

\begin{figure}[!h]
    \thispagestyle{empty}
    \tikzstyle{state} = [rectangle, thick,rounded corners, minimum width = 0.74cm, minimum height=0.5cm,text centered, draw = black,font=\fontsize{6}{6}\selectfont]
    \tikzstyle{arrow} = [->,>=stealth]
    %
\end{figure} 
\vspace{-0.6cm}
\begin{figure}[!h]
    \thispagestyle{empty}
    \tikzstyle{state} = [rectangle, thick,rounded corners, minimum width = 0.74cm, minimum height=0.5cm,text centered, draw = black,font=\fontsize{6}{6}\selectfont]
    \tikzstyle{arrow} = [->,>=stealth]
    %
    \label{a21}
\end{figure} 
\vspace{-0.6cm}

\begin{figure}[!h]
    \thispagestyle{empty}
    \tikzstyle{state} = [rectangle, thick,rounded corners, minimum width = 0.74cm, minimum height=0.5cm,text centered, draw = black,font=\fontsize{6}{6}\selectfont]
    \tikzstyle{arrow} = [->,>=stealth]
    %
    \label{a23}
\end{figure} 
\vspace{-0.6cm}      

\begin{figure}[!h]
    \thispagestyle{empty}
    \tikzstyle{state} = [rectangle, thick,rounded corners, minimum width = 0.74cm, minimum height=0.5cm,text centered, draw = black,font=\fontsize{6}{6}\selectfont]
    \tikzstyle{arrow} = [->,>=stealth]
    %
    \label{a25}
\end{figure} 
\vspace{-0.6cm}
\begin{figure}[!h]
    \thispagestyle{empty}
    \tikzstyle{state} = [rectangle, thick,rounded corners, minimum width = 0.74cm, minimum height=0.5cm,text centered, draw = black,font=\fontsize{6}{6}\selectfont]
    \tikzstyle{arrow} = [->,>=stealth]
    %
    \label{a26}
\end{figure} 
\vspace{-0.6cm}
\begin{figure}[!h]
    \thispagestyle{empty}
    \tikzstyle{state} = [rectangle, thick,rounded corners, minimum width = 0.74cm, minimum height=0.5cm,text centered, draw = black,font=\fontsize{6}{6}\selectfont]
    \tikzstyle{arrow} = [->,>=stealth]
    %
    \label{a27}
\end{figure} 
\vspace{-0.6cm}

\begin{figure}[!h]
    \thispagestyle{empty}
    \tikzstyle{state} = [rectangle, thick,rounded corners, minimum width = 0.74cm, minimum height=0.5cm,text centered, draw = black,font=\fontsize{6}{6}\selectfont]
    \tikzstyle{arrow} = [->,>=stealth]
    %
    \label{a30}
\end{figure} 
\vspace{-0.6cm}
        
\begin{figure}[!h]
    \thispagestyle{empty}
    \tikzstyle{state} = [rectangle, thick,rounded corners, minimum width = 0.74cm, minimum height=0.5cm,text centered, draw = black,font=\fontsize{6}{6}\selectfont]
    \tikzstyle{arrow} = [->,>=stealth]
    %
    \label{a31}
\end{figure} 
\vspace{-0.6cm}
\begin{figure}[!h]
    \thispagestyle{empty}
    \tikzstyle{state} = [rectangle, thick,rounded corners, minimum width = 0.74cm, minimum height=0.5cm,text centered, draw = black,font=\fontsize{6}{6}\selectfont]
    \tikzstyle{arrow} = [->,>=stealth]
    %
    \label{a32}
\end{figure} 
\vspace{-0.6cm}

\begin{figure}[!h]
    \thispagestyle{empty}
    \tikzstyle{state} = [rectangle, thick,rounded corners, minimum width = 0.74cm, minimum height=0.5cm,text centered, draw = black,font=\fontsize{6}{6}\selectfont]
    \tikzstyle{arrow} = [->,>=stealth]
    %
    \label{a33}
\end{figure} 
\vspace{-0.6cm}

\begin{figure}[!h]
    \thispagestyle{empty}
    \tikzstyle{state} = [rectangle, thick,rounded corners, minimum width = 0.74cm, minimum height=0.5cm,text centered, draw = black,font=\fontsize{6}{6}\selectfont]
    \tikzstyle{arrow} = [->,>=stealth]
    %
    \label{a34}
\end{figure} 
\vspace{-0.6cm}
\begin{figure}[!h]
    \thispagestyle{empty}
    \tikzstyle{state} = [rectangle, thick,rounded corners, minimum width = 0.74cm, minimum height=0.5cm,text centered, draw = black,font=\fontsize{6}{6}\selectfont]
    \tikzstyle{arrow} = [->,>=stealth]
    %
    \label{a35}
\end{figure} 
\vspace{-0.6cm}

\begin{figure}[!h]
    \thispagestyle{empty}
    \tikzstyle{state} = [rectangle, thick,rounded corners, minimum width = 0.74cm, minimum height=0.5cm,text centered, draw = black,font=\fontsize{6}{6}\selectfont]
    \tikzstyle{arrow} = [->,>=stealth]
    %
    \label{a36}
\end{figure} 
\vspace{-0.6cm}

\begin{figure}[!h]
    \thispagestyle{empty}
    \tikzstyle{state} = [rectangle, thick,rounded corners, minimum width = 0.74cm, minimum height=0.5cm,text centered, draw = black,font=\fontsize{6}{6}\selectfont]
    \tikzstyle{arrow} = [->,>=stealth]
    %
    \label{a38}
\end{figure} 
\vspace{-0.6cm}

\begin{figure}[!h]
    \thispagestyle{empty}
    \tikzstyle{state} = [rectangle, thick,rounded corners, minimum width = 0.74cm, minimum height=0.5cm,text centered, draw = black,font=\fontsize{6}{6}\selectfont]
    \tikzstyle{arrow} = [->,>=stealth]
    %
    \label{a39}
\end{figure} 
\vspace{-0.6cm}
\begin{figure}[!h]
    \thispagestyle{empty}
    \tikzstyle{state} = [rectangle, thick,rounded corners, minimum width = 0.74cm, minimum height=0.5cm,text centered, draw = black,font=\fontsize{6}{6}\selectfont]
    \tikzstyle{arrow} = [->,>=stealth]
    %
    \label{a47}
\end{figure}






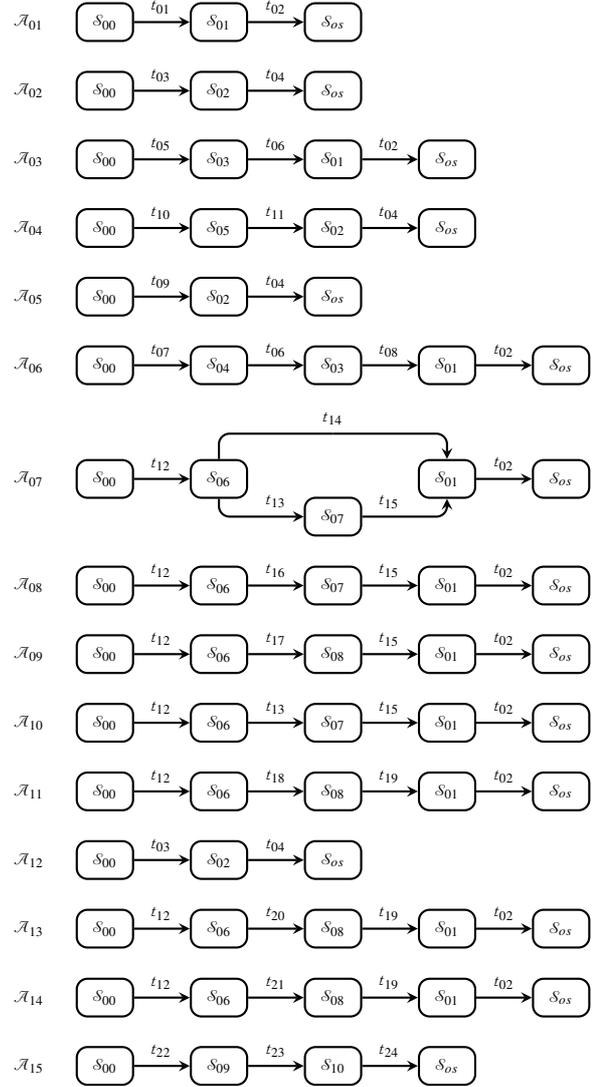
\end{document}